\documentclass[11pt,tightenlines,aps,prd,notitlepage,nofootinbib,superscriptaddress]{revtex4-1}
\usepackage[utf8]{inputenc}
\usepackage[table,dvipsnames]{ xcolor}

\usepackage{multirow}
\usepackage{caption}
\usepackage{subcaption}
\usepackage{hyperref}
\usepackage{graphicx}
\usepackage{mathtools}
\usepackage{amsmath,amssymb,amsfonts,amsthm,latexsym,stmaryrd,physics,mathrsfs}
\numberwithin{figure}{section}
\allowdisplaybreaks[4]
\usepackage{marginnote}
\usepackage{color}
\usepackage{bm}
\usepackage{float}
\usepackage{nicefrac}
\usepackage{microtype} 
\usepackage{booktabs}
\usepackage{verbatim}
\usepackage{setspace}
\usepackage[T1]{fontenc}
\usepackage[utf8]{inputenc}
\usepackage{stfloats}
\usepackage{diagbox}
\usepackage{dsfont}
\usepackage{tikz}
\usepackage{geometry}

\hypersetup{
	colorlinks=true
,urlcolor=blue
,anchorcolor=blue
,citecolor=blue
,filecolor=blue
,linkcolor=blue
,menucolor=blue
,pagecolor=blue
,linktocpage=true
,pdfproducer=medialab
}

\def\beq{\begin{equation}}
\def\eeq{\end{equation}}
\newcommand{\bea}{\begin{eqnarray}}
\newcommand{\eea}{\end{eqnarray}}
\def\bi{\begin{itemize}}
\def\ei{\end{itemize}}
\def\ba{\begin{array}}
\def\ea{\end{array}}
\def\bfig{\begin{figure}}
\def\efig{\end{figure}}

\newtheorem{theorem}{Theorem}[section]

\def\be{\begin{eqnarray}}
\def\ee{\end{eqnarray}}

\newcommand\sjs[6]{\left\{ \begin{array}{ccc}
    #1 & #2 & #3\\
    #4 & #5 & #6\\
\end{array}\right\}}

\usepackage{geometry}
\begin{document}

\title{Bridging Quantum and Semiclassical Volume: A Numerical Study of Coherent State Matrix Elements in Loop Quantum Gravity
}

\author{Haida Li}
\email{eqwaplay@scut.edu.cn}
\affiliation{School of Physics and Optoelectronics, South China University of Technology, Guangzhou 510641, China}
\affiliation{Institute for Theoretical Sciences and Department of Physics, Westlake University, Hangzhou 310024, Zhejiang, China}

\author{Hongguang Liu} 
\email{Corresponding author: liuhongguang@westlake.edu.cn}
\affiliation{Institute for Theoretical Sciences and Department of Physics, Westlake University, Hangzhou 310024, Zhejiang, China}

\begin{abstract}

In Loop Quantum Gravity, the quantum action of the volume operator is crucial in understanding quantum dynamics. In this work, we implement a generalized numerical algorithm that can compute the quantum action of the volume operator on a broad class of gauge-variant and gauge-invariant spin-network states. This algorithm is later used to calculate the coherent state expectation value and coherent state matrix elements of the volume operator. By comparing the results generated by our numerical model with the analytical results in various scenarios at the near-semiclassical region, not only is our numerical model validated with high accuracy, but it also provides a complete picture of how the full quantum action of the volume operator connects with its semiclassical approximations. We further find that the maximal eigenvalue approaches the classical polyhedral volume in the semiclassical regime. For irregular geometries, we also observe that the relative volume magnitudes can change in the deep quantum regime.
\end{abstract}

\maketitle

\begin{spacing}{-0.1}
    \tableofcontents
\end{spacing}

\section{Introduction}

Loop Quantum Gravity (LQG) is a background-independent approach to quantum gravity \cite{Thiemann:2007pyv,Rovelli:2014ssa,Ashtekar:2017yom,Perez:2012wv}.
Its canonical approach is founded on the Hamiltonian formulation of General Relativity, where gravity is initially described as a constrained system governed by the spatial diffeomorphism and Hamiltonian constraints. Quantization is subsequently performed on the holonomy-flux algebra \cite{Ashtekar:1996eg,Ashtekar:2004eh}. A consistent framework to represent these constraints as quantum operators on a kinematical Hilbert space $\mathcal{H}_{kin}$ has been established \cite{Thiemann:1996aw,Thiemann:1996av}, leading to the quantum constraint equations and the physical Hilbert space $\mathcal{H}_{phy}$. While extracting physical predictions from the full 4-dimensional theory remains extremely challenging, significant progress has been made in symmetry-reduced models. Loop Quantum Cosmology (LQC) \cite{Bojowald:2001xe,Ashtekar:2006rx,Banerjee:2011qu} and the study of spherically symmetric black holes \cite{Ashtekar:2005qt,Gambini:2013ooa,Ashtekar:2018lag}, where symmetric conditions are imposed classically before quantization, have yielded profound results. Most notably, the effective dynamics in these models resolve classical singularities by replacing them with quantum bounces \cite{Ashtekar:2008zu,Taveras:2008ke,Ashtekar:2011ni,Agullo:2016tjh}.

At the heart of the quantum dynamics in full LQG lies the volume operator. It is an indispensable building block not only in the standard Dirac quantization program for regularizing and solving the Hamiltonian constraint \cite{Thiemann:1996aw}, but also in the deparametrization scheme using the reduced phase space approach \cite{giesel2010algebraic,thiemann2006reduced,Brown:1994py,Husain:2011tk,Li:2022dei}. In the latter, classical deparametrization is achieved by coupling gravity to external reference matter fields, constructing gauge-invariant Dirac observables, and generating a physical, non-vanishing Hamiltonian.
The regularization and quantization of this physical Hamiltonian heavily involve the volume operator \cite{Thiemann:2007pyv}. Furthermore, within the broader context of Algebraic Quantum Gravity (AQG) \cite{giesel2010algebraic}, the properties of the volume operator are fundamentally essential for defining consistent semi-classical perturbative expansions. 

In this work, we employ the volume operator proposed by Ashtekar and Lewandowski (AL) \cite{Ashtekar:1997fb}, and the numerical methodologies developed here for our numerical package \cite{LQG-Volume} are equally applicable to the Rovelli-Smolin proposal \cite{Rovelli:1994ge}. While the explicit analytical formula for the square of AL volume operator was elegantly simplified by Brunnemann and Thiemann \cite{Brunnemann:2004xi}, rendering its action more tractable, a fundamental obstacle remains.

The definition of the volume operator involves taking the square root of a complicated operator composed of flux operators. Analytically computing this square root is notoriously difficult \cite{Brunnemann:2007ca} and has historically been restricted to the simplest cases, such as its action on 4-valent vertices \cite{DePietri:1996tvo, Brunneman:2007as, Bianchi:2010gc,Zhang:2019dgi}.

The primary objective of this work is to investigate the quantum properties of the volume operator by introducing a numerical algorithm that accurately converges to analytical expansions in the semi-classical regime, paving the way for the study of full LQG dynamics. Our approach bypasses the analytical square root bottleneck: we first construct the matrix elements of the operator inside the square root in the spin-network basis, and then compute the square root numerically by diagonalizing the matrix in the eigenbasis of the volume operator. 

However, constructing a powerful numerical algorithm immediately raises the question of validation, particularly when exploring the semi-classical regime. To bridge the deep quantum regime of spin networks and semiclassical asymptotics, coherent states must be employed. The construction of heat kernel complexifier coherent states—initially inspired by weave states \cite{PhysRevLett.69.237}—has been extensively developed for compact Lie groups \cite{HALL1994103} and adapted for canonical LQG \cite{Thiemann:2000bw,Thiemann:2000ca,Thiemann:2000bx,Thiemann:2000by,Sahlmann:2001nv,Thiemann:2002vj,Bahr:2007xa,Bianchi:2010gc}. These coherent states inherently capture semi-classical spatial geometries and serve as the foundation for modern techniques such as the coherent state path integral method \cite{Han:2019vpw,Han:2019feb,Han:2020iwk,Han:2020chr,Bodendorfer:2020ovt,Han:2020uhb,Han:2021cwb}. 

Crucially, previously known asymptotic properties of the volume operator on coherent states relied heavily on operator expansion techniques developed by Giesel and Thiemann \cite{Giesel:2006um}. However, their perturbative expansion was predominantly formulated using expectation values and suffers from poor convergence when applied directly to matrix elements. To the best of our knowledge, this work presents the first accurate numerical computation of the coherent state matrix elements of the volume operator from the deep quantum regime to the near semiclassical regime. In a companion paper \cite{Liliu:202603A}, we propose a novel analytical expansion method tailored specifically for matrix elements and demonstrate improved convergence.

More fundamentally, this work constructs a combined framework of numerics and analytical expansions that spans from the semi-classical regime to the deep quantum regime. In principle, this novel framework has the capability to connect all semi-classical asymptotic calculations with deep quantum numerical computations, providing an instrumental pathway for studying the full quantum dynamics of LQG.

Moreover, we show that the maximal eigenvalue of the discrete volume operator approaches the classical volume of the corresponding polyhedron, with the associated coherent-state overlap becoming increasingly concentrated on the maximal-volume eigenstate, especially for highly symmetric networks. We also find that, in the deep quantum regime, the relative magnitudes of volumes associated with different geometries can change, with strongly deformed configurations sometimes exceeding more symmetric ones.

The structure of this paper is as follows: In Section 2, we briefly review the definition of the volume operator in both the spin-network and coherent state representations. Section 3 introduces our core numerical algorithm, which consists of four main components: computing the transformation matrix between representations, calculating the exact action of the volume operator, performing the summation over the spin-network basis to obtain coherent state expectation values as well as matrix elements, and validating the algorithm against semiclassical asymptotics. Section 4 presents our numerical results across several geometrically significant scenarios, including gauge-variant 3-bridges, regular and irregular 4-bridges (corresponding to classical tetrahedra), and gauge-variant 6-valent vertices. As we will demonstrate, our computations provide good accuracy in verifying the $\hbar^n$-order asymptotics of the volume operator matrix elements. Finally, Section 5 concludes with a summary and a discussion of future applications for studying the full dynamics.

\section{Volume Operator in Canonical LQG}

In this section, we review the foundational results regarding the definition and properties of the volume operator in canonical Loop Quantum Gravity (LQG). We begin with the formal definition of the volume operator and the construction of both gauge-variant and gauge-invariant spin-network states using $3N$-$j$ symbols, followed by the explicit closed-form formula for the matrix elements of the volume operator. Subsequently, we introduce the framework of heat kernel coherent states—encompassing both gauge-variant and gauge-invariant cases—and detail the action of the volume operator within this representation. Finally, we establish a novel semi-classical expansion scheme specifically tailored for the coherent state matrix elements of the volume operator.

\subsection{Volume Operator acting on gauge-variant and gauge-invariant spin-network states}

In the kinematical framework of LQG, quantum states are initially described by $SU(2)$ cylindrical functions $f_{\Gamma}$ defined on graphs $\Gamma$. A graph $\Gamma$ consists of a collection $E(\Gamma)$ of oriented edges $e$, intersecting at their respective endpoints, which form the set of vertices $V(\Gamma)$. These cylindrical functions depend on $SU(2)$ group elements $h_e$, physically representing the holonomies of the Ashtekar-Barbero connection along the edges $e\in E(\Gamma)$. By invoking the Peter-Weyl theorem, one can decompose the space of square-integrable functions on $SU(2)^{E(\Gamma)}$, naturally transitioning to the spin-network basis. In this representation, the states are fundamentally labeled by $SU(2)$ spin representations $j_e$ on the edges and intertwiners $\iota_v$ at the vertices.

Following the standard regularization techniques established in LQG, the volume operator $\widehat{V}(R)_{\Gamma}$ corresponding to a spatial region $R$ is defined as \cite{Thiemann:1996au,Brunnemann:2004xi}:
\begin{equation}\label{VolumeC}
    \widehat{V}(R)_{\Gamma}=\int_R d^3p\sqrt{\mathrm{det} q(p)_{\Gamma}}=\int_R d^3p\,\widehat{V}(p)_{\Gamma},
\end{equation}
where its localized action at the vertices is given by:
\begin{equation}\label{vdef}
    \begin{split}
    \widehat{V}(p)_{\Gamma}&=\sum\limits_{v\in V(\Gamma)}\delta^3(p,v)\widehat{V}_{v,\Gamma},\\
    \widehat{V}_{v,\Gamma}&=\sqrt{\widehat{Q}_v}:=\sqrt{\left|\sum\limits_{I<J<K}\epsilon(e_I,e_J,e_K)\widehat{Q}_{IJK}\right|},\\
    \widehat{Q}_{IJK}&\equiv \frac{1}{64}\epsilon_{ijk}\widehat{p}^i_I\widehat{p}^j_J\widehat{p}^k_K \equiv \frac{(it)^3}{64}\epsilon_{ijk}\widehat{J}^i_I\widehat{J}^j_J\widehat{J}^k_K.
    \end{split}
\end{equation}
The numerical coefficient $\frac{1}{64}$ in the expression for $\widehat{Q}_{IJK}$ guarantees the proper correspondence with the classical volume (see, e.g., \cite{Han:2019vpw}, eqn. (7.1)). The flux operator is defined as $\widehat{p}^b_{I} = t \widehat{J}^b_I = i t R^b_I/2$, where $R^b_I$ denotes the $SU(2)$ self-adjoint right-invariant vector fields acting on the edge $e_I$. Here, the index $b \in \{1,2,3\}$ labels the internal $SU(2)$ algebra, while $I$ enumerates the edges converging at the vertex $v$. The semi-classical parameter is defined as $t = l_P^2/a^2$, where $l_P = \sqrt{\hbar G}$ is the Planck length and $a$ is a length scale introduced to render the flux operator $\widehat{p}^i_I$ dimensionless. Crucially, $t$ also serves as the fundamental semi-classical parameter (functioning as a diffusion time in the heat kernel) for the construction of coherent states \cite{Thiemann:2000bw,Thiemann:2000ca,Thiemann:2000bx,Thiemann:2000by,Sahlmann:2001nv,Thiemann:2002vj}.

For any given edge $e_I$, the generators $\widehat{J}^b_I$ precisely satisfy the standard angular momentum commutation relations of quantum mechanics. The summation in the volume operator is performed over all vertices $v \in V(\Gamma)$ and, subsequently, over all ordered triples $(e_I,e_J,e_K)$ of distinct edges adjacent to $v$. The geometric factor $\epsilon(e_I,e_J,e_K)$ represents the sign of the scalar triple product of the tangent vectors of the respective edges evaluated at the vertex $v$.

The matrix elements of the volume operator acting on spin-network states can be systematically calculated by decomposing the states using $3N$-$j$ symbols at each vertex $v$. A generic local state at an $N$-valent vertex is denoted by:
\begin{equation}\label{3nj1}
    |\vec{\iota},\vec{j}, M\rangle,
\end{equation}
where $\vec{j}=(j_1,j_2,\dots ,j_N)$ characterizes the $SU(2)$ spin representations on all incident edges, and $M$ is the total magnetic quantum number. $\vec{\iota}=(\iota_1,\iota_2,\dots ,\iota_N)$ represents the intertwiner, which is structurally defined by the sequential recoupling of angular momenta. For instance, $\iota_1=j_1$, $\iota_2$ is the intermediate coupling between $j_1$ and $j_2$ (governed by the standard selection rules $|j_1-j_2|\le \iota_2 \le |j_1+j_2|$), and iteratively, $\iota_N$ is the final coupling between $\iota_{N-1}$ and $j_N$.

Local $SU(2)$ gauge invariance is strictly enforced by requiring the total angular momentum at the vertex to vanish, i.e., $J=\iota_N=0$, along with $M=0$. This constraint deterministically fixes $\iota_{N-1}=j_N$, as it is the unique configuration capable of yielding a scalar singlet upon coupling with $j_N$. Consequently, a pure gauge-invariant state at an $N$-valent vertex is concisely denoted as:
\begin{equation}\label{3nj12}
    |\vec{\iota},\vec{j}\rangle := |\vec{\iota}=(\iota_1=j_1,\dots,\iota_{N-1}=j_N,\iota_N=0),\vec{j},M=0\rangle.
\end{equation}
Extending this to an arbitrary graph $\Gamma$ comprising multiple vertices, the global gauge-variant spin-network state is encapsulated by the notation $|\Gamma,\{\vec{\iota}\},\{\vec{j}\},\{M\}\rangle$. Here, $\{\vec{\iota}\}$ compiles the intertwiner channels across all vertices, $\{\vec{j}\}$ designates the spin representations traversing the edges, and $\{M\}$ collects the total magnetic quantum numbers at the vertices. Imposing global gauge invariance annihilates all local magnetic degrees of freedom ($M=0$ identically at every vertex) and prohibits the existence of open edges. Thus, the corresponding global gauge-invariant spin-network state simplifies to $|\Gamma,\{\vec{\iota}\},\{\vec{j}\}\rangle$.

Gauge-variant graphs are analogously formulated utilizing $3N$-$j$ symbols without imposing restrictions on $J=\iota_N$ and $M$, permitting $-\iota_N\leq M\leq \iota_N$. In such cases, the magnetic indices of the $SU(2)$ representations on internal edges are fully contracted, leaving open degrees of freedom exclusively at uncontracted open edges. Based on this robust algebraic formulation, the matrix elements of the fundamental spatial geometry operator $\widehat{Q}_{IJK}$ can be explicitly computed as \cite{Brunnemann:2004xi}:
\begin{equation}\label{vol1}
    \begin{split}
        &\langle\vec{\iota},\vec{j}|\widehat{Q}_{IJK}|\vec{\iota}',\vec{j}\rangle=\\
        &\frac{it^3}{256}(-1)^{j_K+j_I+\iota_{I-1}+\iota_K}(-1)^{\iota_I-\iota_I'}(-1)^{\sum\limits^{J-1}_{n=I+1}j_n}(-1)^{-\sum\limits^{K-1}_{p=J+1}j_P} \times\\
        &\times X(j_I,j_J)^{\frac{1}{2}}X(j_J,j_K)^{\frac{1}{2}}\sqrt{(2\iota_I+1)(2\iota_I'+1)}\sqrt{(2\iota_J+1)(2\iota_J'+1)}\times\\
        &\times \sjs{\iota_{I-1}}{j_I}{\iota_I}{1}{\iota_I'}{j_I}\left[\prod\limits^{J-1}_{n=I+1}\sqrt{(2\iota_n'+1)(2\iota_n+1)}(-1)^{\iota_{n-1}'+\iota_{n-1}+1}\sjs{j_n}{\iota_{n-1}'}{\iota_n'}{1}{\iota_n}{\iota_{n-1}}\right]\times\\
        &\times\left[\prod\limits^{K-1}_{n=J+1}\sqrt{(2\iota_n'+1)(2\iota_n+1)}(-1)^{\iota_{n-1}'+\iota_{n-1}+1}\sjs{j_n}{\iota_{n-1}'}{\iota_n'}{1}{\iota_n}{\iota_{n-1}}\right]\sjs{\iota_K}{j_K}{\iota_{K-1}}{1}{\iota_{K-1}'}{j_K}\times\\
        &\times\left[(-1)^{\iota_J'+\iota_{J-1}'}\sjs{\iota_J}{j_J}{\iota_{J-1}'}{1}{\iota_{J-1}}{j_J}\sjs{\iota_{J-1}'}{j_J}{\iota_J'}{1}{\iota_J}{j_J}-(-1)^{\iota_J+\iota_{J-1}}\sjs{\iota_J'}{j_J}{\iota_{J-1}'}{1}{\iota_{J-1}}{j_J}\sjs{\iota_{J-1}}{j_J}{\iota_J'}{1}{\iota_J}{j_J}\right]\times\\
        &\times\prod\limits^{I-1}_{n=2}\delta_{\iota_n\iota_n'}\prod\limits^N_{n=K}\delta_{\iota_n\iota_n'},
    \end{split}
\end{equation}
where $X(j_1,j_2)=2j_1(2j_1+1)(2j_1+2)2j_2(2j_2+1)(2j_2+2)$, and the symbols $\sjs{a}{b}{c}{1}{d}{e}$ denote standard Wigner $6j$-symbols. The indices are bounded by $I\ge 2$ and $I\le J\le K$; complementary permutations for arbitrary positive integers $I, J, K$ are detailed in \cite{Brunnemann:2004xi}. Eq. (\ref{vol1}) distinctly demonstrates that the evaluation of the volume operator factorizes ultra-locally, isolating its action purely to individual vertices. The resulting matrix elements are solely governed by the internal intertwiner channels $\vec{\iota}$ and the specific spin labels $\vec{j}$ incident to that vertex, rendering $\widehat{Q}_{IJK}$ explicitly computable for arbitrary spin-network configurations.

\subsection{Coherent States in LQG}

To probe the semi-classical asymptotics of the volume operator, we evaluate its action upon the heat kernel complexifier coherent states pioneered by Thiemann \cite{Thiemann:2000bw,Thiemann:2000ca,Thiemann:2000bx,Thiemann:2000by,Sahlmann:2001nv,Thiemann:2002vj}. These states serve as the paramount tool in canonical LQG for extracting continuum macroscopic physics from the underlying discrete quantum geometry. In this framework, we systematically employ both gauge-variant and gauge-invariant coherent states, projecting them onto the exact intertwiner basis of a specified graph. This projection mechanism is what ultimately grants us the analytical leverage to map out the semi-classical transition of the volume operator.

We start by introducing coherent states $|g(e)\rangle$ associated with a single edge $e\in E(\Gamma)$ written as a function of $h(e)\in SU(2)$ \cite{Thiemann:2000ca}:
\begin{equation}\label{coherent1}
    \psi^{t}_{g(e)}(h(e))\equiv\langle h(e)|g(e)\rangle=\sum\limits_{j_e\in\mathbb{Z}_+/2\cup\{0\}}d_{j_e}e^{-t\lambda_{j_e}/2}\chi_{j_e}(g(e)h^{-1}(e)),
\end{equation}
where $\chi_j$ is the SU(2) character at spin-$j$, $d_j:=2j+1$, and $\lambda_j:=j(j+1)$. $t$ is the semiclassical parameter which goes to zero in the semi-classical region. The coherent state label $g(e)\in SL(2,\mathbb{C})$ is the complexified holonomy parametrized in the following way:
\begin{equation}\label{groupg}
    g(e)=e^{-ip^a(e)\tau_a/2}e^{z^a(e)\tau_a/2},\quad p^a(e),z^a(e)\in\mathbb{R}^3,
\end{equation}
where $\tau_a=-i\sigma_a$ and $\sigma^a,a=1,2,3$ are the Pauli matrices. The spin-network representation of these states is:
\begin{equation}
    \psi^t_{g(e)}(j,m,n)\equiv\langle j,m,n|g(e)\rangle=\sqrt{d_j}e^{-t\lambda_j/2}D^{(j)}_{mn}(g(e)).
\end{equation}
It is worth noting that the above construction is not normalized. The normalized coherent state is denoted by:
\begin{equation}
    \tilde{\psi}^t_{g(e)}:=\frac{\psi^t_{g(e)}}{|| \psi^t_{g(e)} ||}.
\end{equation}
After the construction of coherent states on a single edge, the generalization to a graph with $L$ edges is straightforward:
\begin{equation}\label{stateL}
    \psi^t_{\Gamma,\{g\}}(h(e_1),\dots ,h(e_L))=\prod\limits_{l=1}^{L}\left(\sum\limits_{j_l\in\mathbb{Z}_+/2\cup\{0\}}d_{j_l}e^{-t\lambda_{j_l}/2}\chi_{j_l}(g(e_l)h^{-1}(e_l))\right),
\end{equation}
where $\{g\}$ denotes the set of complexified $SL(2,\mathbb{C})$ holonomies on every edge contained in the graph $\Gamma$. The normalized multi-edge coherent state is denoted as $\tilde{\psi}^t_{\Gamma,\{g\}}(h(e_1),\dots ,h(e_L))$.

While the product state $\tilde{\psi}^t_{\Gamma,\{g\}}$ succinctly describes a semi-classical geometry on the graph, it inherently violates local $SU(2)$ gauge invariance at the vertices. To remedy this, a global gauge-invariant coherent state $\Psi^t_{\Gamma,\{g\}}$ is constructed by performing group averaging integrations (integrating out the gauge degrees of freedom) over the Haar measure at all $N$ vertices \cite{Bahr:2007xn,PhysRevD.92.104023}:
\begin{equation}
    \Psi^t_{\Gamma,\{g\}}(h(e_1),\dots ,h(e_L))=\int_{SU(2)^N}d\mu_H(\tilde{g}_1)\dots d\mu_H(\tilde{g}_N)\psi^t_{\{g\}}(\tilde{g}_{t(1)}h(e_1)\tilde{g}^{-1}_{s(1)},\dots ,\tilde{g}_{t(L)}h(e_L)\tilde{g}^{-1}_{s(L)}),
\end{equation}
where $s(l)$ and $t(l)$ map the source and target vertices of edge $l$. Expanding this integral using the Peter-Weyl theorem yields:
\begin{equation}\label{coherent_int}
    \begin{split}
    \Psi^t_{\Gamma,\{g\}}(h(e_1),\dots ,h(e_L))&=\prod\limits_l\sum\limits_{j_l}d_{j_l}e^{-t\lambda_{j_l}/2}D^{(j_l)}_{m_ln_l}(g_l)D^{(j_l)}_{\mu_l\nu_l}(h^{-1}(e_l))\int_{SU(2)}d\mu_H(\tilde{g}_l)D^{(j_l)}_{n_l\mu_l}(\tilde{g}_{s(l)})D^{(j_l)}_{\nu_lm_l}(\tilde{g}_{t(l)}^{-1}).
    \end{split}
\end{equation}
The integral in (\ref{coherent_int}) is conducted on every vertex in $\Gamma$, it corresponds to $\sum\limits_{\iota}|\iota\rangle\langle\iota|$ in the spin-network representation, which by construction preserves the SU(2) gauge invariance of spin-network states. 

Using the orthogonality relation of the spin-network basis, the transformation matrix between the gauge-invariant coherent state and gauge-invariant spin-network state can be computed as:
\begin{equation}\label{transform123}
    \langle\vec{\iota},\vec{j}|\Psi^t_{\Gamma,\{g\}} \rangle=e^{-t(\lambda_{j_1}+\cdots+\lambda_{j_L})/2}\overline{\Phi_{\Gamma,\{j_l\},\{\iota_n\}}(g_1,\dots ,g_L)},
\end{equation}
where we introduce the notation:
\begin{equation}\label{spin-networkG}
    \Phi_{\Gamma,\{j_l\},\{\iota_n\}}(h(e_1),\dots ,h(e_l)):=\left(\prod\limits_n\iota_n\right)^{n_1\dots n_l}_{m_{1}\dots m_l}\left((-1)^{j_1+\dots +j_l-n_1-\dots -n_l}\prod\limits_l\sqrt{d_{j_l}}\overline{D^{(j_l)}_{m_l,-n_l}(h(e_l))}\right),
\end{equation}
where $(\iota_n)^{n_1\dots n_l}_{m_1\dots m_l}$ is the intertwiner at the $n$-th vertex, with upper and lower indices representing that the intertwiner is connected with an outgoing or ingoing edge, respectively. The bar denotes complex conjugation. $|\vec{\iota},\vec{j}\rangle:=\Phi_{\Gamma,\{j_l\},\{\iota_n\}}(h(e_1),\dots ,h(e_l))$ is the gauge-invariant spin-network state labeled by two collections of labels ${\iota_n}$ and ${j_l}$. 

Now, we discuss the normalization of gauge-variant and gauge-invariant coherent states. Using the Poisson summation formula, the normalization factor on one edge can be calculated as \cite{Thiemann:2000ca}:
\begin{equation}\label{Normalization}
    ||\psi^t_g||^2=\psi^{2t}_{H^2}(1)=\frac{2\sqrt{\pi}e^{t/4}}{t^{3/2}}\frac{1}{\sqrt{y^2-1}}\sum\limits^{\infty}_{n=-\infty}(\textup{arcosh}(y)-2\pi in)e^{-\frac{(2\pi n+i \textup{arcosh}(y))^2}{t}},
\end{equation}
where $H=e^{-ip^a(e)\tau^a/2}$ is the boost part of $g$ and $y:=\textup{tr}(H^2)/2$. In the meantime, the normalization factor for gauge-invariant coherent states can also be calculated, as shown in \cite{Bahr:2007xn}. For example, if we consider a simple 2-flower graph, i.e., a single gauge-invariant vertex with 2 self-connected edges, the normalization factor can be calculated as:
\begin{equation}\label{Normalization2}
    \begin{split}
        ||\Psi^t_{[g_1,g_2]}||^2=\langle \Psi^t_{[g_1,g_2]}|\Psi^t_{[g_1,g_2]}\rangle=\frac{4\pi e^{t/2}}{t^{3}}\int_{SU(2)}&d\mu_H(k)\sum\limits_{n_1,n_2\in \mathbb{Z}}\frac{f_1(k)-2\pi in_1}{\sinh(f_1(k)-2\pi in_1)}\frac{f_2(k)-2\pi in_2}{\sinh(f_2(k)-2\pi in_2)}\\
        \times&\exp(\frac{(f_1(k)-2\pi in_1)^2+(f_2(k)-2\pi i n_2)^2}{t}),
    \end{split}
\end{equation}
where:
\begin{equation}
    \begin{split}
        \cosh f_1(k)&=\frac{1}{2}\tr(g_1^{\dagger}kg_1k^{-1})\\
        \cosh f_2(k)&=\frac{1}{2}\tr(g_2^{\dagger}kg_2k^{-1}),\\
    \end{split}
\end{equation}
and $k$ is an $SU(2)$ group element using the following parametrization:
\begin{equation}\label{paramg}
    \begin{split}
            k=\exp(i\vec{\sigma}\cdot \vec{\phi})=:\begin{pmatrix}
            \cos(\theta)+ i\cos(\phi)\sin(\theta) & (i\cos(\varphi)+\sin(\varphi))\sin(\theta)\sin(\phi)\\
            (i\cos(\varphi)-\sin(\varphi))\sin(\theta)\sin(\phi) & \cos(\theta)- i\cos(\phi)\sin(\theta)\\
        \end{pmatrix},
    \end{split}
\end{equation}
where $\vec{\phi}=:\phi \cdot(\cos\varphi\sin\theta,\sin\varphi\sin\theta,\cos\theta)$, and the $SU(2)$ integral of functions of $k$ is given by:
\begin{equation}
    I=:\frac{1}{2\pi^2}\int^{2\pi}_0\int^{\pi}_0\int^{\pi}_0 f(k)\sin^2\phi \sin\theta d\phi d\theta d\varphi.
\end{equation}

The action of $\widehat{Q}_{IJK}$ on coherent states can also be computed. For a coherent state $\psi^t_{g}(h)$ defined on a single edge $e$ , we have:
\begin{equation}\label{paction}
    \widehat{p}^j_e\psi^t_{g}(h):=t\widehat{J}^j_e\psi^t_{g}(h)=\frac{it}{2}(\frac{d}{ds})_{s=0}\psi_g^t(e^{s\tau_j}h),
\end{equation}
then we have:
\begin{equation}
    \langle\psi^t_{g},\widehat{p}^j_e\psi^t_{g'}\rangle=\frac{it}{2}(\frac{d}{ds})_{s=0}\psi^{2t}_{e^{-s\tau_j}g'\bar{g}^T}(1).
\end{equation}
By defining: $\cosh(z_0)=:\frac{1}{2}\trace(g'\bar{g}^T)$, it can be calculated that \cite{Thiemann:2000bx}:
\begin{equation}
    \begin{split}
        &\langle\psi^t_g,\widehat{p}^j_e\psi^t_{g'}\rangle=-\frac{it}{4}\frac{\trace(\tau_jg'\bar{g}^T)}{\sinh(z_0)}\frac{\sqrt{\pi}e^{t/4}}{4T^3}\\
        &\times\sum\limits_{n_c}e^{\frac{(z_0-2\pi in_c)^2}{t}}[2\frac{(z_0-2\pi in_c)^2}{t\sinh(z_0)}-(z_0-2\pi in_c)\frac{\cosh(z_0)}{\sinh^2(z_0)}+\frac{1}{\sinh(z_0)}],
    \end{split}
\end{equation}
where $T:=\sqrt{t}/2$. Since the action of $\widehat{p}^j_e$ is restricted only to one edge, a straightforward generalization to a vertex with multiple edges using equation (\ref{stateL}) and the definition of $\widehat{Q}_{IJK}$ in equation (\ref{vdef}) will produce the action of $\widehat{Q}_{IJK}$ in the coherent state representation.

It is not difficult to generalize this result for arbitrary graphs (by taking products similar to (\ref{stateL}) for all edges) and also to gauge-invariant coherent states (by integrating over gauge transformations similar to (\ref{Normalization2}) on every vertex). The same method is used to calculate the action of $(\widehat{p}^j_e)^N$ operators:
\begin{equation}\label{pn1}
    \langle\psi^t_{g},\widehat{p}^{j_1}_e\dots \widehat{p}^{j_N}_e\psi^t_{g'}\rangle=(\frac{it}{2})^N(\frac{d^N}{ds_1\dots ds_N})_{\{s\}=0}\psi^{2t}_{e^{-s\tau_{j_1}}\dots e^{-s\tau_{j_N}}g'\bar{g}^T}(1).
\end{equation}
Defining the $\widehat{Q}^q_{IJK}$ operators ($q$ labels the $q$-th power of operator $\widehat{Q}_{IJK}$) on coherent states is crucial in computing the semiclassical expansions described below.

\subsection{Semiclassical expansion of the volume operator matrix elements}
As originally formulated by Giesel and Thiemann \cite{Giesel:2006um}, the coherent state matrix elements of the volume operator $\langle\psi|\widehat{V}_v|\psi'\rangle$ purportedly admit a semi-classical expansion in powers of the classicality parameter (proportional to $\hbar$), truncated at order $\hbar^{k+1}$:
\begin{equation}\label{Vexpand}
    \langle\psi|\widehat{V}_v|\psi'\rangle \approx\langle\tilde{\psi}|\widehat{Q}_v|\tilde{\psi}\rangle^{\frac{1}{2}}\langle\psi|\left[1+\sum\limits^{2k+1}_{N=1}(-1)^{N+1}\frac{(1-1/4)\dots (2k-1/4)}{4N!}\left(\frac{\widehat{Q}^2_v}{\langle\tilde{\psi}|\widehat{Q}_v|\tilde{\psi}\rangle^2}-1\right)^N\right]|\psi'\rangle,
\end{equation}
However, a critical vulnerability in this standard approach is that its convergence properties are heavily optimized solely for expectation values (where $\psi = \psi'$), rendering it ill-equipped for non-diagonal matrix elements and deeply problematic for complex gauge-invariant coherent configurations. The physical intuition underlying this breakdown is straightforward: when the coherent states $\psi$ and $\psi'$ are macroscopically separated in phase space, the overlap terms within the standard expansion resist regular perturbative expansion in $\hbar$. 
Consequently, the traditional series is not optimized for off-diagonal matrix elements and can become quantitatively unreliable at finite $\hbar$ when the coherent-state labels are sufficiently separated.

To systematically circumvent this obstruction, we employ a new semiclassical expansion specifically organized for matrix elements $\langle\psi|\widehat{V}_v|\psi'\rangle$:
\begin{eqnarray}\label{VexpandNew}
   \langle\psi |\widehat{V}_v | \psi'\rangle\approx \left(\frac{\langle\psi|\widehat{Q}_v|\psi'\rangle}{\langle\psi|\psi'\rangle} \right)^{\frac{1}{2}}\langle\psi|\left[1+\sum\limits^{2k}_{N=1}(-1)^{N+1}\frac{(1-1/4)\dots (2k-1/4)}{4N!} \left(\frac{\langle\psi|\psi'\rangle^2 \widehat{Q}^2_v}{\langle\psi|\widehat{Q}_v|\psi'\rangle^2}-1 \right)^N \right]|\psi'\rangle
\end{eqnarray}
This formalism extends naturally to a broader class of polynomially generated operators on the Hilbert space $\mathcal{H}$; the corresponding derivation and convergence analysis are presented in our companion paper \cite{Liliu:202603A}. In the present article, our main focus is on building the computational framework required to evaluate these quantities from the deep quantum discrete scale and to test the resulting expansion against exact numerical data.

 \section{Computational Framework and Algorithm}
In this section, the detailed computational framework developed to study the volume operator across both the deep quantum regime and the semi-classical limit is presented. In order to achieve the required computational performance, the algorithm is divided into two distinct parts. First, the numerical state-sum evaluation is implemented using Julia, which fully utilizes its high-performance vectorization capabilities. Second, the analytical algebraic expansions are calculated using Python's SymPy and SymEngine libraries. Both programs \cite{LQG-Volume} are strongly optimized for parallel computing, scaling efficiently across 40 to 80 CPU cores during our test computations.

The matrix elements of the volume operator acting on normalized coherent states can be computed by inserting the resolution of identity using the volume operator eigenbasis $\{|\lambda\rangle\}$, the equation for the gauge-invariant states takes the following form: 
\begin{equation}\label{sum111}
    \begin{split}
        \langle \tilde{\Psi}^t_{\Gamma,\{g\}}|\widehat{V}_{v}|\tilde{\Psi}^t_{\Gamma,\{g'\}} \rangle=&\sum\limits_{\{\vec{j}\}}\bigg(\sum\limits_{\lambda,\lambda'}\sum\limits_{\{\vec{\iota}\},\{\vec{\iota}'\}}\langle \tilde{\Psi}^t_{\Gamma,\{g\}}|\Gamma,\{\vec{\iota}\},\{\vec{j}\}\rangle \\
        &\qquad \langle\Gamma,\{\vec{\iota}\},\{\vec{j}\}|\lambda\rangle\langle\lambda|\widehat{V}_v|\lambda'\rangle\langle\lambda'|\Gamma,\{\vec{\iota}'\},\{\vec{j}\}\rangle\langle\Gamma,\{\vec{\iota}'\},\{\vec{j}\}|\tilde{\Psi}^t_{\Gamma,\{g'\}} \rangle\bigg),
    \end{split}
\end{equation}
while the corresponding equation for the gauge-variant coherent states is given by:
\begin{equation}\label{sum333}
    \begin{split}
        \langle \tilde{\psi}^t_{\Gamma,\{g\}}|\widehat{V}_v|\tilde{\psi}^t_{\Gamma,\{g'\}} \rangle=&\sum\limits_{\{\vec{j}\},\{M\},\{M'\}}\bigg(\sum\limits_{\lambda,\lambda'}\sum\limits_{\{\vec{\iota}\},\{\vec{\iota}'\}}\langle \tilde{\psi}^t_{\Gamma,\{g\}}|\Gamma,\{\vec{\iota}\},\{\vec{j}\},\{M\}\rangle\langle\Gamma,\{\vec{\iota}\},\{\vec{j}\},\{M\}|\lambda\rangle \\
        & \times \langle\lambda|\widehat{V}_v|\lambda'\rangle\langle\lambda'|\{\vec{\iota}'\},\{\vec{j}\},\{M'\}\rangle\langle\{\vec{\iota}'\},\{\vec{j}\},\{M'\}|\tilde{\psi}^t_{\Gamma,\{g'\}} \rangle\bigg) .\\
    \end{split}
\end{equation}
Here, the matrix $\langle\lambda'|\{\vec{\iota}'\},\{\vec{j}\},\{M'\}\rangle$ acts as the unitary transformation matrix that relates the eigenbasis back to the spin-network intertwiner basis. Because the operator $\langle\lambda|\widehat{V}_v|\lambda'\rangle$ is strictly diagonal in this specific subspace, the analytical difficulty of taking the square root of the operator $\widehat{Q}_v$ is completely avoided by employing numerical diagonalization.

Due to the Gaussian peakedness feature of the coherent states, the transition amplitudes $\langle\Gamma,\{\vec{\iota}\},\{\vec{j}\},\{M\}|\Psi^t_{\Gamma,\{g\}} \rangle$ naturally decay to zero when $|\vec{j}|$ becomes large. Therefore, it is physically justified to introduce a specific cut-off value $j_{\mathrm{cap}}$ for the numerical summation. This parameter safely truncates the summation over both the boundary edge spins and the internal intertwiner angular momenta constrained by the triangular inequalities.

 \subsection{Calculation of the Transformation Matrix}
 
The numerical calculation of Eq. (\ref{transform123}) is computationally expensive. To solve this numerical difficulty, our Julia-based program utilizes three main optimization techniques. Firstly, it is well-known that constructing intertwiners requires calculating a massive amount of Wigner $3$-$j$ symbols. Instead of computing them repeatedly during the loop using standard libraries such as \texttt{WignerSymbols.jl} \cite{WignerSymbols.jl}, a complete array of all possible $3$-$j$ symbols up to $j_{\mathrm{cap}}$ is pre-calculated and stored in the system memory. By using this hash-map pre-calculation method, the computation time is vastly reduced because fetching values from memory only requires $\mathcal{O}(1)$ time.

Secondly, by making use of the Single Instruction, Multiple Data (SIMD) feature available in modern CPUs, the numerical algorithm is fully vectorized. Specifically, a one-dimensional array containing all possible internal intertwiner indices $\{\vec{\iota}\}$ for a fixed $\{\vec{j}\}$ configuration is generated initially. Then, the tensor contractions needed to compute $\langle\Gamma,\{\vec{\iota}\},\{\vec{j}\},\{M\}|\Psi^t_{\Gamma,\{g\}} \rangle$ are broadcasted globally across this list, which further increases the computation speed by a factor of 10. 

Finally, the calculation of the complexified $\mathrm{SL}(2,\mathbb{C})$ holonomy matrices is optimized by using the built-in fast Kronecker product function (\texttt{kron!}) in Julia, thereby minimizing the nested iteration loops during the network contraction.

\subsection{Computing the Action of the Volume Operator}
The main difficulty of evaluating the volume operator is related to calculating the square root of $\widehat{Q}_v$. Usually, this mathematical problem prevents researchers from studying the matrix elements of the volume operator on complicated graphs. Our proposed numerical framework avoids this problem. Namely, for each fixed $\{j\}$ combination, the dimensions of the volume matrix is fixed as only internal intertwiner degrees of freedom need to be counted. Moreover, for large $j$ values where the numerical matrix dimension becomes extremely large, the calculation can be safely truncated due to the peakedness properties of the input coherent states. By performing convergence tests comparing the numerical and analytical results in the intermediate region, the numerical error can be strictly controlled. Therefore, by combining the numerical and analytical results, the volume operator (and subsequently the Hamiltonian operator) can be evaluated across the entire parameter space.

 \subsection{The Final $j$-Summation and Parallelization}
 
After the matrix elements of the volume operator and the corresponding transition matrices are computed, a final summation over all possible $j$ indices up to $j_{\mathrm{cap}}$ must be performed. In order to optimize the computational performance, only this outermost summation loop is parallelized. The projection and diagonalization steps are packaged as internal functions that only take the semiclassical parameter $t$ and specific $\{\vec{j}\}$ arrays as isolated inputs. Then, the $j$-summation tasks are randomly allocated to the available computing threads using Julia's distributed mapping environment. This ensures that the computational workload is dynamically balanced across the 40-80 CPU cores.

 \subsection{Analytical Computation of the $(\widehat{Q}_v)^q$ Operators}

In order to validate the state-sum numerical model and to perform the semi-classical expansion, the numerical results are compared with the exact expectation values and matrix elements of the $(\widehat{Q}_v)^q$ operators computed purely analytically in the coherent state representation. Furthermore, these analytically calculated $(\widehat{Q}_v)^q$ terms are used to explicitly construct the $t$-order and $t^2$-order Taylor expansions of the volume operator via Eq. (\ref{VexpandNew}).

However, the exact calculation of $(\widehat{Q}_v)^q$ operators (especially for $q > 2$) is a difficult task because it causes severe combinatorial explosions during the symbolic derivation. To solve this problem, a specialized symbolic algorithm is written using the \texttt{SymPy} and \texttt{SymEngine} packages in Python. Specifically, three different mathematical simplifications are implemented to reduce the calculation time to a practical level:

\textbf{1. Decoupling of the edge derivatives:} It should be noted that a direct expansion of $(\widehat{Q}_{IJK})^q = (\epsilon_{ijk}\widehat{p}^i_{(e_I)}\widehat{p}^j_{(e_J)}\widehat{p}^k_{(e_K)})^q$ generates a total of $6^q$ non-zero differential terms. Nevertheless, because the specific flux operators $\widehat{p}^i_{(e_I)}$, $\widehat{p}^j_{(e_J)}$, and $\widehat{p}^k_{(e_K)}$ act strictly independently on their corresponding edges (as shown in Eq. \ref{paction}), only $3\times 3^q$ unique derivatives actually need to be calculated. Therefore, the program first computes these isolated derivatives as functions of the continuous variable $s$, and saves them into an index list. The condition $s=0$ is only substituted at the very end. By reading directly from this pre-calculated list, redundant symbolic derivations are avoided.

This first simplification is sufficient to evaluate the expressions up to $q \le 6$ for simple gauge-variant graphs. However, as will be shown in Section \ref{sectionIV}, calculating the second-order volume expansion for gauge-invariant states requires the exact results for $q=8$. This implies that extremely complicated terms like the following must be computed:
\begin{equation}\label{pinv}
    \begin{split}
        &\langle \Psi^t_{[g_1,g_2,g_3,g_4]}|(\epsilon_{i_1j_1k_1}\widehat{p}^{i_1}_{(e_1)}\widehat{p}^{j_1}_{(e_2)}\widehat{p}^{k_1}_{(e_3)})\dots(\epsilon_{i_8j_8k_8}\widehat{p}^{i_8}_{(e_1)}\widehat{p}^{j_8}_{(e_2)}\widehat{p}^{k_8}_{(e_3)})|\Psi^t_{[g_1,g_2,g_3,g_4]}\rangle\\
        &=t^{24}\epsilon_{i_1j_1k_1}\dots\epsilon_{i_8j_8k_8}\int_{SU(2)}d\mu_H(k)d\mu_H(h)\left[\frac{d}{dw_1\dots dw_8}\left(\psi^{2t}_{ke^{w_1\tau_{i_1}}\dots e^{w_8\tau_{i_8}}g_1h^{-1}\bar{g}^T_1}(1)\right) \Bigg|_{\{u\}=0}\right.\\
        &\times\left.\frac{d}{dr_1\dots dr_8}\left(\psi^{2t}_{ke^{r_1\tau_{j_1}}\dots e^{r_8\tau_{j_8}}g_2h^{-1}\bar{g}^T_2}(1)\right) \Bigg|_{\{r\}=0}\times\frac{d}{ds_1\dots ds_8}\left(\psi^{2t}_{ke^{s_1\tau_{k_1}}\dots e^{s_8\tau_{k_8}}g_3h^{-1}\bar{g}^T_3}(1)\right) \Bigg|_{\{s\}=0}\psi^{2t}_{kg_4h^{-1}\bar{g}^T_4}(1)\right],\\
    \end{split}
\end{equation}

\textbf{2. Implementation of auxiliary variables:} Computing the 8-th order derivatives directly on the infinite series presented in Eq. (\ref{pinv}) will immediately cause a memory overflow in SymPy. To prevent this, an abstract scalar variable is defined as $g_z(\{w\}):=\frac{1}{2}\mathrm{tr}(ke^{w_1\tau_{i_1}}\dots e^{w_8\tau_{i_8}}g_1h^{-1}\bar{g}^T_1)$. By utilizing the general chain rule, the required derivation of the Poisson summation can be rewritten with respect to $g_z$:
\begin{equation}\label{dgzdq}
\begin{split}
    &\frac{d}{dw_1\dots dw_8}\left(\psi^{2t}_{ke^{w_1\tau_{i_1}}\dots e^{w_8\tau_{i_8}}g_1h^{-1}\bar{g}^T_1}(1)\right) \Bigg|_{\{w\}=0}\\
    &=\frac{dg_z}{dw_1}\frac{d}{dg_z}\left(\cdots\left(\frac{dg_z}{dw_8}\frac{d}{dg_z}\left(\frac{\sqrt{\pi}e^{t/4}}{4\sinh{(\mathrm{arccosh} (g_z))}T^3}\sum\limits_n(\mathrm{arccosh} (g_z)-2\pi i n)e^{\frac{(\mathrm{arccosh} (g_z)-2\pi i n)^2}{t}}\right)\right)\right).
\end{split}
\end{equation}
By treating $g_z$ as a simple variable during the symbolic operation, the expression size is drastically minimized. The actual derivatives of $g_z$ (up to the 8-th order of $w_i$) are independently computed and then substituted back into Eq. (\ref{dgzdq}) only at the final step.

\textbf{3. Operator reordering using commutation relations:} Finally, it is known that directly computing all possible mixed indices for an 8-th order operator requires handling $3^8 = 6561$ different expressions. However, by strictly applying the $SU(2)$ commutation relation:
\begin{equation}
[\widehat{p}^i_I,\widehat{p}^j_J]=it\epsilon_{ijk}\widehat{p}^k_I\delta_{IJ},
\end{equation}
any arbitrarily ordered expectation value $\langle \widehat{p}^{i_1}\dots\widehat{p}^{i_8} \rangle$ can be systematically rewritten into a standard ordered sequence $\langle \widehat{p}^{i_1'}\dots\widehat{p}^{i_8'} \rangle$ satisfying $i_1'\leq\dots\leq i_8'$, together with a finite number of lower-order $t$ corrections. As summarized in Table \ref{table123n}, this mathematical property implies that only 164 unique derivatives (where only 45 are strictly of the 8-th order) need to be explicitly evaluated by the computer. This method alone increases the computation speed by about 100 times.

\begin{table}[h!]
 \begin{center}
\footnotesize
\renewcommand{\arraystretch}{1.5}
\begin{tabular}{ |c||c|c|c|c|c|c|c|c| } 
 \hline
 derivative order& 1 & 2 & 3 & 4 & 5 & 6 & 7 & 8 \\ 
 \hline
 Original & 0 & 0 & 0 & 0 & 0 & 0 & 0 & 6561\\ 
 \hline
  Optimized & 3 & 6 & 10 & 15 & 21 & 28 & 36 & 45\\ 
 \hline
\end{tabular}
\caption{The number of terms before and after the optimization for $q=8$}\label{table123n}
\renewcommand{\arraystretch}{1}
\end{center}
\end{table}

After all the algebraic integrands are successfully computed, the continuous integration over the $SU(2)$ gauge variables $d\mu_H(k)$ and $d\mu_H(h)$ must be evaluated. Because the analytical integration is generally impossible, the saddle-point approximation is systematically applied according to Hörmander’s theorem \cite{hormander2015analysis}:

\begin{theorem}[Hörmander’s theorem 7.7.5]\label{theorem111}
Let K be a compact subset in $\mathbb{R}^n$, X an open neighborhood of K, and k a positive integer. If: (1) the complex functions $u \in C_0^{2 k}(K)$, $f \in C^{3 k+1}(X)$ and $\operatorname{Im} f(x) \geq 0 \quad\forall x\in X$, (2) there is a unique point $x_0 \in K$ satisfying $\operatorname{Im}\left(S\left(x_0\right)\right)=0$, $f^{\prime}\left(x_0\right)=0$, $\operatorname{det}\left(f^{\prime \prime}\left(x_0\right)\right) \neq 0$ (where $f''$ denotes the Hessian matrix), and $f^{\prime}(x) \neq 0 \quad\forall x\in K \backslash\left\{x_0\right\}$, then we have the following estimation:
\begin{equation}\label{saddle1}
\left|\int_K u(x) e^{i \lambda f(x)} d x-e^{i \lambda f \left(x_0\right)}\left[\operatorname{det}\left(\frac{\lambda f^{\prime \prime}\left(x_0\right)}{2 \pi i}\right)\right]^{-\frac{1}{2}} \sum_{s=0}^{k-1}\left(\frac{1}{\lambda}\right)^s L_s u\left(x_0\right)\right| \leq C\left(\frac{1}{\lambda}\right)^k \sum_{|\alpha| \leq 2 k} \sup \left|D^\alpha u\right| .
\end{equation}
Here the constant $C$ is bounded when $f$ stays in a bounded set in $C^{3 k+1}(X)$. We use the standard multi-index notation $\alpha=\left\langle\alpha_1, \ldots, \alpha_n\right\rangle$ and:
\begin{equation}
D^\alpha=(-i)^\alpha \frac{\partial^{|\alpha|}}{\partial x_1^{\alpha_1} \ldots \partial x_n^{\alpha_n}}, \quad \text { where } \quad|\alpha|=\sum_{i=1}^n \alpha_i,
\end{equation}
 $L_s u\left(x_0\right)$ is defined as:
\begin{equation}\label{LSU}
L_s u\left(x_0\right)=i^{-s} \sum_{l-m=s} \sum_{2 l \geq 3 m} \frac{(-1)^l 2^{-l}}{l ! m !}\left[\sum_{a, b=1}^n H_{a b}^{-1}\left(x_0\right) \frac{\partial^2}{\partial x_a \partial x_b}\right]^l\left(g_{x_0}^m u\right)\left(x_0\right),
\end{equation}
where $H(x)=f^{\prime \prime}(x)$ denotes the Hessian matrix and the function $g_{x_0}(x)$ is given by:
\begin{equation}
g_{x_0}(x)=f(x)-f\left(x_0\right)-\frac{1}{2} H^{a b}\left(x_0\right)\left(x-x_0\right)_a\left(x-x_0\right)_b,
\end{equation}
satisfying $g_{x_0}\left(x_0\right)=g_{x_0}^{\prime}\left(x_0\right)=g_{x_0}^{\prime \prime}\left(x_0\right)=0$. For each $s$, $L_s$ is a differential operator of order $2s$ acting on $u(x)$.
\end{theorem}

In order to obtain the $t^2$ order expansion, the total integrand in Eq. (\ref{pinv}) is formally separated into a measure term $u(x)$ and an exponential action term $g(x)$. Both functions are defined over the 6-dimensional parameters $(x_1, \dots, x_6)$, which correspond directly to the Euler angles $(\phi,\theta,\varphi)$ of the $k$ and $h$ gauge transformations defined in Eq. (\ref{paramg}).

According to Eq. (\ref{saddle1}), all required derivatives of $u(x)$ up to the 4-th order, and derivatives of $g(x)$ up to the 6-th order, are computed at the critical point $k=h=\mathbb{I}$ (i.e., $x_i=0$). Then, the results for the full operator $(\widehat{Q}_{IJK})^8$ are generated by contracting the previously computed individual $\widehat{p}^{i}$ derivatives with the corresponding Levi-Civita symbols. Finally, substituting these evaluated terms into the general saddle-point formula yields the analytical approximation polynomial.

In conclusion, the numerical procedures involving the spin-network basis projection, combined with the analytical simplifications presented above, provide an efficient tool for calculating the expectation values of the volume operator. In the next section, the numerical results for different geometric graphs will be thoroughly discussed and validated by comparing them with these analytical small-$t$ expansions.

\section{Main Results}\label{sectionIV}

In this section, several interesting scenarios are studied in detail, where both the numerical full-quantum results and analytical semi-classical asymptotics can be computed. First, it is necessary to validate whether the numerical state-sum model works accurately. This is done by making a direct comparison between the numerically computed normalization factors and expectation values of $\widehat{Q}^q_{v}$ operators with the analytical results obtained purely in the coherent state representation.

\begin{figure}[h!]
 \centering
    \begin{subfigure}[b]{0.3\textwidth}
         \centering
          \includegraphics[height=3cm]{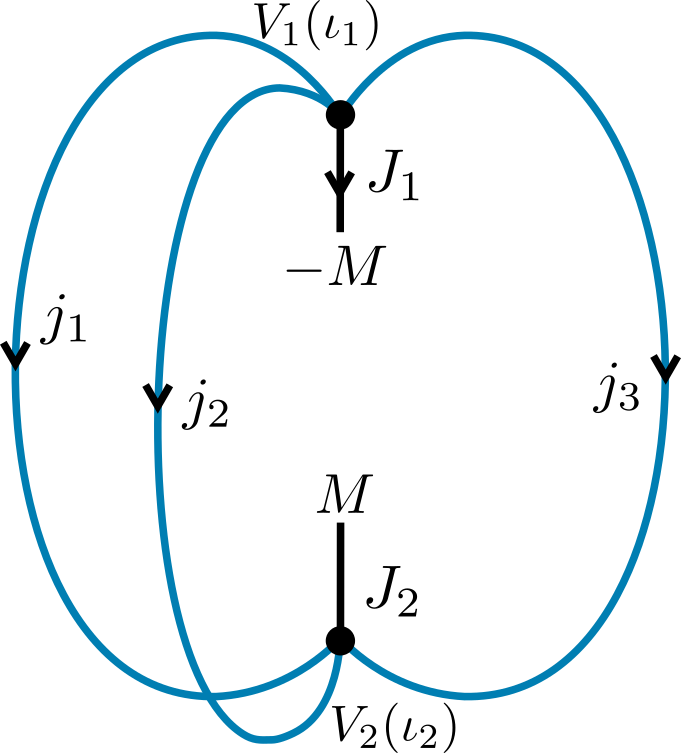} 
         \caption{gauge-variant 3-bridges}
     \end{subfigure}
     \begin{subfigure}[b]{0.3\textwidth}
         \centering
          \includegraphics[height=3cm]{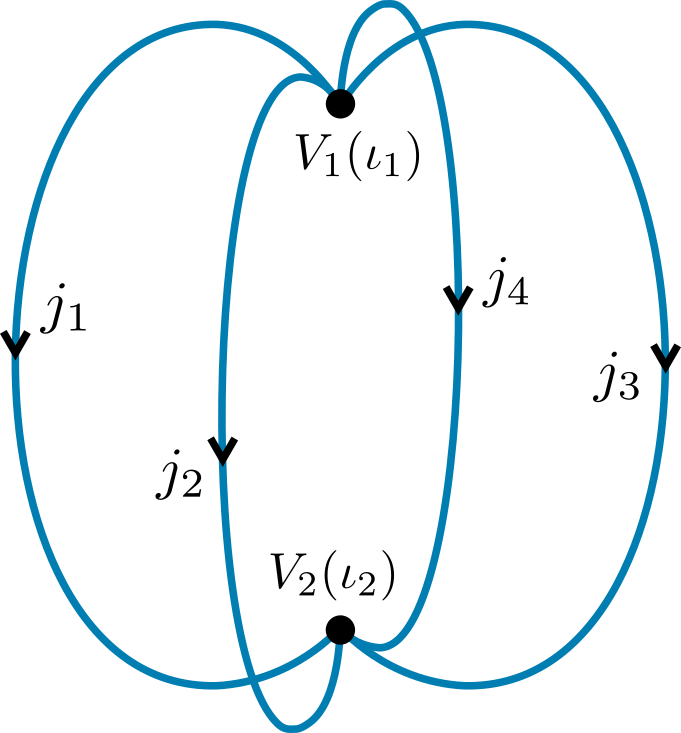} 
         \caption{Gauge-invariant 4-bridges}
     \end{subfigure}
      \begin{subfigure}[b]{0.3\textwidth}
     \centering
     \includegraphics[height=3.5cm]{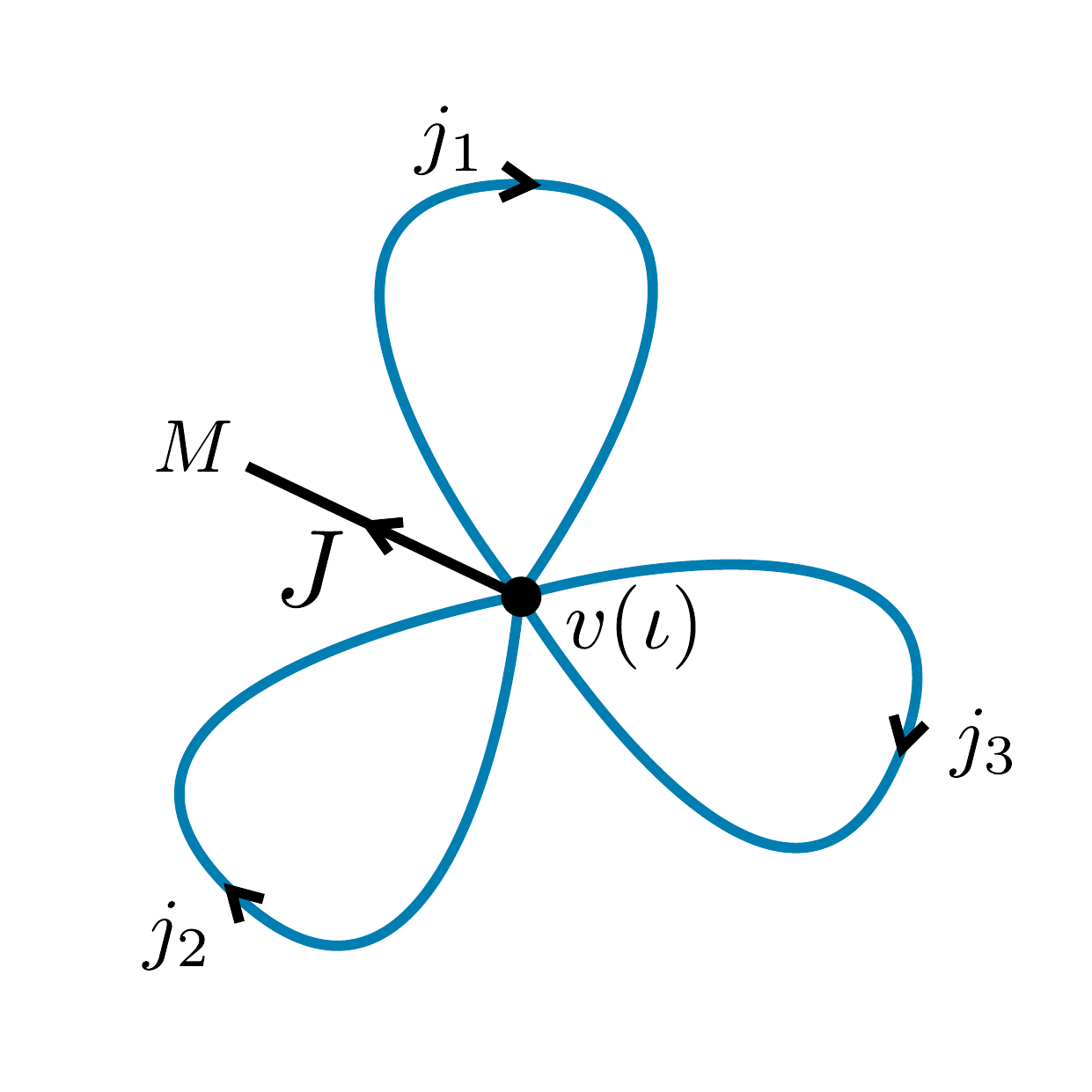} 
     \caption{Gauge-variant 3-flower}
     \end{subfigure}
 \caption{Graphical illustration of the spin-networks of: (a) Gauge variant 3-bridges, (b) gauge-invariant 4-bridges (each vertex corresponds to a tetrahedron), and (c) Gauge variant 3-flower (the vertex corresponds to a parallelepiped).}\label{fig3-1}
 \end{figure}

Secondly, we will check whether the newly proposed semi-classical operator expansion (\ref{VexpandNew}) is consistent with our numerical data for both gauge-variant and gauge-invariant cases. Finally, the computational performance of our algorithm is evaluated, and the specific range of the semi-classical parameter $t$ where the calculation is practically effective will be established. These points will be investigated across different geometries, namely the expectation values and matrix elements of the volume operator on gauge-variant 3-bridges (Fig. \ref{fig3-1} (a)), the expectation value of the volume operator on gauge-invariant 4-bridges (Fig. \ref{fig3-1} (b)) acting as regular and irregular tetrahedra, and the gauge-variant 3-flower graph (Fig. \ref{fig3-1} (c)). Furthermore, the eigensystem of the volume operator is studied by mapping the distribution of quantum states onto the basis of eigenvectors, and the correspondence between the maximum eigenvalue and the classical continuous volume is also discussed.

Before going into the details, since there are multiple key parameters used in this computation, and some of them involve numerical cut-offs, we provide a summary of these physical parameters here for clarity:
 
\begin{itemize}
    \item \textbf{$t$}: The dimensionless semi-classical parameter (also known as the heat-time) used in Thiemann coherent states to govern the phase-space dispersion. The limit $t\rightarrow 0$ corresponds to the classical continuum limit of LQG. On the other hand, the region where $t \sim 1$ generally represents the deep quantum level, where higher-order $\hbar$ corrections become strictly important.
    \item \textbf{$q$}: The exponent of the $\widehat{Q}^q_v$ operators. As formulated in Eq. (\ref{Vexpand}) and (\ref{VexpandNew}), the behavior of the non-analytic volume operator $\widehat{V}_v$ can be perturbatively calculated using a finite series expansion of these analytical $\widehat{Q}^q_v$ operators.
    \item \textbf{$n$}: The topological winding number introduced by the Poisson resummation formula (e.g., Eq. \ref{Normalization}). The $n=0$ term gives the main Gaussian peak in the semi-classical regime, while the $n\neq0$ terms represent the corrections related to the compactness of the $SU(2)$ group. These corrections are usually small, but their precise effects for larger $t$ need to be carefully bounded.
    \item \textbf{$j_{\mathrm{cap}}$}: The maximum $SU(2)$ spin representation cut-off on the edges. Because a full sum over $j$ is required when resolving the identity with the spin-network basis, an optimized cut-off $j_{\mathrm{cap}}$ must be applied so that the numerical program does not run into memory overflows.
\end{itemize}

 \subsection{Expectation value of the volume operator on the gauge-variant 3-bridges}
In this subsection, we first discuss the expectation value of the volume operator acting on the gauge-variant 3-bridges graph. The corresponding Thiemann coherent states are parametrized by the complex variables $\vec{z}_I:=(z^1_I,z^2_I,z^3_I)$ and the momentum fluxes $\vec{p}_I:=(p^1_I,p^2_I,p^3_I)$, with $I=1,2,3$ labeling the three edges. To represent a flat 3D geometry, the edges are oriented along the three orthogonal directions: $\vec{n}_1=(1,0,0)$, $\vec{n}_2=(0,1,0)$, and $\vec{n}_3=(0,0,1)$.

\subsubsection{Numerical Accuracy and $n$-Winding Corrections}

Basically, the normalization factor squared $||\psi||^2$ of a coherent state on graph $\Gamma$ can be computed analytically by Poisson resummation: one uses Eq. (\ref{Normalization}) for the gauge-variant case $\psi^t_{\Gamma,\{g\}}$, and Eq. (\ref{Normalization2}) for the gauge-invariant case $\Psi^t_{\Gamma,\{g\}}$. On the other hand, the same value must be recovered numerically when we perform the summation $\sum\limits_{\{\vec{j}\},\{M\}}\sum\limits_{\{\vec{\iota}\}}\langle\psi^t_{\Gamma,\{g\}}|\Gamma,\{\vec{\iota}\},\{\vec{j}\},\{M\}\rangle\langle \Gamma,\{\vec{\iota}\},\{\vec{j}\},\{M\} | \psi^t_{\Gamma,\{g\}} \rangle$ in the full spin-network basis. Thus, a consistency check can be constructed to validate our numerical method. This is achieved by comparing the normalization factors and expectation values of $\widehat{Q}^q_v$ calculated numerically against the analytic results in the small-$t$ limit.

\begin{figure}[h!]
 \centering
    \begin{subfigure}[b]{0.45\textwidth}
         \centering
 \includegraphics[height=5cm]{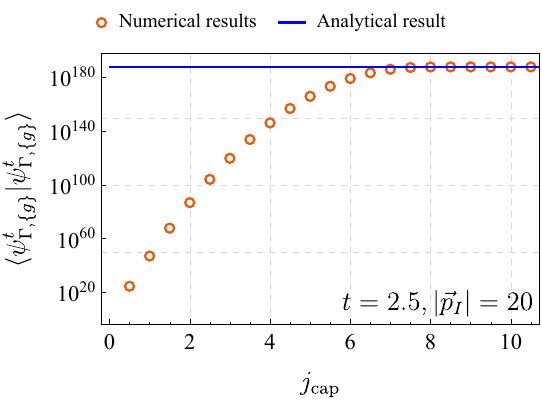} 
         \caption{Consistency check}
     \end{subfigure}
     \begin{subfigure}[b]{0.45\textwidth}
         \centering
 \includegraphics[height=4.45cm]{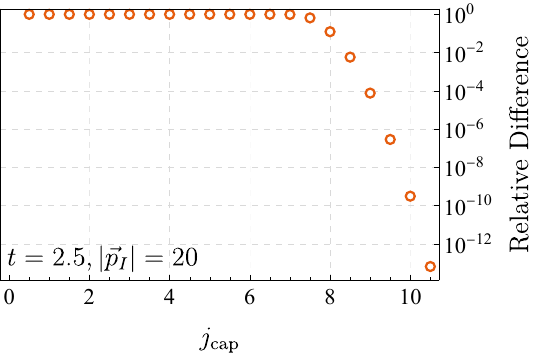} 
         \caption{Relative difference between computed results}
     \end{subfigure}
         \begin{subfigure}[b]{0.45\textwidth}
         \centering
 \includegraphics[height=5.5cm]{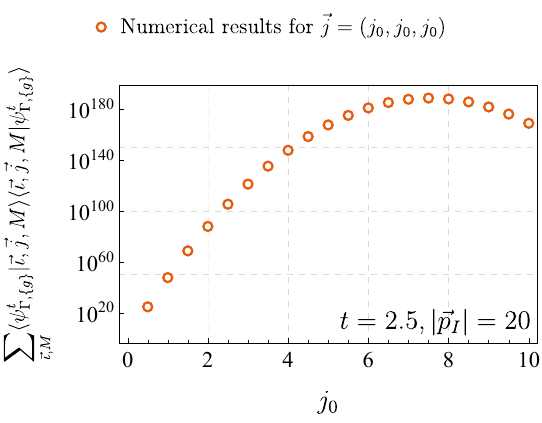} 
         \caption{Peakedness along $\vec{j}$ direction}
     \end{subfigure}
         \begin{subfigure}[b]{0.45\textwidth}
         \centering
 \includegraphics[height=5.15cm]{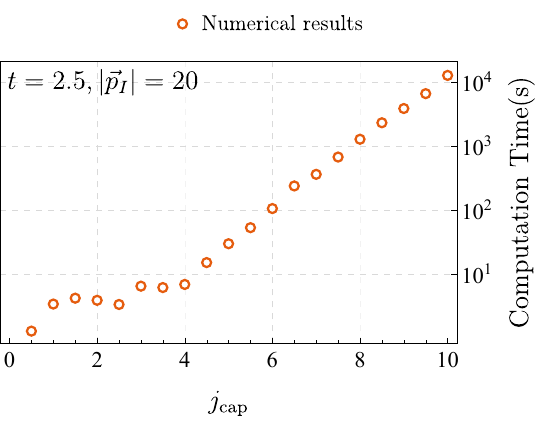} 
         \caption{Computation time versus $j$-cutoff}
     \end{subfigure}
 \caption{Consistency check of the normalization factor. (a) The normalization factor is obtained numerically (circles, using spin-network representation) and analytically (blue line, computed purely in coherent state representation) for different $j_{\mathrm{cap}}$. The parameters are set to be $t=2.5$, $\vec{z}_I=\vec{0}$ and $\vec{p}_I=20\vec{n}_I$. (b) The relative difference between analytical results and numerical results (computed as a percentage over the classical result). High accuracy can be achieved by raising $j_{\mathrm{cap}}$ higher. (c) Peakedness along $\vec{j}=(j_0,j_0,j_0)$ direction of the intermediate value before the final $j$-summation (where $\iota$ and $M$ indices are already summed). For $j_0\leq 5$ and $j_0\geq 10$, the contribution to the final result is negligible, which coincides with the results shown in (a) and (b). (d) The computation time of our numerical algorithm corresponds to each circle on the left. An exponential increase can be observed as $j_{cap}$ (the limit of $j$ summation) increases.}\label{fig1}
 \end{figure}

In the meantime, because of the Gaussian peakedness of the complexifier coherent state, this checking process also helps us to find the optimized $j$-cutoff. This means we can capture almost all the non-zero contributions from the spin-network basis (restricting the relative error under $10^{-10}$) without wasting too much computation time. In Fig. \ref{fig1}, the numerical test of the normalization factor is shown for $t=2.5$, with purely spatial momenta $\vec{p}_I=20\vec{n}_I$ and $\vec{z}_I=\vec{0}$.

It can be seen from Fig. \ref{fig1}(a) and (b) that, as long as the $j_{\mathrm{cap}}$ is set sufficiently large, the state-sum numerical output will match the analytical curve. For the chosen parameters, the analytical integral yields $7.625172013072415 \times 10^{187}$, and our numerical code yields $7.625172013071715 \times 10^{187}$ at $j_{\mathrm{cap}} = 21/2$. This small error validates our numerical architecture.

Furthermore, from Fig. \ref{fig1}(c), one can see that if we plot the intermediate values (where internal $\iota$ and $M$ indices are already summed out) along the $\vec{j}= j \cdot (1,1,1)$ symmetric direction, the distribution has a clear peak at $j_0 = 7.5$. This value is very close to the expected macroscopic classical momentum $j_0 \approx |\vec{p}_I|/t = 20/2.5 = 8$, which agrees with the physical property of the Thiemann coherent states. Fig. \ref{fig1}(d) shows that the CPU time scales exponentially as $j_{\mathrm{cap}}$ becomes large. Because $j_{\mathrm{cap}}$ must be set larger than $|\vec{p}_I|/t$ to make the sum converge, exploring very large flux domains purely using numerical methods is very hard, making our analytical expansion framework (discussed later) highly necessary.

Coming back to Eq. (\ref{Normalization}), the analytical formula requires an infinite sum over the topological winding number $n$. Since the analytical integration becomes highly complicated for $n \neq 0$, it is very important to test how the choice of $n_{\mathrm{cap}}$ affects the final error, so that we can optimize the SymPy algebraic speed.

\begin{figure}[h!]
 \centering
    \begin{subfigure}[b]{0.45\textwidth}
         \centering
\includegraphics[height=5cm]{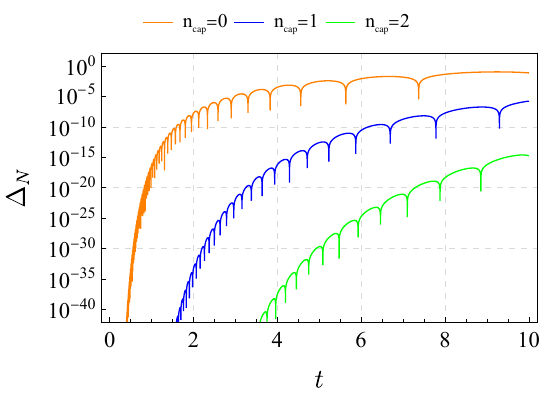} 
         \caption{Difference between $n_{\mathrm{cap}}=0,1,2$ and $n_{\mathrm{cap}}=3$}
     \end{subfigure}
     \begin{subfigure}[b]{0.45\textwidth}
         \centering
\includegraphics[height=5cm]{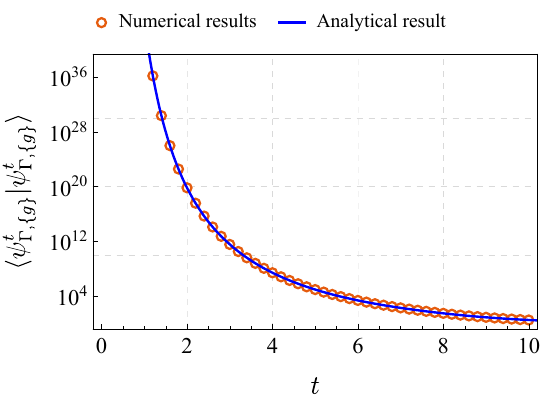} 
         \caption{Comparison between results for $1\leq t\leq10$}
     \end{subfigure}
\caption{(a) Relative difference between the normalization factor of $n_{\mathrm{cap}}=0,1,2$ and $n_{\mathrm{cap}}=3$. (b) Difference on the normalization factor between the numerical result for $\vec{p}_a=6\vec{n}_a$ and the analytical result on the same setting for $n_{\mathrm{cap}}=3$.}\label{fig2}
 \end{figure}

Figure \ref{fig2}(a) computes the relative differences caused by applying different cut-offs $n_{\mathrm{cap}}=0, 1, 2$, where the $n_{\mathrm{cap}}=3$ case is used as the baseline. It can be seen that for the region $0 \le t \le 10$, the difference between $n_{\mathrm{cap}}=2$ and $n_{\mathrm{cap}}=3$ is negligible. If we use the simplest Gaussian approximation ($n_{\mathrm{cap}}=0$), there is an obvious deviation at large $t$. However, as $t$ goes to the semi-classical region, namely $t \le 2$, the difference between $n_{\mathrm{cap}}=0$ and the $n_{\mathrm{cap}}=3$ case gradually becomes negligible.

Similarly, Fig. \ref{fig2}(b) shows that our numerical calculation can easily reproduce the $n_{\mathrm{cap}}=3$ analytic results across the whole range $0 \le t \le 10$. Because the main physics discussed in this paper focuses on the semi-classical transition where $t < 1$, we can safely set $n_{\mathrm{cap}}=0$ for all the following high-order analytical expansions without losing any real physical accuracy.

\subsubsection{Matrix Elements of Higher-Order $\widehat{Q}^q_{V_1}$ Operators}

Now we move forward to study the behavior of the polynomial operators $\widehat{Q}^q_{V_1}$. We begin by checking the consistency between the numerical and analytical methods for $q=1$ and $q=2$:

\begin{figure}[h!]
 \centering
    \begin{subfigure}[b]{0.45\textwidth}
         \centering
 \includegraphics[height=5cm]{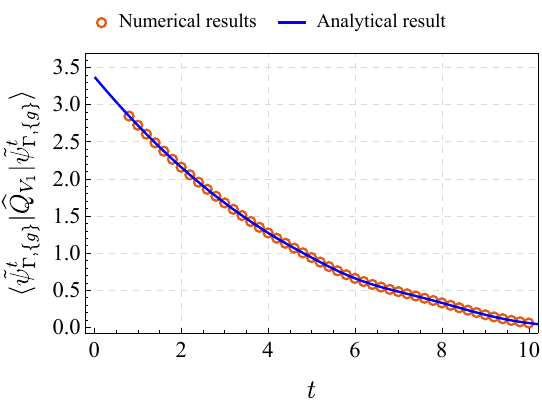} 
         \caption{Expectation value of $\widehat{Q}_{V_1}$}
     \end{subfigure}
     \begin{subfigure}[b]{0.45\textwidth}
         \centering
 \includegraphics[height=5cm]{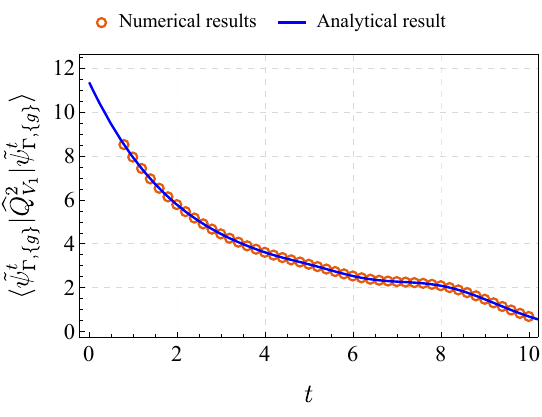}
         \caption{Expectation value of $\widehat{Q}^2_{V_1}$}
     \end{subfigure}
 \caption{Comparison between the analytical results ($n_{\mathrm{cap}}=0$) and numerically computed results of the expectation value of (a) $\widehat{Q}_{V_1}$, (b) $\widehat{Q}^2_{V_1}$.}\label{Qn3}
 \end{figure}

  \begin{figure}[h!]
 \centering
    \begin{subfigure}[b]{0.45\textwidth}
         \centering
 \includegraphics[height=5cm]{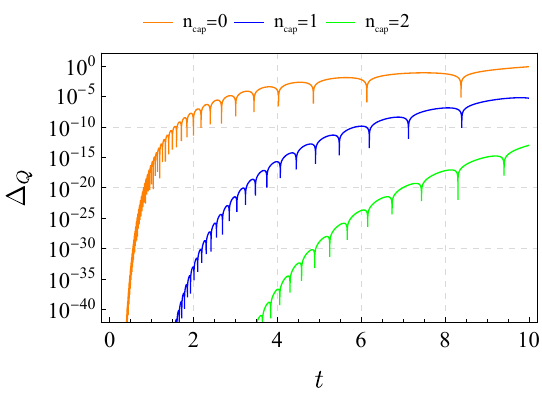} 
         \caption{$n_{\mathrm{cap}}$ difference of $\langle\tilde{\psi}|\widehat{Q}_{V_1}|\tilde{\psi}\rangle$}
     \end{subfigure}
     \begin{subfigure}[b]{0.45\textwidth}
         \centering
 \includegraphics[height=5cm]{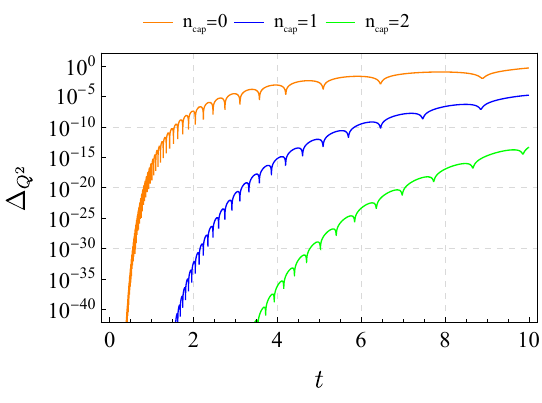} 
         \caption{$n_{\mathrm{cap}}$ difference of $\langle\tilde{\psi}|\widehat{Q}^2_{V_1}|\tilde{\psi}\rangle$}
     \end{subfigure}
 \caption{the $n_{\mathrm{cap}}$ difference between $n_{\mathrm{cap}}=0$, $n    _{\mathrm{cap}}=1$, $n_{\mathrm{cap}}=2$ cases and $n_{\mathrm{cap}}=3$ for the numerically computed (normalized) expectation value of (a) $\widehat{Q}_{V_1}$, (b) $\widehat{Q}^2_{V_1}$ operators for $\vec{p}_I=6\vec{n}_I$.}\label{Qnnc}
 \end{figure}

As shown in Fig. \ref{Qn3}, just like the previous normalization tests, very good agreement is found for both $\widehat{Q}_{V_1}$ and $\widehat{Q}^2_{V_1}$ up to $t=10$, with $\vec{p}_I=6\vec{n}_I$. By pushing $j_{\mathrm{cap}}$ high enough, the relative error between the numerical state-sum and the analytical calculation is suppressed to below $10^{-10}$.

In Fig. \ref{Qnnc}, the $n$-winding corrections for the operators are plotted. We can clearly conclude that for $0 < t \le 1.5$, the $n_{\mathrm{cap}}=0$ terms are already sufficient to guarantee a $10^{-10}$ precision. In this small-$t$ domain, the highly complex exact analytical expressions of $\widehat{Q}^q_{V_1}$ can be further expanded into classical polynomials of $t$.

This analytical simplification can be clearly understood if one looks at the explicit equation for the $n_{\mathrm{cap}}=0$ main term of $\widehat{Q}_{V_1}$:
\begin{equation}
    \begin{split}
        &\langle\psi^t_{\Gamma,\{g\}}|\widehat{Q}_{V_1}|\psi^t_{\Gamma,\{g'\}}\rangle=\frac{(it)^3}{64}\langle\psi^t_{\Gamma,\{g\}}|\widehat{J}^1\widehat{J}^2\widehat{J}^3|\psi^t_{\Gamma,\{g'\}}\rangle\\
        &=-\frac{(it)^3}{64}\prod\limits_{i=1,2,3}\left(\frac{\trace(\tau_ig_i'\bar{g}_i^T)}{\sinh(z_i)}\left[2\frac{z_i^2}{t\sinh(z_i)}-z_i\frac{\cosh(z_i)}{\sinh^2(z_i)}+\frac{1}{\sinh(z_i)}\right]\right),
    \end{split}
\end{equation}
where $\cosh(z_i):=\frac{1}{2}\trace(g_i'\bar{g}_i^T)$. Because the internal $1/t$ cancels the leading coefficient $t^3$, the whole $\widehat{Q}_{V_1}$ expectation value behaves as a polynomial function truncated up to $\mathcal{O}(t^3)$.

\begin{figure}[h!]
 \centering         
 \includegraphics[height=5cm]{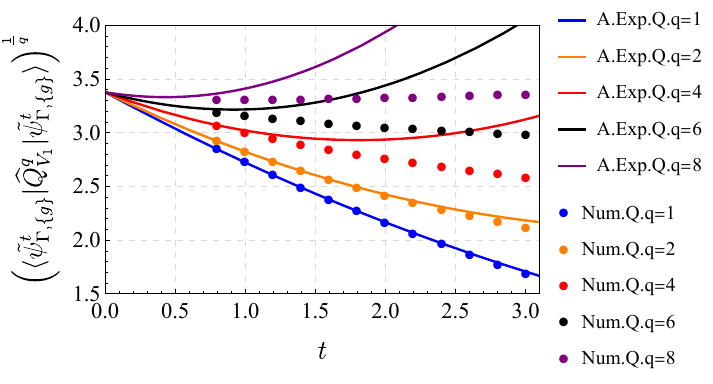} 
 \caption{Comparison between the $t^2$ order expansion of $\left(\langle\tilde{\psi}^t_{\Gamma,\{g\}}|\widehat{Q}^q_{V_1}|\tilde{\psi}^t_{\Gamma,\{g\}}\rangle\right)^{\frac{1}{q}}$ ($n_{\mathrm{cap}}=0$, $q=1,2,4,6,8$, labeled as "A.Exp.Q") in coherent state representation and numerical results in spin-network representation (labeled as "Num.Q").}\label{nc3}
 \end{figure}

In Fig. \ref{nc3}, we compare the analytical Taylor expansion up to $\mathcal{O}(t^2)$ (labeled as "A.Exp.Q") with the numerical outputs (labeled as "Num.Q") for $\left(\langle\tilde{\psi}|\widehat{Q}^q_{V_1}|\tilde{\psi}\rangle\right)^{1/q}$, where $q=1,2,4,6,8$. The physical reason to compute the $1/q$ root is very straightforward: at the semiclassical limit ($t \to 0$), these quantum expectation values should give the same semiclassical limit. Thus, the splitting between different $q$ curves at $t>0$ is a direct representation of the quantum fluctuations embedded in the canonical LQG theory.

Because higher powers of $q$ enlarge the non-Gaussian quantum fluctuations at the tail of the wave function, we can see that for larger $q$, the curves deviate from the main $q=1$ line more significantly around $t \sim 1$. However, for $t \le 1$, all numerical results converge perfectly onto their corresponding $t^2$ analytical series. This confirms that the higher-order inputs required for Eq. (\ref{VexpandNew}) are well-controlled.

 \subsubsection{Convergence of the Volume Operator Expansion}

Based on the successful calculation of the $\widehat{Q}^q_{V_1}$ operators, we can now investigate the actual physical volume operator $\widehat{V}_{V_1}$. By applying the semi-classical formula (\ref{Vexpand}), one can obtain the perturbative series of the volume operator up to $\hbar^{k+1}$ order.

 \begin{figure}[h!]
 \centering
    \begin{subfigure}[b]{0.45\textwidth}
         \centering
 \includegraphics[height=5cm]{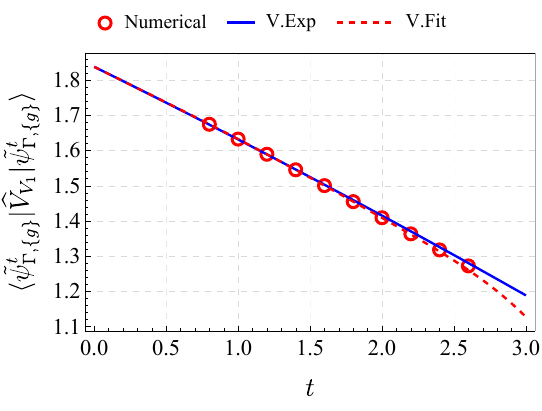}
         \caption{Expectation value of $\widehat{V}_{V_1}$}
     \end{subfigure}
    \begin{subfigure}[b]{0.45\textwidth}
         \centering
 \includegraphics[height=5cm]{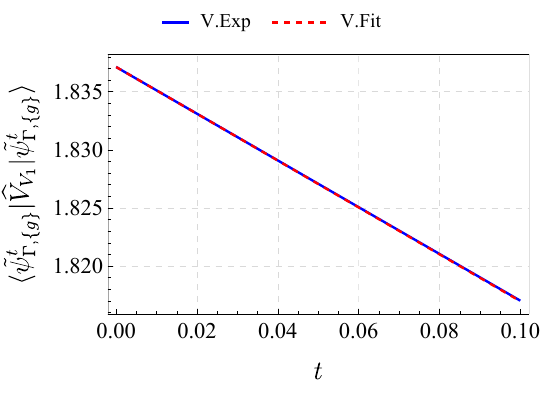}
         \caption{Close-up image for $0\leq t\leq 0.1$}
     \end{subfigure}
 \caption{Comparison between the fitted results obtained from the numerical calculation and analytical result for the $\bar{h}$-order expansion of the volume operator. (a) The red circles labeled as "Numerical" are the actual numerical results for the expectation value of the volume operator, while the orange dashed line labeled as "V.Fit" represents the Non-linearly fitted results from the numerically computed values of the volume operator from our numerical model. Also, in the same graph, the blue line labeled as "V.Exp" represents the analytical results obtained by calculating the operator expansion of the volume operator up to $\hbar$ order, using the analytically calculated exact expectation value of $\widehat{Q}^q_{V_1}$ operators up to $q=8$. (b) The close-up image of the linear order terms obtained using non-linear fitting of the numerical data ("V.Fit") and the analytical expansion ("V.Exp").}\label{ExpV3}
 \end{figure}

In Fig. \ref{ExpV3}(a), it is shown that the eigenvalues of $\widehat{V}_{V_1}$ (obtained numerically by matrix diagonalization) agree well with the analytical curve derived from summing the $\widehat{Q}^q_{v_1}$ operators in the small-$t$ regime.

To see the accuracy more clearly, Fig. \ref{ExpV3}(b) provides a close-up view for very small $t \in (0, 0.1)$. We performed a polynomial fit based on the raw numerical data points, and compared its coefficients with the exact analytical Taylor expansion. The analytical expression truncated to the second order gives:
\begin{equation*}
    V_{\text{analytic}} \approx 1.83712 - 0.200938\, t - 0.00176928\, t^2.
\end{equation*}
At the same time, the pure numerical fit yields:
\begin{equation*}
    V_{\text{numeric-fit}} \approx 1.83712 - (0.201091 \pm 0.000837)\, t - (0.005889 \pm 0.003371)\, t^2.
\end{equation*}
The good matching between these two polynomials—especially the fact that the analytical first-order coefficient sits well within the statistical error bounds of the numerical fit—proves that the semi-classical operator expansion (\ref{Vexpand}) captures the correct quantum geometrical physics at leading $\hbar$ orders.

To summarize our discussion on the gauge-variant 3-bridges, we want to emphasize three points. Firstly, the numerical results for $\widehat{Q}^q_{V_1}$ operators match the analytical results across the whole $t \le 10$ region if the $n \neq 0$ windings are included. Secondly, setting $n_{\mathrm{cap}}=0$ is practically sufficient for evaluating the small-$t$ semi-classical area. Finally, the polynomial fitting from our independent state-sum algorithm produces the same linear order corrections as the theoretical analytical expansion. This means that, at least for simple diagonal expectation values, the traditional expansion method proposed by Giesel and Thiemann \cite{Giesel:2006um} is indeed reliable.

\subsection{Matrix elements of the volume operator on gauge-variant 3-bridges}

Next, the matrix elements (non-diagonal expectation values) of the volume operator on the gauge-variant 3-bridges graph are discussed. First, two different sets of $SL(2,\mathbb{C})$ group elements on the three edges, denoted as $\{g\}$ and $\{g'\}$, are generated. Namely, $\{g'\}$ is defined by setting the extrinsic curvature to zero ($\vec{z}'_I=0$) and rotating the first momentum vector by an angle $\theta$: $\vec{p}'_1=\left(6\cos\theta,6\sin\theta,0\right)$, with $\vec{p}'_2=\left(0,6,0\right)$ and $\vec{p}'_3=\left(0,0,6\right)$. The reference set $\{g\}$ is defined by setting $\vec{z}_I=0$, $\vec{p}_1=\left(6,0,0\right)$, $\vec{p}_2=\left(0,6,0\right)$, and $\vec{p}_3=\left(0,0,6\right)$. Subsequently, by constructing the corresponding Thiemann coherent states $\psi^t_{\Gamma,\{g\}}$ and $\psi^t_{\Gamma,\{g'\}}$, the matrix elements of the $\widehat{Q}^q_{V_1}$ operators can be calculated both analytically and numerically. Based on these results, the matrix elements of the volume operator are investigated.

\subsubsection{Matrix elements of the $\widehat{Q}^q_{V_1}$ operators}

First, the numerical results for the $\widehat{Q}^q_{V_1}$ operators are validated by making a direct comparison with the analytical formulas. It is well-known that in the pure coherent state representation, the matrix elements of the $\widehat{Q}^q_{V_1}$ operators can be explicitly calculated analytically \cite{Thiemann:2000bx}. Here, the normalized matrix elements for the $\widehat{Q}^q_{V_1}$ operators are defined as:
\begin{equation}\label{ExpDecay}
    \langle\tilde{\psi}^t_{\Gamma,\{g\}}|\widehat{Q}^q_{V_1}|\tilde{\psi}^t_{\Gamma,\{g'\}}\rangle =\frac{\langle\psi^t_{\Gamma,\{g\}}|\widehat{Q}^q_{V_1}|\psi^t_{\Gamma,\{g'\}}\rangle}{{||\psi^t_{\Gamma,\{g\}}||\cdot||\psi^t_{\Gamma,\{g'\}}||}}.
\end{equation}

\begin{figure}[h!]
 \centering         
 \includegraphics[height=5cm]{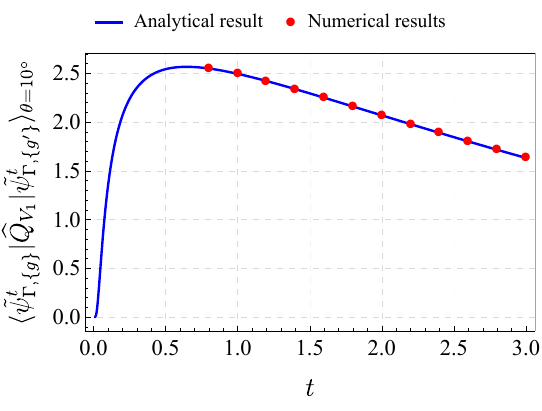}
 \includegraphics[height=5cm]{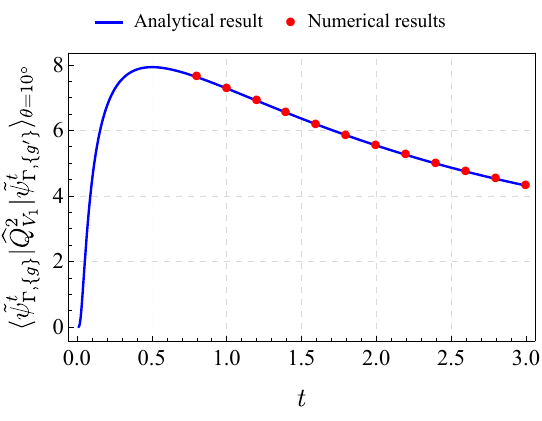}
 \includegraphics[height=5cm]{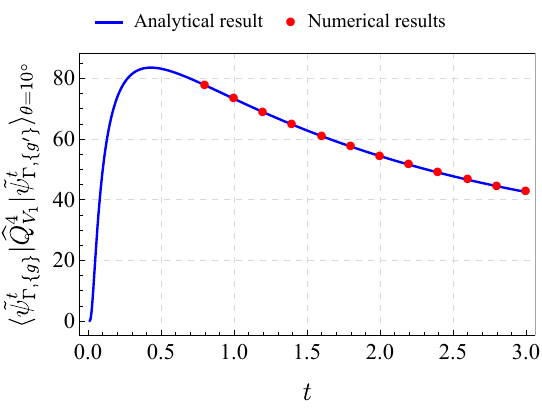}
 \includegraphics[height=5cm]{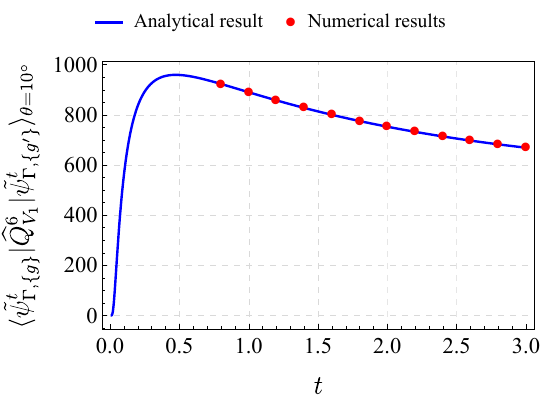}
 \caption{Comparison between analytical and numerical results for $\widehat{Q}_{V_1}$, $\widehat{Q}^2_{V_1}$, $\widehat{Q}^4_{V_1}$ and $\widehat{Q}^6_{V_1}$ operators, $\theta=10^{\circ}$. The blue line labeled "Analytical result" represents the $n_{\mathrm{cap}}=0$ analytical result computed in pure coherent state representation. The orange dots are exact results obtained numerically in the spin-network representation}\label{QMat}
 \end{figure}

Following the same procedure discussed in the previous subsection, a consistency check is performed. It can be seen from Fig. \ref{QMat} that the numerically obtained matrix elements of the $\widehat{Q}_{V_1}$, $\widehat{Q}^2_{V_1}$, $\widehat{Q}^4_{V_1}$, and $\widehat{Q}^6_{V_1}$ operators are in good agreement with the purely analytical coherent state evaluations.

Furthermore, one can observe from Fig. \ref{QMat} that as the semi-classical parameter $t$ approaches zero, all the matrix elements for the $\widehat{Q}^q_{V_1}$ operators vanish. This is because the exponential term involved in equation (\ref{ExpDecay}) decays to zero much faster than any polynomial function of $t$ in the $t \to 0$ limit. To solve this problem, we need to consider instead the Berezin symbol \cite{Berezin:1974du} (also see our companion paper \cite{Liliu:202603A} for details) of $\widehat{Q}^q_{V_1}$, namely the matrix element divided by the overlapping function:
 \begin{equation}
     [\widehat{Q}^q_{V_1}](\{g\},\{g'\}):=\frac{\langle\psi^t_{\Gamma,\{g\}}|\widehat{Q}^q_{V_1}|\psi^t_{\Gamma,\{g'\}}\rangle}{\langle\psi^t_{\Gamma,\{g\}}|\psi^t_{\Gamma,\{g'\}}\rangle}.
 \end{equation}

\begin{figure}[h!]
 \centering
    \begin{subfigure}[b]{0.45\textwidth}
         \centering
 \includegraphics[height=4cm]{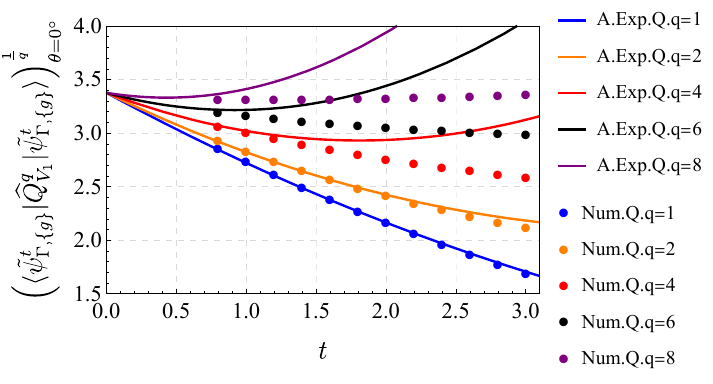} 
         \caption{$\theta=0^\circ$}
     \end{subfigure}
     \begin{subfigure}[b]{0.45\textwidth}
         \centering
 \includegraphics[height=4cm]{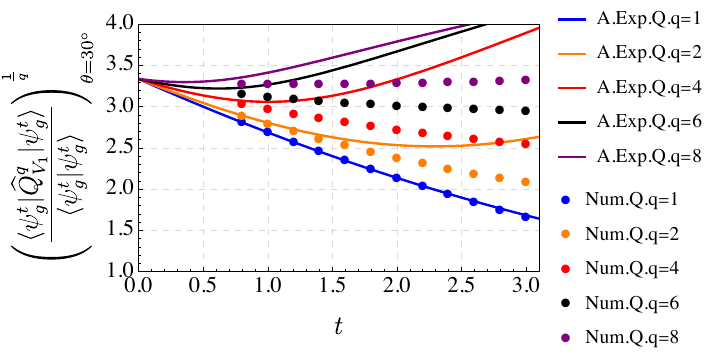} 
         \caption{$\theta=30^\circ$}
     \end{subfigure}
 \caption{Comparison between (a) $\left(\langle\tilde{\psi}^t_{\Gamma,\{g\}}|\widehat{Q}^q_{V_1}|\tilde{\psi}^t_{\Gamma,\{g\}}\rangle\right)^{\frac{1}{q}}$ for $\theta=0^\circ$ and (b) $\left(\frac{\langle\psi^t_{\Gamma,\{g\}}|\widehat{Q}^q_{V_1}|\psi^t_{\Gamma,\{g'\}}\rangle}{\langle\psi^t_{\Gamma,\{g\}}|\psi^t_{\Gamma,\{g'\}}\rangle}\right)^{1/q}$ for $\theta=30^\circ$.}\label{QnncMat}
 \end{figure}

Similar to the case of the expectation value, here we also compare $\left(\frac{\langle\psi^t_{\Gamma,\{g\}}|\widehat{Q}^q_{V_1}|\psi^t_{\Gamma,\{g'\}}\rangle}{\langle\psi^t_{\Gamma,\{g\}}|\psi^t_{\Gamma,\{g'\}}\rangle}\right)^{1/q}$ for $q=1,2,4,6,8$ as they all converge to the same semi-classical limit and is thus directly comparable. Fig. \ref{QnncMat} plots the comparison between the $\mathcal{O}(t^2)$ analytical expansion (labeled as "A.Exp.Q") and the numerical results (labeled as "Num.Q") for $\left(\frac{\langle\psi^t_{\Gamma,\{g\}}|\widehat{Q}^q_{V_1}|\psi^t_{\Gamma,\{g'\}}\rangle}{\langle\psi^t_{\Gamma,\{g\}}|\psi^t_{\Gamma,\{g'\}}\rangle}\right)^{1/q}$. In Fig. \ref{QnncMat}(a), the $\theta=0^\circ$ case is plotted, in which case $\left(\frac{\langle\psi^t_{\Gamma,\{g\}}|\widehat{Q}^q_{V_1}|\psi^t_{\Gamma,\{g\}}\rangle}{\langle\psi^t_{\Gamma,\{g\}}|\psi^t_{\Gamma,\{g\}}\rangle}\right)^{1/q}=\left(\langle\tilde{\psi}^t_{\Gamma,\{g\}}|\widehat{Q}^q_{V_1}|\tilde{\psi}^t_{\Gamma,\{g\}}\rangle\right)^{1/q}$ for normalized coherent states is identical to the diagonal expectation value previously presented in Fig. \ref{nc3}. In Fig. \ref{QnncMat}(b), the Berezin symbols are plotted for the separated $\theta=30^\circ$ configuration. By comparing the two sub-figures, it is concluded that even when the classical orientation of $g'_1$ is chosen to be substantially deviated from $g_1$ ($\theta=30^\circ$), the actual numerical deviation in the polynomial moments remains remarkably small in the semi-classical small-$t$ domain.

Furthermore, the numerical data points converge perfectly to the corresponding analytical series up to $q=8$ for $t\leq 0.8$. This indicates that the second-order $\hbar$ expansion of the matrix element of the non-analytic volume operator is reliable, at least within the geometric boundary $0^\circ\leq\theta\leq30^\circ$.

\subsubsection{Evaluation of the full volume operator matrix elements}

\begin{figure}[h!]
 \centering
    \begin{subfigure}[b]{0.45\textwidth}
         \centering
 \includegraphics[height=5cm]{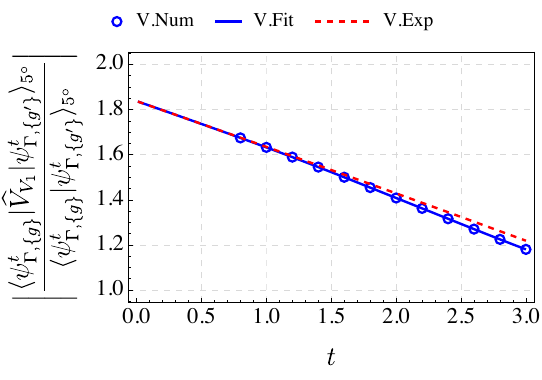} 
         \caption{$\theta=5^\circ$}
     \end{subfigure}
     \begin{subfigure}[b]{0.45\textwidth}
         \centering
 \includegraphics[height=5cm]{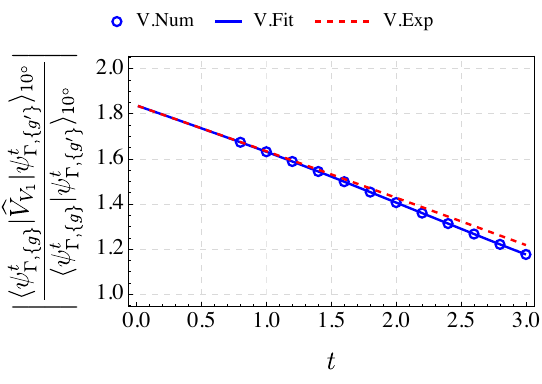} 
         \caption{$\theta=10^\circ$}
     \end{subfigure}
          \begin{subfigure}[b]{0.45\textwidth}
         \centering
 \includegraphics[height=5cm]{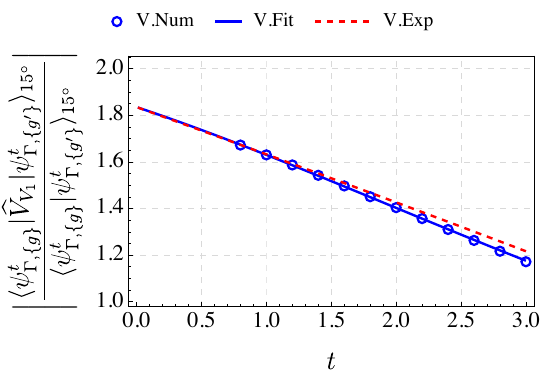} 
         \caption{$\theta=15^\circ$}
     \end{subfigure}
          \begin{subfigure}[b]{0.45\textwidth}
         \centering
 \includegraphics[height=5cm]{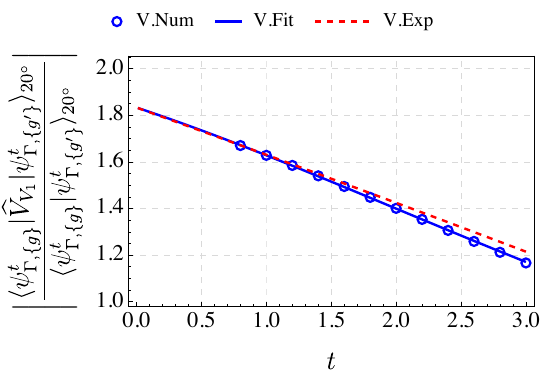} 
         \caption{$\theta=20^\circ$}
     \end{subfigure}
          \begin{subfigure}[b]{0.45\textwidth}
         \centering
 \includegraphics[height=5cm]{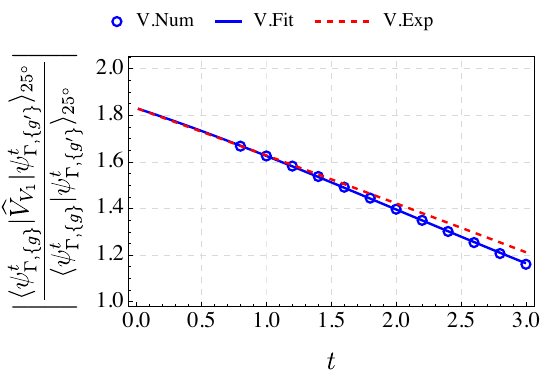} 
         \caption{$\theta=25^\circ$}
     \end{subfigure}
          \begin{subfigure}[b]{0.45\textwidth}
         \centering
 \includegraphics[height=5cm]{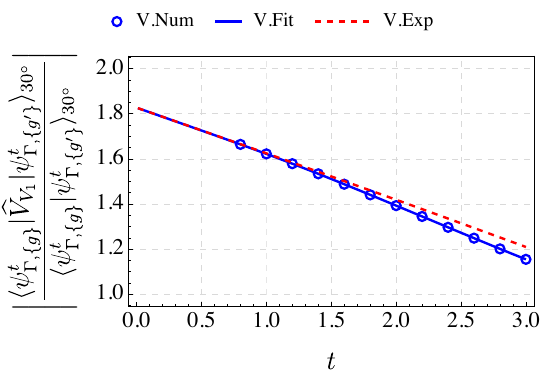} 
         \caption{$\theta=30^\circ$}
     \end{subfigure}
 \caption{Comparison between results of the normalized matrix element of the volume operator with the intersecting angle $\theta=5^{\circ}$, $\theta=10^{\circ}$, $\theta=15^{\circ}$, $\theta=20^{\circ}$, $\theta=25^{\circ}$, $\theta=30^{\circ}$ obtained both numerically and analytically. "V.Num" labels the matrix elements of the volume operator obtained numerically in the spin-network representation, "V.Fit" is the power series fitting of the numerical results, and "V.Exp" stands for the second-order expansions obtained in pure coherent state representation.}\label{VolumeM2}
 \end{figure}

Now, the final results regarding the matrix elements of the volume operator on the gauge-variant 3-bridges are presented. Fig. \ref{VolumeM2} shows the Berezin symbol of the volume operator computed for various angles ($\theta=5^{\circ}, 10^{\circ}, \dots, 30^{\circ}$). The blue circles represent the discrete state-sum outputs (labeled as "V.Num"). The red line ("V.Exp") corresponds to the analytical formula utilizing the newly proposed operator expansion given in Eq. (\ref{VexpandNew}). Finally, the solid blue line represents a pure numerical polynomial fit ("V.Fit"). It is shown that all curves converge into a single trajectory as $t \to 0$.

\begin{figure}[h!]
 \centering
    \begin{subfigure}[b]{0.45\textwidth}
         \centering
 \includegraphics[height=4.3cm]{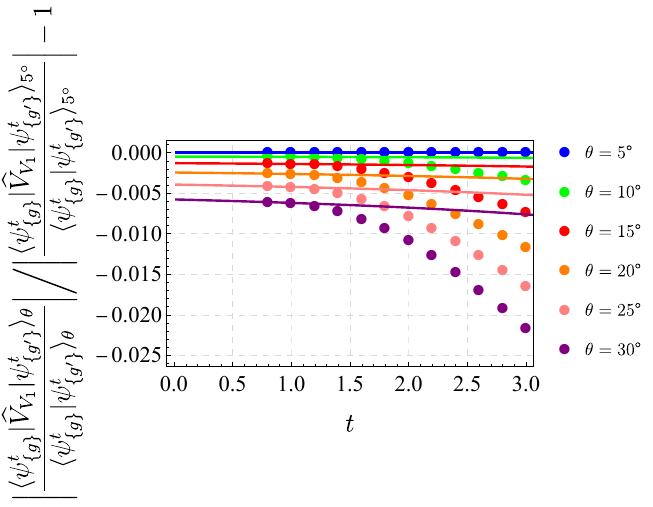} 
     \end{subfigure}
     \begin{subfigure}[b]{0.45\textwidth}
         \centering
 \includegraphics[height=3.7cm]{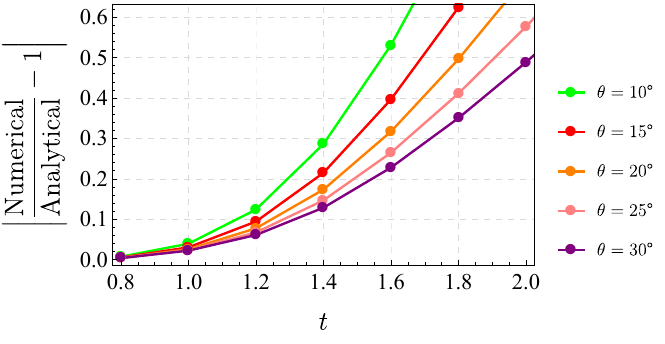} 
     \end{subfigure}
 \caption{On the left, the difference between the matrix elements of $\widehat{V}_{V_1}$ (after removing the exponential term $M$) with the intersecting angle $\theta=10^{\circ},15^{\circ},20^{\circ},25^{\circ},30^{\circ}$ and $\theta=5^{\circ}$ obtained both numerically and analytically. On the right, the deviation by percentage between analytical and numerical results shown on the left}\label{VolumeM3Mat}
 \end{figure}

To quantify the geometric dependence, Fig. \ref{VolumeM3Mat} measures the numerical differences between matrix elements evaluated at different angles $\theta$. In Fig. \ref{VolumeM3Mat}(a), the absolute difference relative to the baseline $\theta=5^\circ$ case is calculated. It can be seen that these deviations are negligibly small (on the order of $10^{-3}$). The continuous lines map the analytical $\mathcal{O}(t^2)$ Taylor series, while the points trace the numerical data. This alignment is consistent with the visual overlap previously observed in Fig. \ref{VolumeM2}. Furthermore, Fig. \ref{VolumeM3Mat}(b) computes the relative difference between the numerical outputs and the analytical approximations, which originate from the truncation up to second order terms in equation (\ref{VexpandNew}), proving that sufficient convergence is robustly achieved for $t\leq 1$.

\subsection{Expectation value of the volume operator on gauge-invariant 4-bridges}

\begin{figure}[h!]
 \centering
 \includegraphics[height=6cm]{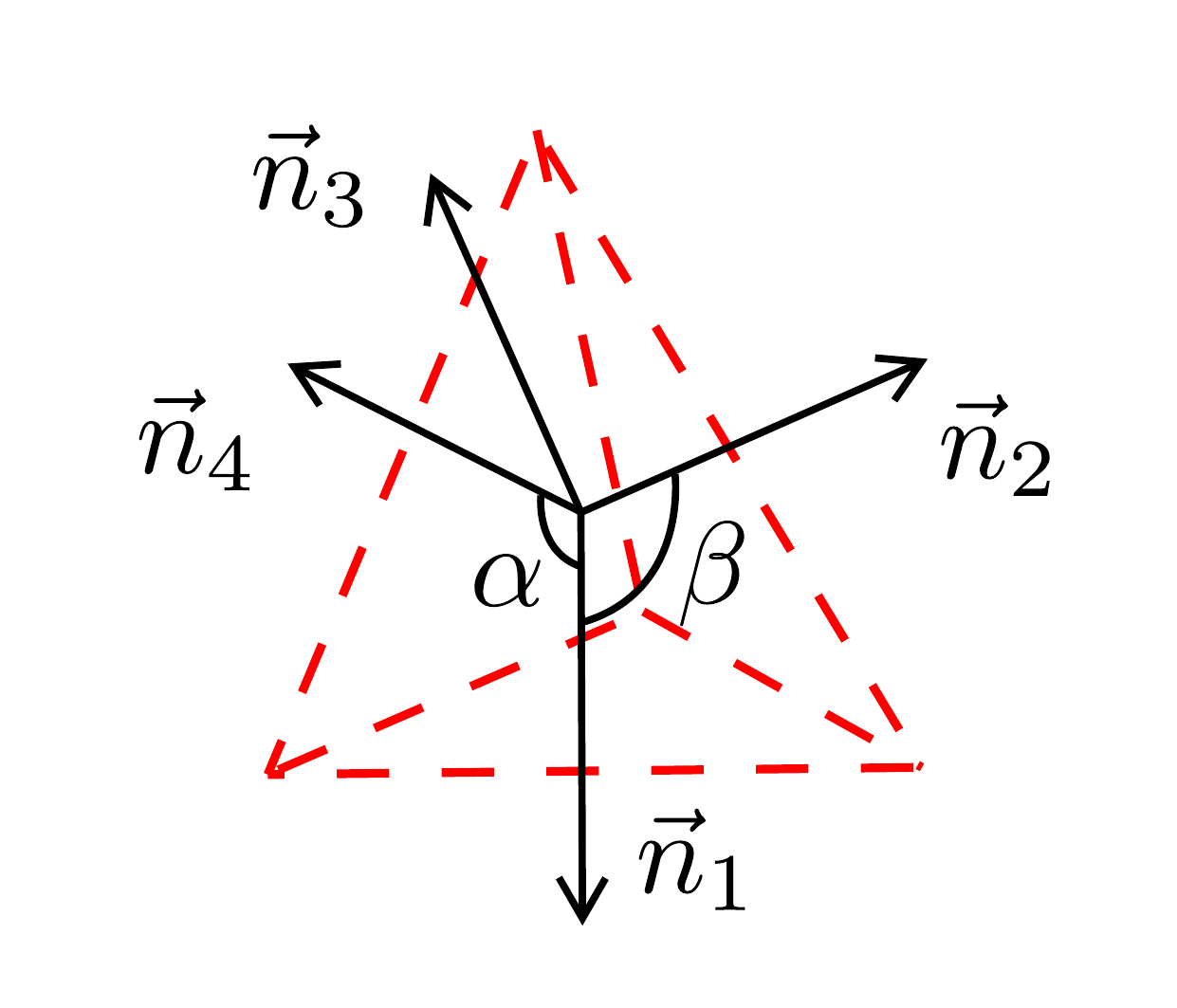} 
 \caption{The illustration of a tetrahedron as being determined by its four face normals and two angles $\alpha$ and $\beta$ between normals.}\label{V4Angle}
 \end{figure}

In this subsection, the expectation value of the physical volume operator evaluated over gauge-invariant 4-bridges is computed. Namely, this corresponds to a closed fundamental spin-network where two 4-valent vertices are connected by four shared edges. Because each vertex, together with its four attached edges, classically defines a 3D tetrahedron, this topology provides a direct platform to compare the continuous classical volume against exact canonical LQG quantum expectation values. For these gauge-invariant states, the classical flux closure condition $\sum_{I=1}^{4}\vec{p}_I=0$ needs to be enforced. Once this closure relation is satisfied, it is well-known from classical geometry that the spatial shape of a tetrahedron is uniquely determined (as parameterized in Fig. \ref{V4Angle}) by fixing the 4 areas of the triangles (as the length of the flux vector) and the two internal angles: $\alpha$, defined as the angle between normal vectors $\vec{n}_1$ and $\vec{n}_2$, and $\beta$, the angle between $\vec{n}_1$ and $\vec{n}_3$.

Hence, two distinct geometric configurations are studied. The first setting is the regular tetrahedron, where the face normals are symmetrically distributed (e.g., $\vec{n}^{\mathrm{Tet}}_1=(1,0,0)$), the lengths of all four fluxes are identical, and the characteristic angles are fixed to $\alpha=\beta=\pi-\arccos(1/3)$. The second setting models a set of irregular tetrahedra. In this deformed case, the lengths of all four fluxes are preserved, but the angles are given according to $\alpha=\pi-\arccos(1/3)-a$ and $\beta=\pi-\arccos(1/3)+a$, where $a$ acts as a continuous deformation parameter bounded such that the classical tetrahedron does not degenerate.

\subsubsection{Regular gauge-invariant 4-bridges}

\begin{figure}[h!]
 \centering         
 \includegraphics[height=5cm]{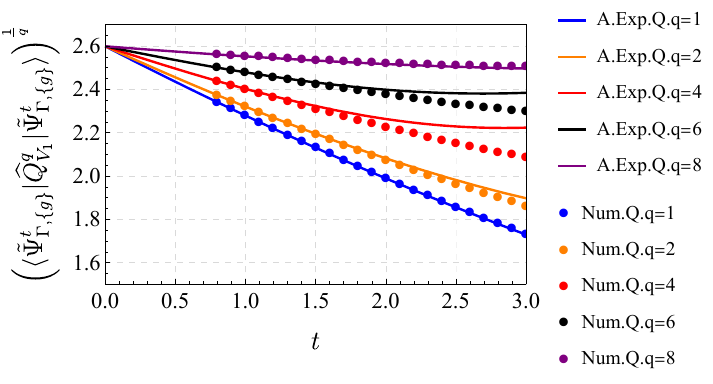} 
 \caption{Comparison between the numerical results (labeled as"Num.Q") and analytical expansions (labeled as "A.Exp.Q") of $\left(\langle\tilde{\Psi}^t_{\Gamma,\{g\}}|\widehat{Q}^q_{V_1}|\tilde{\Psi}^t_{\Gamma,\{g\}}\rangle\right)^{\frac{1}{q}}$ for the regular gauge-invariant 4-bridges graph.}\label{V4Q}
 \end{figure}
 
For the fully symmetric regular 4-bridges, the coherent state parameters are defined as $\vec{p}_I=6\vec{n}^{\mathrm{Tet}}_I$ for $I \in \{1,2,3,4\}$. In the classical limit, this maps to a regular tetrahedron with uniform macroscopic face areas.

Fig. \ref{V4Q} presents the direct comparison between the $\mathcal{O}(t^2)$ analytical series (labeled as "A.Exp.Q") and the non-perturbative numerical results ("Num.Q") mapped for the operator roots $\left(\langle\tilde{\Psi}|\widehat{Q}^q_{V_1}|\tilde{\Psi}\rangle\right)^{1/q}$ ($q=1,2,4,6,8$), evaluated purely on the normalized gauge-invariant basis $|\tilde{\Psi}^t_{\Gamma,\{g\}}\rangle$.

As can be clearly seen from the graph, all the raw numerical data points converge identically onto their corresponding analytical expansions up to $q=8$ for $t \le 1$. Thus, the second-order $\hbar$ Taylor series of the gauge-invariant volume operator becomes reliable in the semi-classical domain via Eq. (\ref{VexpandNew}). Furthermore, when compared with the earlier observations in Fig. \ref{nc3} and Fig. \ref{QnncMat}, it seems that the series convergence for the regular 4-bridges topology is better than that of the simpler gauge-variant 3-bridges.
   
   \begin{figure}[h!]
 \centering
    \begin{subfigure}[b]{0.3\textwidth}
         \centering
 \includegraphics[height=3.5cm]{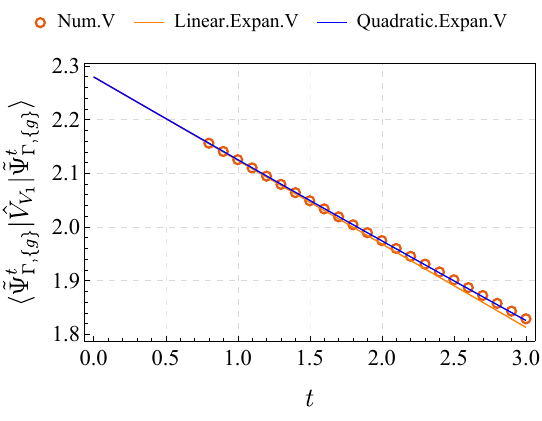} 
         \caption{Expectation value of $\widehat{V}_{V_1}$}
     \end{subfigure}
    \begin{subfigure}[b]{0.3\textwidth}
         \centering
 \includegraphics[height=3.5cm]{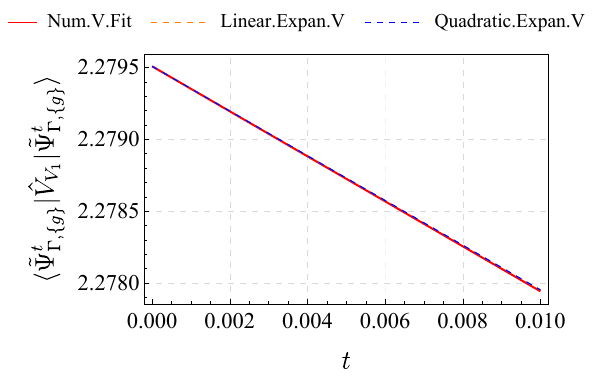} 
         \caption{Close-up image for $0\leq t\leq 0.01$}
     \end{subfigure}
     \begin{subfigure}[b]{0.3\textwidth}
         \centering
 \includegraphics[height=3.5cm]{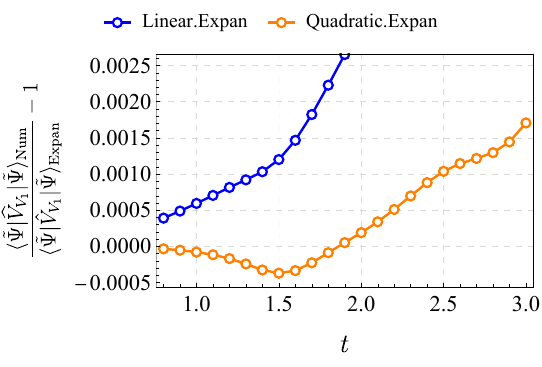} 
         \caption{Expansion convergence}
     \end{subfigure}
 \caption{(a) Comparison between the expectation value of the volume operator obtained directly (red circles, labeled as "Num.V") and via linear/quadratic operator expansion techniques (red/blue line, labeled as "Linear.Expan.V"/"Quadratic.Expan.V"). (b) Close-up image of (a) showing the near-perfect overlap of fitted numerical result, linear analytical expansion, and quadratic analytical expansion for $0\leq t\leq 0.01$. (c) the relative difference between the numerical results and linear/quadratic analytical expansion (labeled as "Linear.Expan"/"Quadratic.Expan").}\label{V4V}
    \end{figure}

Fig. \ref{V4V}(a) shows the cross-validation of the full volume expectation value $\langle\widehat{V}_{V_1}\rangle$ computed numerically ("Num.V"), contrasted with the linear $\mathcal{O}(t)$ analytical approximation (red line) and the quadratic $\mathcal{O}(t^2)$ approximation (blue line). Fig. \ref{V4V}(b) gives a localized close-up view restricted to $0 \le t \le 0.01$, where a good overlap among all three methods is observed. Specifically, the quadratic analytical expansion derived using our symbolic framework is 
\begin{align}
2.27951 - 0.155648 t + 0.0014293 t^2 \ .
\end{align} Simultaneously, an independent non-linear fit performed on the numerical result gives 
\begin{align} 2.27951 - (0.156577\pm0.00044) t + (0.00175132\pm0.00043) t^2 + \mathcal{O}(t^3)\ .
    \end{align}
This high level of consistency validates our numerical algorithm. Moreover, Fig. \ref{V4V}(c) plots the relative errors, indicating that the quadratic term accelerates the functional convergence towards the quantum matrix trace as $t$ grows towards $0.8$.

\subsubsection{Irregular gauge-invariant 4-bridges}

For the irregular geometry test cases, the internal angles of the four face normals are deformed as follows:
\begin{equation}
    \begin{split}
        \alpha&=\pi-\arccos(1/3)+a,\\
        \beta&=\pi-\arccos(1/3)-a,\\
        a&\in \left\{\frac{1}{6}, \frac{1}{3}, \frac{1}{2}, \frac{2}{3}, \frac{5}{6}, 1\right\}.
    \end{split}
\end{equation}
The detailed momentum parametrization defining each specific tetrahedron is formulated in TABLE \ref{tablep}. Similar to the regular scheme, the uniform macroscopic condition $|\vec{p}_{I}|=6$ is enforced across all four edges during the construction of the $SL(2,\mathbb{C})$ variables via Eq. (\ref{groupg}).

\begin{table}[h!]
\caption{List of 4-bridges considered}\label{tablep}
\begin{center}
\footnotesize
\renewcommand{\arraystretch}{1.5}
\begin{tabular}{ |c|c|c|c|c|c|c|c| } 
 \hline
 $\#$ & type & $\alpha$ & $\beta$ & $\vec{p}^1$ & $\vec{p}^2$ & $\vec{p}^3$ & $\vec{p}^4$\\ 
 \hline
 I & regular & $\pi - \mathrm{arccos}(\frac{1}{3})$ & $\pi - \mathrm{arccos}(\frac{1}{3})$ & $(6, 0, 0)$ & $(-2,5.66, 0)$ & $(-2, -2.83, -4.90)$ & $(-2, -2.83, 4.90)$\\ 
 \hline
 II & irregular & $\alpha_0+\frac{1}{6}$ & $\beta_0-\frac{1}{6}$ & $(6, 0, 0)$ & $(-2.91,5.25, 0)$ & $(-1.03, -2.92, -5.14)$ & $(-2.06, -2.32, 5.14)$\\ 
 \hline
 III & irregular & $\alpha_0+\frac{1}{3}$ & $\beta_0-\frac{1}{3}$ & $(6, 0, 0)$ & $(-3.74, 4.69, 0)$ & $(-0.04, -2.87, -5.27)$ & $(-2.22, -1.82, 5.27)$\\ 
 \hline
 IV & irregular & $\alpha_0+\frac{1}{2}$ & $\beta_0-\frac{1}{2}$ & $(6, 0, 0)$ & $(-4.47, 4.01, 0)$ & $(0.96, -2.66, -5.29)$ & $(-2.49, -1.34, 5.29)$\\ 
 \hline
 V & irregular & $\alpha_0+\frac{2}{3}$ & $\beta_0-\frac{2}{3}$ & $(6, 0, 0)$ & $(-5.07, 3.21, 0)$ & $(1.92, -2.30, -5.20)$ & $(-2.86, -0.91, 5.20)$\\ 
 \hline
  VI & irregular & $\alpha_0+\frac{5}{6}$ & $\beta_0-\frac{5}{6}$ & $(6, 0, 0)$ & $(-5.53, 2.32, 0)$ & $(2.84, -1.78, -4.97)$ & $(-3.31, -0.54, 4.97)$\\ 
  \hline
   VII & irregular & $\alpha_0+1$ & $\beta_0-1$ & $(6, 0, 0)$ & $(-5.84, 1.37, 0)$ & $(3.68, -1.12, -4.60)$ & $(-3.84, -0.25, 4.60)$\\ 
 \hline
\end{tabular}
\renewcommand{\arraystretch}{1}
\end{center}
\end{table}

\begin{figure}[h!]
 \centering
    \begin{subfigure}[b]{0.45\textwidth}
         \centering
 \includegraphics[height=3.8cm]{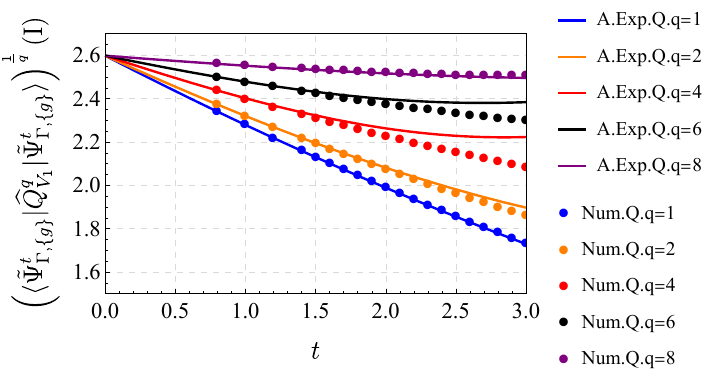} 
         \caption{Case I}
     \end{subfigure}
    \begin{subfigure}[b]{0.45\textwidth}
         \centering
 \includegraphics[height=3.8cm]{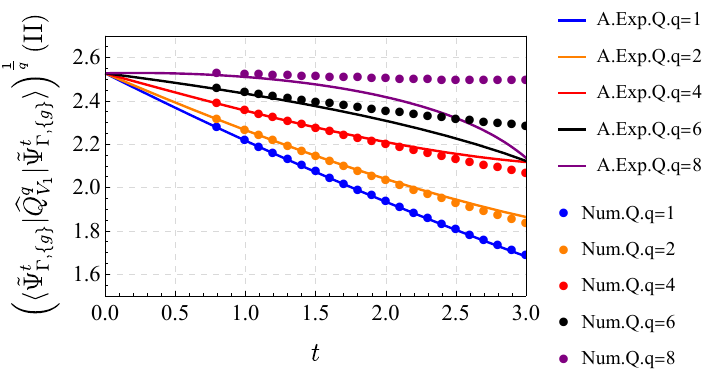} 
         \caption{Case II}
     \end{subfigure}
     \begin{subfigure}[b]{0.45\textwidth}
         \centering
 \includegraphics[height=3.8cm]{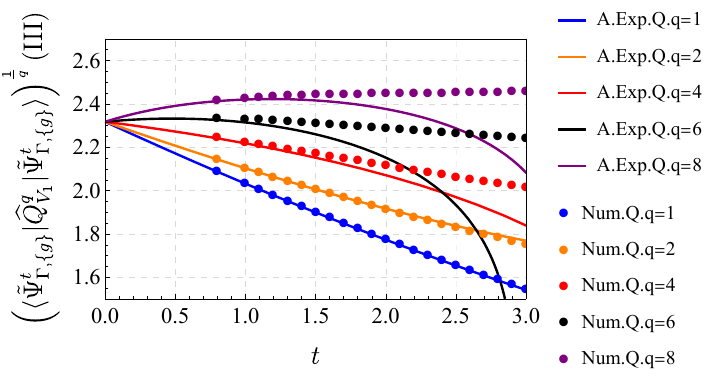} 
         \caption{Case III}
     \end{subfigure}
          \begin{subfigure}[b]{0.45\textwidth}
         \centering
 \includegraphics[height=3.8cm]{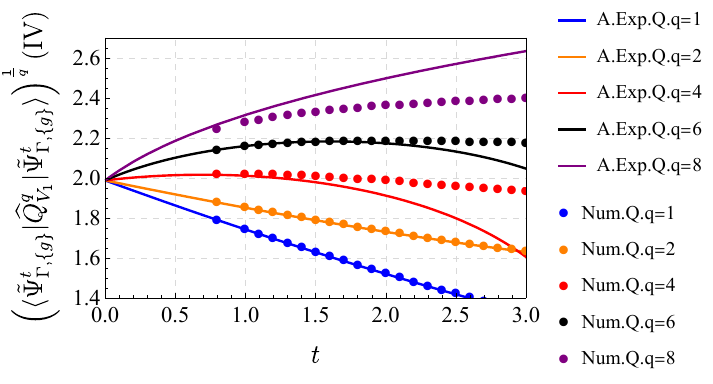} 
         \caption{Case IV}
     \end{subfigure}
          \begin{subfigure}[b]{0.45\textwidth}
         \centering
 \includegraphics[height=3.8cm]{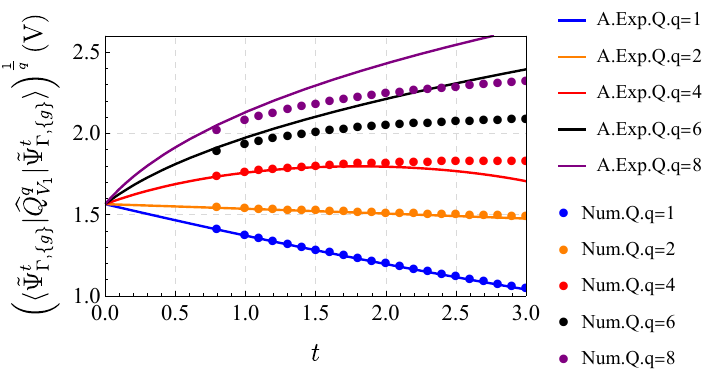} 
         \caption{Case V}
     \end{subfigure}
          \begin{subfigure}[b]{0.45\textwidth}
         \centering
 \includegraphics[height=3.8cm]{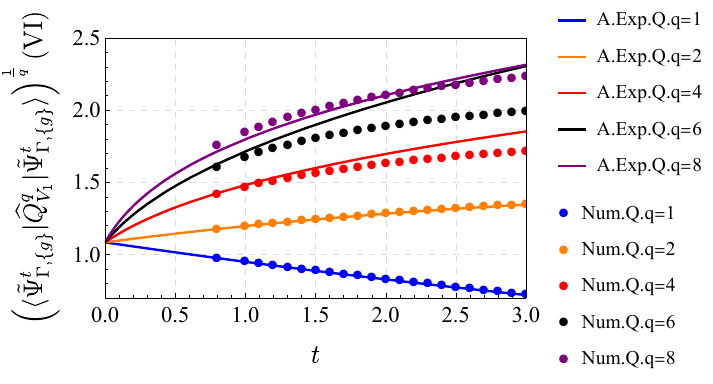} 
         \caption{Case VI}
     \end{subfigure}
          \begin{subfigure}[b]{0.45\textwidth}
         \centering
 \includegraphics[height=3.8cm]{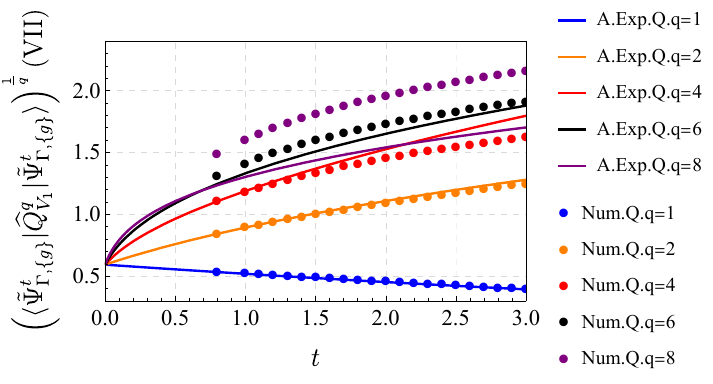} 
         \caption{Case VII}
     \end{subfigure}
 \caption{The comparison of $\left(\langle\tilde{\Psi}^t_{\Gamma,\{g\}}|\widehat{Q}^q|\tilde{\Psi}^t_{\Gamma,\{g\}}\rangle\right)^{\frac{1}{q}}$, $q=1,2,4,6,8$, with $|\tilde{\Psi}^t_{\Gamma,\{g\}}\rangle$ computed in our numerical model (in spin-network representation, dots) and via numerical integration (in coherent state representation, blue line) for seven cases of tetrahedron. The details of the shape and parametrization of all seven tetrahedra are given in TABLE \ref{tablep}.}\label{V4QI}
 \end{figure}

Fig. \ref{V4QI} outlines the comparative curves for the $\mathcal{O}(t^2)$ analytical operator expansions against the numerical data. It can be observed that for the highly deformed irregular geometries (specifically cases IV, V, VI, and VII), the quadratic Taylor expansion struggles to follow the numerical trajectory for heavily non-linear moments ($q=8$) when $t \ge 0.8$. This implies that as the spatial configuration of the vertex strongly deviates from maximum symmetry, the truncation errors associated with neglected higher-order $\hbar$ corrections become rapidly amplified.

To quantify this difference, the numerical results and the analytical expansions at $t=0.8$ are isolated and presented in TABLE \ref{tableexp}. From this table, two observations can be straightforwardly deduced. Firstly, for higher orders like $q=6$ and $q=8$, the evaluated ratio breaks away from 1 very quickly in the highly deformed geometries (cases V, VI, and VII), signaling the breakdown of the low-order perturbation. Secondly, bounded within the domain of $q \le 4$, the analytical series expansion maintains a sufficiently accurate alignment with the numerical values (whose structural errors are guaranteed below $10^{-10}$).

\begin{table}[h!]
\caption{Ratio between analytical expansions and numerical results}\label{tableexp}
\begin{center}
\footnotesize
\renewcommand{\arraystretch}{2}
\begin{tabular}{ |c|c|c|c|c|c|c|c| } 
 \hline
 $\#$ & type & $\alpha$ & $\beta$ & $ \frac{\langle\tilde{\Psi}^t_g|\widehat{Q}^2_{V_1}|\tilde{\Psi}^t_g\rangle_Q}{\langle\tilde{\Psi}^t_g|\widehat{Q}^2_{V_1}|\tilde{\Psi}^t_g\rangle_N}\big|_{t=0.8}$ & $ \frac{\langle\tilde{\Psi}^t_g|\widehat{Q}^4_{V_1}|\tilde{\Psi}^t_g\rangle_Q}{\langle\tilde{\Psi}^t_g|\widehat{Q}^4_{V_1}|\tilde{\Psi}^t_g\rangle_N}\big|_{t=0.8}$ & $ \frac{\langle\tilde{\Psi}^t_g|\widehat{Q}^6_{V_1}|\tilde{\Psi}^t_g\rangle_Q}{\langle\tilde{\Psi}^t_g|\widehat{Q}^6_{V_1}|\tilde{\Psi}^t_g\rangle_N}\big|_{t=0.8}$ & $ \frac{\langle\tilde{\Psi}^t_g|\widehat{Q}^8_{V_1}|\tilde{\Psi}^t_g\rangle_Q}{\langle\tilde{\Psi}^t_g|\widehat{Q}^8_{V_1}|\tilde{\Psi}^t_g\rangle_N}\big|_{t=0.8}$\\ 
 \hline
 I & regular & $\pi - \mathrm{arccos}(\frac{1}{3})$ & $\pi - \mathrm{arccos}(\frac{1}{3})$ & $1.00045$ & $1.002936$ & $1.00235$ & $0.998156$\\ 
 \hline
 II & irregular & $\alpha_0+\frac{1}{6}$ & $\beta_0-\frac{1}{6}$ & $1.00039$ & $1.00073$ & $0.992744$ & $0.981961$\\ 
 \hline
 III & irregular & $\alpha_0+\frac{1}{3}$ & $\beta_0-\frac{1}{3}$ & $1.00021$ & $0.995573$ & $0.981286$ & $0.991722$\\ 
 \hline
 IV & irregular & $\alpha_0+\frac{1}{2}$ & $\beta_0-\frac{1}{2}$ & $0.999938$ & $0.992316$ & $1.00767$ & $1.09235$\\ 
 \hline
 V & irregular & $\alpha_0+\frac{2}{3}$ & $\beta_0-\frac{2}{3}$ & $0.99969$ & $1.000066$ & $1.08456$ & $1.13845$\\ 
 \hline
  VI & irregular & $\alpha_0+\frac{5}{6}$ & $\beta_0-\frac{5}{6}$ & $1.00013$ & $1.02528$ & $1.09682$ & $0.818307$\\ 
  \hline
   VII & irregular & $\alpha_0+1$ & $\beta_0-1$ & $1.00438$ & $1.01956$ & $0.752336$ & $0.233708$\\ 
 \hline
\end{tabular}
\renewcommand{\arraystretch}{1}
\end{center}
\end{table}

     \begin{figure}[h!]
 \centering
    \begin{subfigure}[b]{0.45\textwidth}
         \centering
 \includegraphics[height=4.5cm]{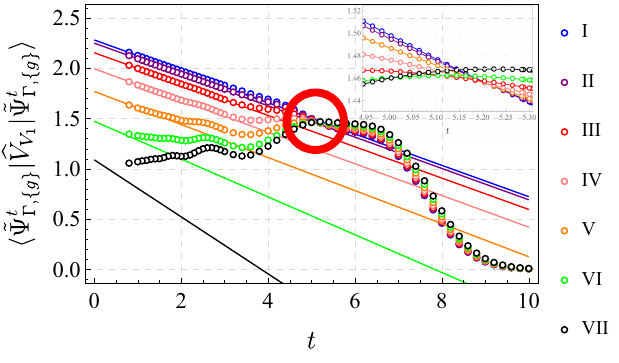} 
         \caption{Comparison with $t$-order expansion}
     \end{subfigure}
    \begin{subfigure}[b]{0.45\textwidth}
         \centering
 \includegraphics[height=4.5cm]{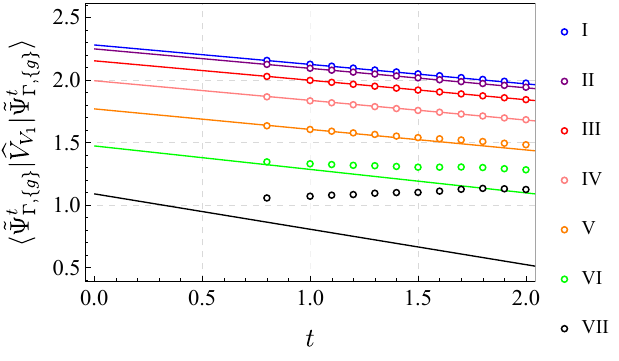} 
         \caption{Close-up image of (a)}
     \end{subfigure}
     \begin{subfigure}[b]{0.45\textwidth}
         \centering
 \includegraphics[height=4.5cm]{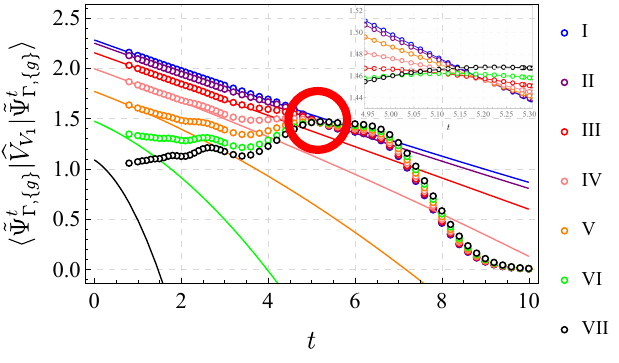} 
         \caption{Comparison with $t^2$-order expansion}
     \end{subfigure}
          \begin{subfigure}[b]{0.45\textwidth}
         \centering
 \includegraphics[height=4.5cm]{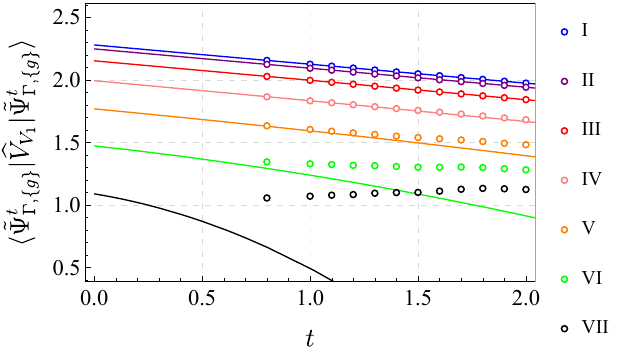} 
         \caption{Close-up image of (d)}
     \end{subfigure}
 \caption{Comparison of the expectation value of the volume operator in all seven cases (as described previously in TABLE \ref{tablep}) between the numerically calculated results and analytical expansion. (a) Comparison with the linear expansion, where the smaller graph in the upper right corner provides a zoom-in photo of the region $4.95\leq t\leq 5.3$ (the same region marked with the thick red circle on the main graph). (b) A close-up image of (a) in the range $0\leq t\leq 2$. (c) Comparison with the quadratic expansion. (d) A close-up image of (c) in the range $0\leq t\leq 2$. }\label{V4VI}
 \end{figure}

Utilizing these preliminary operator behaviors, the expectation value of the volume operator is computed across all seven cases in Fig. \ref{V4VI}. Fig. \ref{V4VI}(a) compares the numerical expectation values directly with their linear semi-classical predictions, while Fig. \ref{V4VI}(b) scales the local window down to the region $0 \le t \le 2$. In these plots, it is visibly demonstrated that for configurations I through V, the state-sum values successfully overlap with the analytical polynomial boundaries at $t=0.8$. Conversely, for cases VI and VII, a clear physical gap persists.

Another interesting phenomenon observed in Fig. \ref{V4VI}(a) and (c) is that all the full-quantum numerical results appear to intersect near the coordinate $t=5$. To dissect this behavior, a magnified sub-plot tracing the interval $4.95 \le t \le 5.3$ is attached to the main figures. It can be seen from this zoomed domain that there is no single absolute converging singularity. Rather, a sequence of localized crossings rapidly occurs throughout this narrow phase-space band. Furthermore, analyzing the asymptotic behavior reveals that for small values of $t$ (classical continuum limit), the expected volume maximizes for the regular tetrahedron (Case I). However, for $t>5$ (deep discrete quantum limit), this ordering is inverted, whereby the maximally irregular state (Case VII) yields the highest quantum geometrical volume. This inversion indicates a non-trivial reshuffling of the quantum volume ordering in the deep quantum regime. At the present stage, we regard it as a robust numerical observation rather than as a fully understood dynamical effect, and a more detailed interpretation will require further analytical investigation.

\begin{figure}[h!]
 \centering
    \begin{subfigure}[b]{0.3\textwidth}
         \centering
 \includegraphics[height=3.8cm]{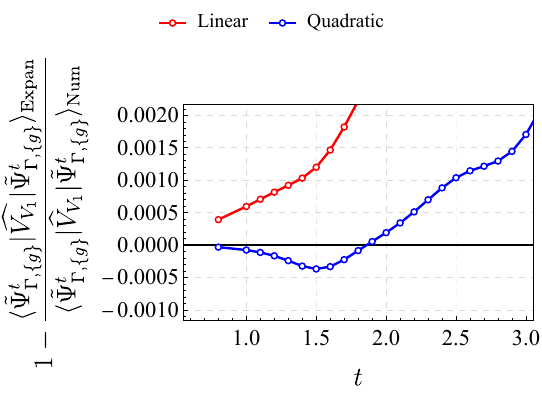} 
         \caption{I}
     \end{subfigure}
    \begin{subfigure}[b]{0.3\textwidth}
         \centering
 \includegraphics[height=3.8cm]{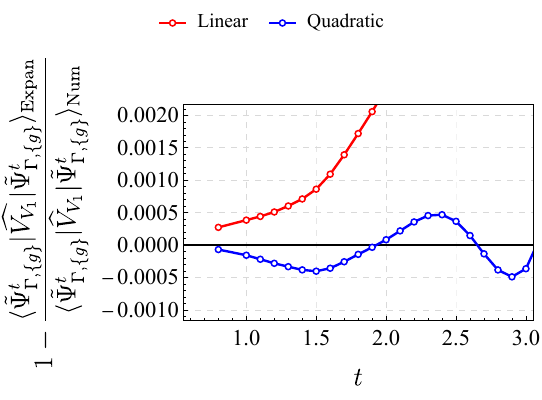} 
         \caption{II}
     \end{subfigure}
     \begin{subfigure}[b]{0.3\textwidth}
         \centering
 \includegraphics[height=3.8cm]{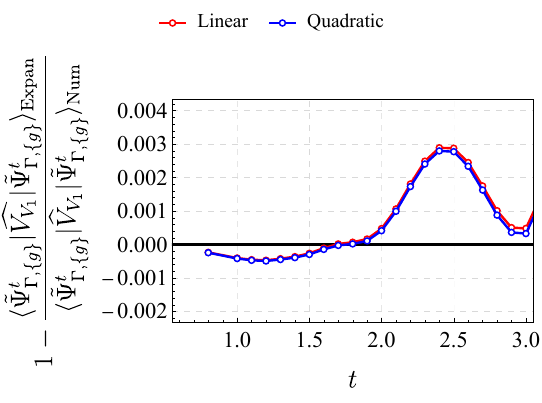} 
         \caption{III}
     \end{subfigure}
          \begin{subfigure}[b]{0.3\textwidth}
         \centering
 \includegraphics[height=3.8cm]{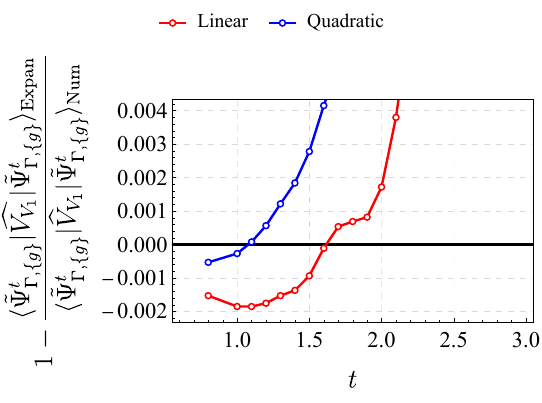} 
         \caption{IV}
     \end{subfigure}
               \begin{subfigure}[b]{0.3\textwidth}
         \centering
 \includegraphics[height=3.8cm]{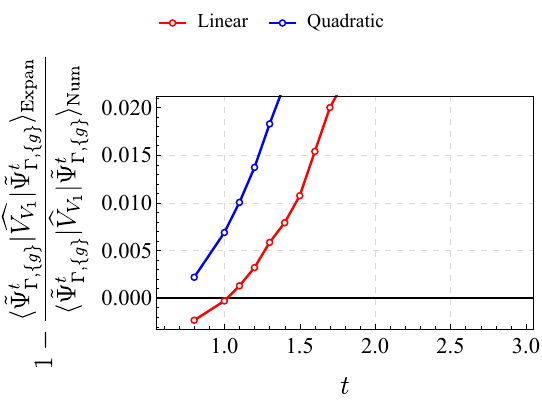} 
         \caption{V}
     \end{subfigure}
               \begin{subfigure}[b]{0.3\textwidth}
         \centering
 \includegraphics[height=3.8cm]{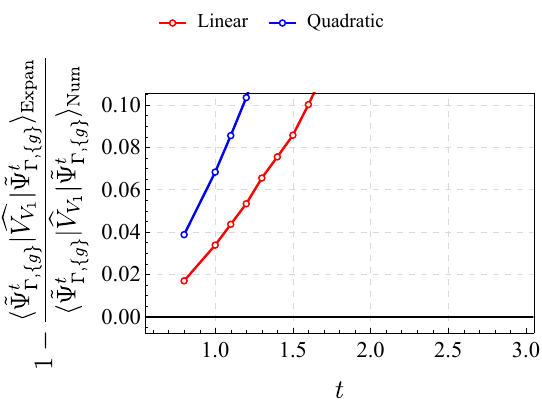} 
         \caption{VI}
     \end{subfigure}
               \begin{subfigure}[b]{0.3\textwidth}
         \centering
 \includegraphics[height=3.8cm]{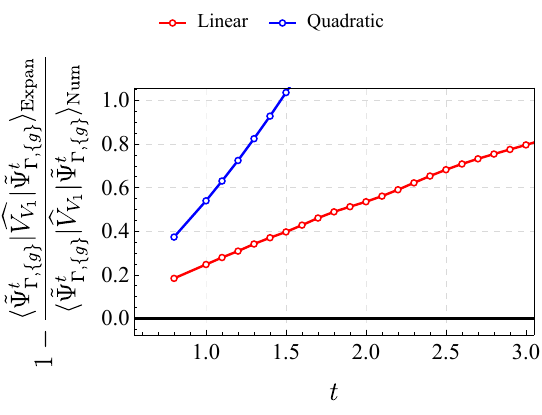} 
         \caption{VII}
     \end{subfigure}
 \caption{Comparison of the convergence of the linear (red) and quadratic (blue) expansions to the numerical results for the cases (I)-(VII).}\label{V4IDiff}
 \end{figure}

To evaluate the differences separating the linear and quadratic approximations, their respective analytical deviations from the numerical results are plotted in Fig. \ref{V4IDiff}. It can be noticed that the specific irregular geometry defining case (III) naturally emerges as a boundary separation point. Namely, for highly regular geometries (cases I and II), the added quadratic correction term evaluates larger than the base linear approximation, whereas for strongly deformed geometries (cases IV-VII), the $\mathcal{O}(t^2)$ value drops below the linear counterpart. Consequently, in relatively stable graphs (cases I, II, and IV), the quadratic curve successfully accelerates convergence to the non-perturbative numerical result. In contrast, for irregular states (cases V-VII), even the second-order series fails to reconstruct the underlying integral locally at $t \sim 1$. Thus, it is shown that canonical semi-classical expansions converge faster around highly symmetric geometric configurations. Interestingly, for case (III), both expansion orders map nearly identical rates of convergence. The origin of this geometric stability point is suggested to lie in the sign reversal zone of the second-order differential coefficient, which is another interesting point worth subsequent exploration. Finally, the second-order Taylor expansion for the volume operator across all cases are documented in TABLE \ref{tabV4I}.

\begin{table}[h!]
 \begin{center}
\footnotesize
\renewcommand{\arraystretch}{1.5}
\caption{Second order expansions of the expectation value of $\widehat{V}_{V_1}$}\label{tabV4I}
\begin{tabular}{ |c|c|c|c|c| } 
 \hline
 $\#$ & type & $\alpha$ & $\beta$ & $ \langle\tilde{\Psi}^t_{\Gamma,\{g\}}|\widehat{V}_{V_1}|\tilde{\Psi}^t_{\Gamma,\{g\}}\rangle$ obtained using operator expansion\\ 
 \hline
 $1$ & regular & $\pi - \mathrm{arccos}(\frac{1}{3})$ & $\pi - \mathrm{arccos}(\frac{1}{3})$ & $2.27951 - 0.155648 t + 0.0014293 t^2$\\ 
 \hline
 $2$ & irregular & $\alpha_0+\frac{1}{6}$ & $\beta_0-\frac{1}{6}$ & $2.24784 - 0.155558 t + 0.00113447 t^2$\\ 
 \hline
 $3$ & irregular & $\alpha_0+\frac{1}{3}$ & $\beta_0-\frac{1}{3}$ & $2.15269 - 0.155648 t + 0.0000288955 t^2$\\ 
 \hline
 $4$ & irregular & $\alpha_0+\frac{1}{2}$ & $\beta_0-\frac{1}{2}$ & $1.99346 - 0.157276 t - 0.00289922 t^2$\\ 
 \hline
 $5$ & irregular & $\alpha_0+\frac{2}{3}$ & $\beta_0-\frac{2}{3}$ & $1.76832 - 0.16413 t - 0.0115095 t^2$\\ 
 \hline
  $6$ & irregular & $\alpha_0+\frac{5}{6}$ & $\beta_0-\frac{5}{6}$ & $1.47213 - 0.187772 t - 0.0458238 t^2$\\ 
  \hline
   $7$ & irregular & $\alpha_0+1$ & $\beta_0-1$ & $1.08893 - 0.282977 t - 0.312581 t^2$\\ 
 \hline
\end{tabular}
\renewcommand{\arraystretch}{1}
\end{center}
\end{table}

 \subsection{Expectation value of the volume operator on a gauge-variant 6-valent vertex}

In this subsection, the expectation value of the volume operator acting on a gauge-variant 6-valent vertex is evaluated. Specifically, as depicted in Fig. \ref{fig3-1}(c), the numerical computation is performed with $|\tilde{\psi}^t_{\Gamma,\{g\}}\rangle$ configured as a gauge-variant 3-flower graph, which classically corresponds to a standard cube. The coherent state $|\tilde{\psi}^t_{\Gamma,\{g\}}\rangle$ is defined by the momenta $\vec{p}_1=(6,0,0)$, $\vec{p}_2=(0,6,0)$, and $\vec{p}_3=(0,0,6)$. Given the geometric correspondence $\mathrm{Area} = |\vec{p}|/2 = 3$, this macroscopic setting maps to a classical cube whose continuous volume equals $3\sqrt{3} \approx 5.19615$.

\begin{figure}[h!]
 \centering
    \begin{subfigure}[b]{0.45\textwidth}
         \centering
 \includegraphics[height=4cm]{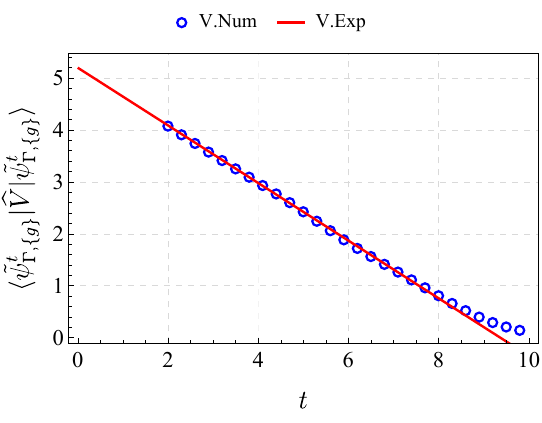} 
 \caption{Comparison result}
     \end{subfigure}
     \begin{subfigure}[b]{0.45\textwidth}
         \centering
 \includegraphics[height=4cm]{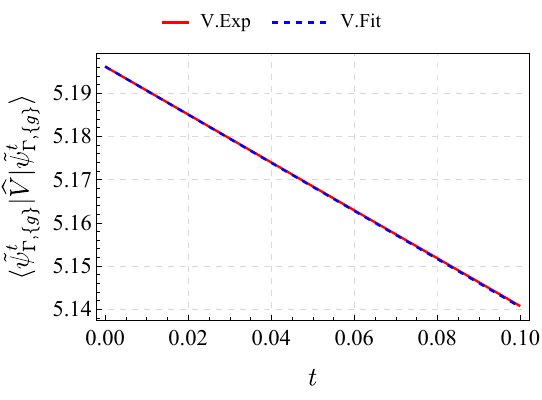} 
 \caption{Close-up image}
     \end{subfigure}
 \caption{The normalization factor and the expectation value of $\widehat{Q}^2$ operator computed in our numerical model (in spin-network representation, circle) and via numerical integration (in coherent state representation, blue line) for the gauge-variant 6-valent vertex.}\label{V6R}
 \end{figure}

Fig. \ref{V6R} presents a comparison between the numerically computed expectation value of the volume operator and the analytical next-to-leading order expansion formulated previously in \cite{Dapor:2017rwv}. In Fig. \ref{V6R}(a), the blue circles (labeled as "V.Num") represent the numerical results generated by our state-sum algorithm, while the red line ("V.Exp") denotes the next-to-leading order Taylor expansion. It can be seen that the linear semi-classical analytical expansion accurately matches the full quantum expectation values up to $t=8$. Furthermore, in Fig. \ref{V6R}(b), an independent polynomial fit of the raw numerical data ("V.Fit") is plotted against the analytical series. In the small-$t$ regime ($0 \leq t \leq 0.1$), the two curves are virtually indistinguishable. Quantitatively, the purely data-driven fitted curve reads $5.19615 - (0.557021\pm 0.000535) \cdot t$, whereas the theoretically derived next-to-leading order expansion is $5.19615 - 0.554805 \cdot t$. This alignment once again confirms the validity of our numerical algorithm.

\subsection{Consistency checks with classical geometric volumes}

At this point, the semi-classical expectation values evaluated at the classical continuum limit ($t \to 0$) are directly compared with their corresponding classical volumetric formulas. It is well-documented in the literature \cite{Flori:2008nw} that for the case of a 4-valent vertex (tetrahedron), the canonical volume operator defined in standard LQG must be artificially rescaled by an overall constant $C=\frac{3}{2\sqrt{2}}$ to correctly recover the classical geometric volume:
\begin{equation}
     \langle\tilde{\psi}|\widehat{V}_v|\tilde{\psi}\rangle_{t=0} = C \cdot V_{\mathrm{classical}}=\frac{3}{2\sqrt{2}} \cdot V_{\mathrm{classical}}.
\end{equation}
Conversely, for a 6-valent cubic graph, it is known that the expectation value inherently matches the classical volume without requiring any supplementary corrections, namely $C=1$.

In the present calculation, the same geometric coefficients also naturally emerge. Following the canonical coherent state parameterization standardized in \cite{Han:2019vpw}, the macroscopic classical face area is dynamically constrained by the momentum flux vector $\vec{p}_I$ via:
\begin{equation}
    \mathrm{Area}=\frac{1}{2}|\vec{p}_I|.
\end{equation}

TABLE \ref{tabVC} summarizes the comparison between the theoretical classical volume $V_{\mathrm{classical}}$ and semiclassically expanded results of the volume operator at $t=0$. The evaluation covers regular tetrahedra (gauge-invariant 4-bridges), multiple irregular tetrahedra, and the cube (gauge-variant 3-flower). It is shown that for all tetrahedral geometries—regardless of their degree of irregularity—the numerical limits yield the proportional factor $C=\frac{3}{2\sqrt{2}}$. Furthermore, for the cubic graph, the factor simplifies to $C=1$. These independent numerical results reproduce the analytical scaling properties previously reported in \cite{Flori:2008nw}, supplying yet another robust, independent validation of our state-sum methodology.

\begin{table}[h!]
 \begin{center}
\footnotesize
\renewcommand{\arraystretch}{1.5}
\caption{Comparison between $V_{\mathrm{classical}}$ and the expectation value of $\widehat{V}_v$ }\label{tabVC}
\begin{tabular}{ |c|c|c|c|c|c|c|c|c| } 
 \hline
 $\#$ & type & $\alpha$ & $\beta$ & $|\vec{p}_I$ & Area & $ \langle\tilde{\Psi}|\widehat{V}|\tilde{\Psi}\rangle_{t=0}$ & $V_{\mathrm{classical}}$ & $C$\\ 
 \hline
 $1$ & regularT & $\pi - \mathrm{arccos}(\frac{1}{3})$ & $\pi - \mathrm{arccos}(\frac{1}{3})$ & $6$ & $3$ & $2.27951$ & $2.14914$ & $\frac{3}{2\sqrt{2}}$\\ 
 \hline
 $2$ & irregularT & $\alpha_0+\frac{1}{6}$ & $\beta_0-\frac{1}{6}$ & $6$ & $3$ & $2.24784$ & $2.11928$ & $\frac{3}{2\sqrt{2}}$\\ 
 \hline
 $3$ & irregularT & $\alpha_0+\frac{1}{3}$ & $\beta_0-\frac{1}{3}$ & $6$ & $3$ & $2.15269$ & $2.02958$ & $\frac{3}{2\sqrt{2}}$\\ 
 \hline
 $4$ & irregularT & $\alpha_0+\frac{1}{2}$ & $\beta_0-\frac{1}{2}$ & $6$ & $3$ & $1.99346$ & $1.87945$ & $\frac{3}{2\sqrt{2}}$\\ 
 \hline
 $5$ & irregularT & $\alpha_0+\frac{2}{3}$ & $\beta_0-\frac{2}{3}$ & $6$ & $3$ & $1.76832$ & $1.38794$ & $\frac{3}{2\sqrt{2}}$\\ 
 \hline
  $6$ & irregularT & $\alpha_0+\frac{5}{6}$ & $\beta_0-\frac{5}{6}$ & $6$ & $3$ & $1.47213$ & $1.38794$ & $\frac{3}{2\sqrt{2}}$\\ 
  \hline
   $7$ & irregularT & $\alpha_0+1$ & $\beta_0-1$ & $6$ & $3$ & $1.08893$ & $1.02665$ & $\frac{3}{2\sqrt{2}}$\\ 
 \hline
 $8$ & Cube & \ & \ & $6$ & $3$ & $5.19615$ & $5.19615$ & $1$\\ 
 \hline
\end{tabular}
\renewcommand{\arraystretch}{1}
\end{center}
\end{table}

\subsection{Asymptotic behavior of the maximum eigenvalue of $\widehat{V}$}

So far, the matrix elements and expectation values of the volume operator have been extensively analyzed within the coherent state representation. However, it is also physically instructive to investigate the intrinsic eigenvalues and corresponding eigenvectors of the bare matrix $\widehat{V}_v$, evaluated within the pure spin-network intertwiner space.

Consequently, the eigensystem of $\widehat{V}_v$ is studied from two different perspectives. Firstly, the distribution of eigenvalues is examined in the large-$j$ limit. It is theoretically expected that for macroscopic spins, the absolute maximum eigenvalue belonging to a local vertex $v$ should asymptotically converge towards the continuous classical volume of its bounding polyhedron. Secondly, the eigenvectors are structurally analyzed by computing their projection overlaps with the semiclassical coherent state. As will be shown, the eigenvector corresponding to the maximum eigenvalue provides the dominant contribution to the overlap distribution.

\subsubsection{Convergence of the maximum eigenvalue to classical geometry}

Analogous to the properties observed for the phase-space expectation values, the maximum eigenvalue extracted from the discrete spin-network matrix diagonalization can be mapped to the classical volumetric formulas by applying the same macroscopic relation $j = 2\mathrm{Area}$.

\begin{figure}[h!]
 \centering
    \begin{subfigure}[b]{0.23\textwidth}
         \centering
 \includegraphics[height=2.5cm]{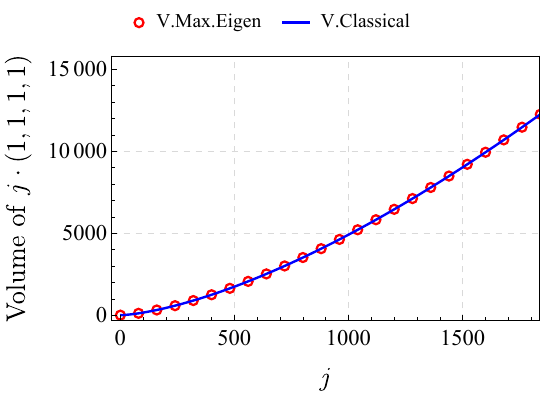} 
         \caption{$\vec{j}=j\cdot(1,1,1,1)$}
     \end{subfigure}
    \begin{subfigure}[b]{0.23\textwidth}
         \centering
 \includegraphics[height=2.5cm]{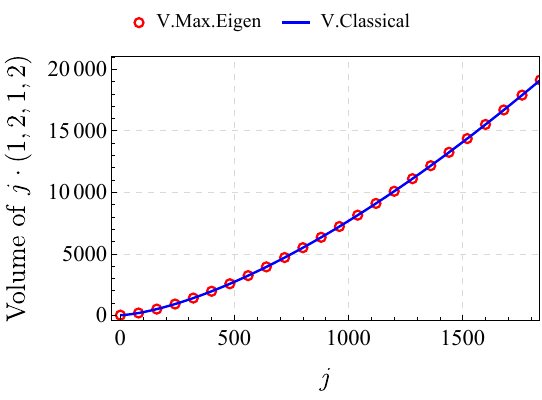} 
         \caption{$\vec{j}=j\cdot(1,2,1,2)$}
     \end{subfigure}
     \begin{subfigure}[b]{0.23\textwidth}
         \centering
 \includegraphics[height=2.5cm]{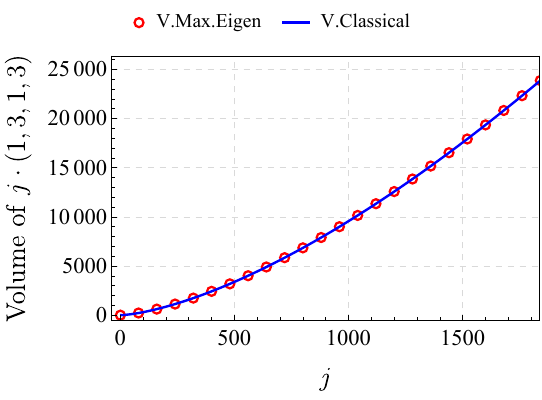} 
         \caption{$\vec{j}=j\cdot(1,3,1,3)$3}
     \end{subfigure}
          \begin{subfigure}[b]{0.23\textwidth}
         \centering
 \includegraphics[height=2.5cm]{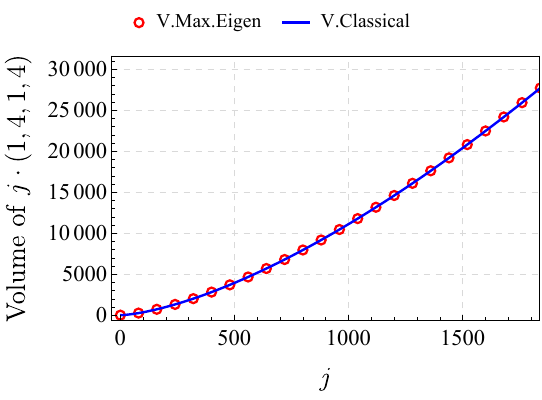} 
         \caption{$\vec{j}=j\cdot(1,4,1,4)$}
     \end{subfigure}
  \begin{subfigure}[b]{0.23\textwidth}
         \centering
 \includegraphics[height=2.5cm]{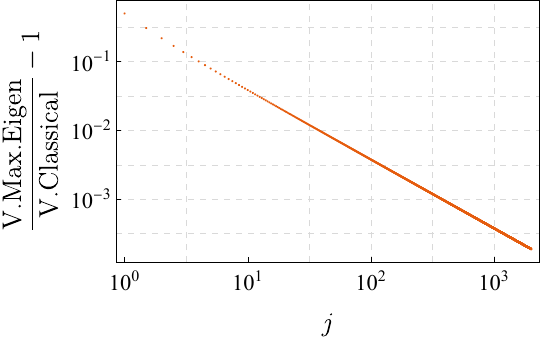} 
         \caption{$\vec{j}=j\cdot(1,1,1,1)$}
     \end{subfigure}
    \begin{subfigure}[b]{0.23\textwidth}
         \centering
 \includegraphics[height=2.5cm]{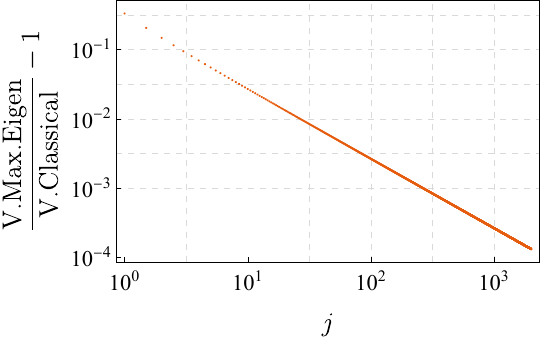} 
         \caption{$\vec{j}=j\cdot(1,2,1,2)$}
     \end{subfigure}
     \begin{subfigure}[b]{0.23\textwidth}
         \centering
 \includegraphics[height=2.5cm]{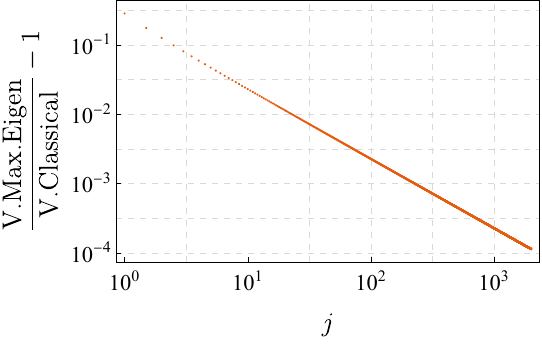} 
         \caption{$\vec{j}=j\cdot(1,3,1,3)$}
     \end{subfigure}
          \begin{subfigure}[b]{0.23\textwidth}
         \centering
 \includegraphics[height=2.5cm]{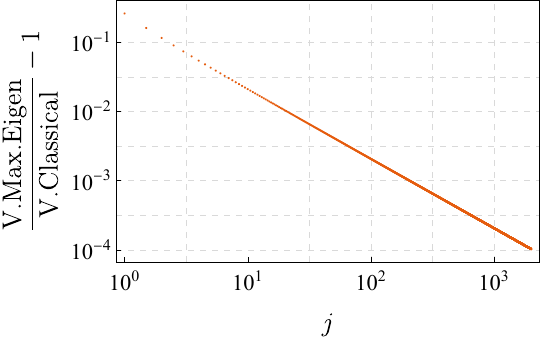} 
         \caption{$\vec{j}=j\cdot(1,4,1,4)$}
     \end{subfigure}
 \caption{The difference between the maximum eigenvalue of the volume operator (labeled as "V.Max.Eigen") acting on a single 4-valent vertex and the corresponding classical volume ("V.Classical") by considering $j\sim 2\mathrm{Area}$.}\label{V4Eig1}
 \end{figure}

Fig. \ref{V4Eig1} demonstrates the scaling relationships computed for the gauge-invariant 4-valent vertex. To maintain geometric consistency, the maximum eigenvalue of $\widehat{V}_v$ is renormalized by the same geometric factor $C$:
\begin{equation}
     V_{\mathrm{Max.Eigen}} = C \cdot V_{\mathrm{classical}} = \frac{3}{2\sqrt{2}} \cdot V_{\mathrm{classical}}.
\end{equation}
In Figs. \ref{V4Eig1}(a) through (d), direct comparisons are made for four different topological assignments, defined by setting the boundary spin vectors to $\vec{j}=j\cdot(1,1,1,1)$, $\vec{j}=j\cdot(1,2,1,2)$, $\vec{j}=j\cdot(1,3,1,3)$, and $\vec{j}=j\cdot(1,4,1,4)$, respectively. Here, the case $\vec{j}=j\cdot(1,1,1,1)$ corresponds to a perfectly regular tetrahedron, while $\vec{j}=j\cdot(1,4,1,4)$ describes a squeezed irregular tetrahedron. The base scale $j$ is evaluated over an extensive domain spanning $[\frac{1}{2},2000]$. It is observed that the maximum eigenvalue approximates the classical volume profile across the entire spectrum, maintaining an alignment even in the deep quantum region (small $j$).

\begin{figure}[h!]
 \centering
 \includegraphics[height=5cm]{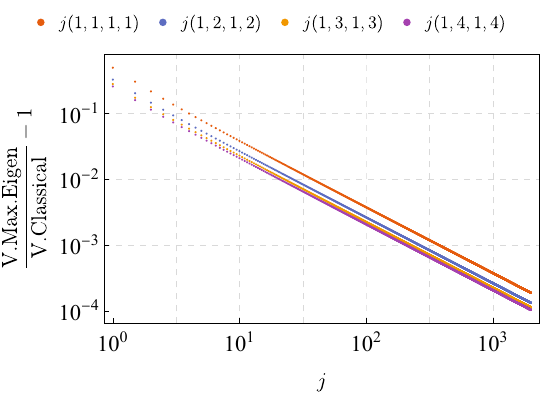} 
 \caption{Comparison between the $\mathrm{Log}_{10}$-$\mathrm{Log}_{10}$ obtained for the cases $\vec{j}=j\cdot(1,1,1,1)$, $\vec{j}=j\cdot(1,2,1,2)$, $\vec{j}=j\cdot(1,3,1,3)$, and $\vec{j}=j\cdot(1,4,1,4)$.}\label{V4EG}
 \end{figure}

To characterize the rate of this convergence, the relative differences between the rescaled maximum eigenvalues and the classical volume are plotted on $\mathrm{Log}_{10}$-$\mathrm{Log}_{10}$ axes in Figs. \ref{V4Eig1}(e) to (h). The asymptotic decay behavior is linear on the logarithmic scale. By performing power-law regressions, the relative ratios are empirically determined to be: 
 \begin{align}
 \frac{\mathrm{V.Max.Eigen}}{\mathrm{V.Classical}} \approx \left(  1+\frac{0.375416}{j^{2.30288}}, \; 1+\frac{0.263663}{j^{2.30281}},  \;  1+\frac{0.220858}{j^{2.29641}},  \; 1+\frac{0.205803}{j^{2.30281}} \right)
 \end{align} for the four respective geometries. Noticeably, the scaling exponent (the power of $j$) remains universally constant ($\approx 2.30$) irrespective of the geometric deformation. However, the leading amplitude coefficients systematically decrease as the state transitions from a regular to a highly irregular configuration. This universal power-law universality is distinctly highlighted in Fig. \ref{V4EG}, where all four logarithmic trajectories are mapped simultaneously.

\begin{figure}[h!]
 \centering
    \begin{subfigure}[b]{0.5\textwidth}
         \centering
 \includegraphics[height=5cm]{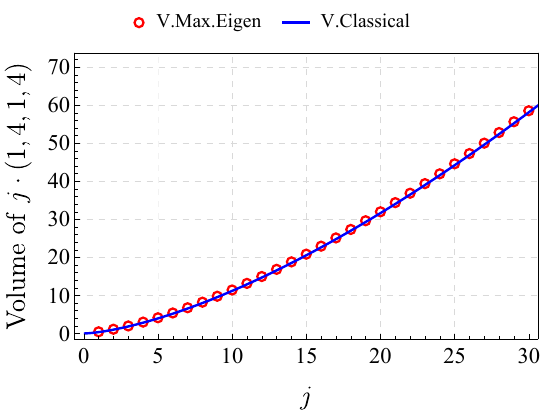} 
         \caption{Comparing "V.Max.Eigen" and "V.Classical"}
     \end{subfigure}
    \begin{subfigure}[b]{0.45\textwidth}
         \centering
 \includegraphics[height=4.3cm]{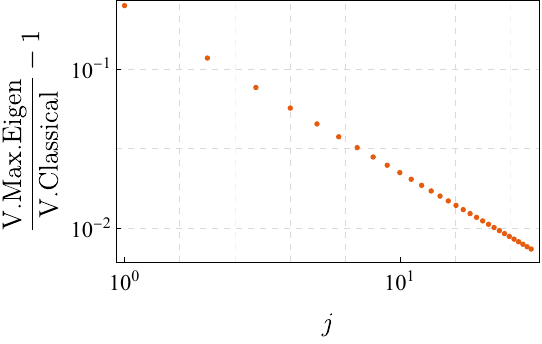} 
         \caption{Relative difference}
     \end{subfigure}
 \caption{The difference between the maximum eigenvalue of the volume operator (labeled as "V.Max.Eigen") acting on a single 6-valent vertex and the corresponding classical volume ("V.Classical") by considering $j\sim 2\mathrm{Area}$.}\label{V6EG1}
 \end{figure}

This analytical strategy is readily generalized to the 6-valent geometry. Fig. \ref{V6EG1} maps the scaling of the maximum eigenvalue for a symmetric 6-valent vertex bounded by $\vec{j}=j\cdot\left(1,1,1,1,1,1\right)$, contrasted against a classical regular cube defined by $\mathrm{Area}=j/2$. As shown in the fitting, the large-$j$ decay for the cubic geometry follows a fundamentally different characteristic power law, specifically given by 
 \begin{align} \frac{\mathrm{V.Max.Eigen}}{\mathrm{V.Classical}} \approx 1+\frac{0.223799}{j^{1.0043}} \ . 
 \end{align}

In summary, the purely discrete non-perturbative evaluation of the maximum eigenvalues inherently recovers the continuous classical volume limits. This scaling behavior complements the semi-classical characteristics derived previously from the complexifier coherent state formulations, thereby providing an exhaustive verification of the underlying spectral properties of the canonical LQG volume operator.

\subsubsection{Eigenstate probability overlap with the coherent state phase-space}

\begin{figure}[h!]
 \centering
    \begin{subfigure}[b]{0.45\textwidth}
         \centering
 \includegraphics[height=5cm]{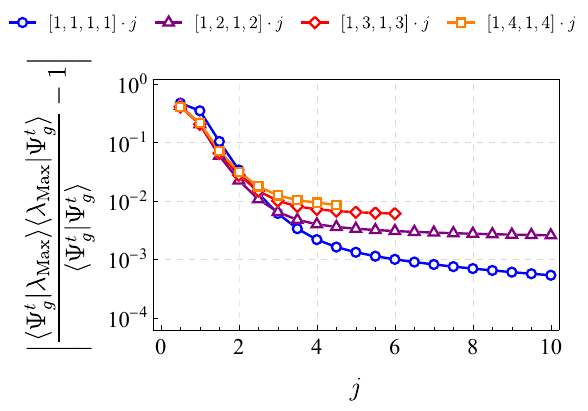} 
         \caption{Relative difference}
     \end{subfigure}
    \begin{subfigure}[b]{0.45\textwidth}
         \centering
 \includegraphics[height=5cm]{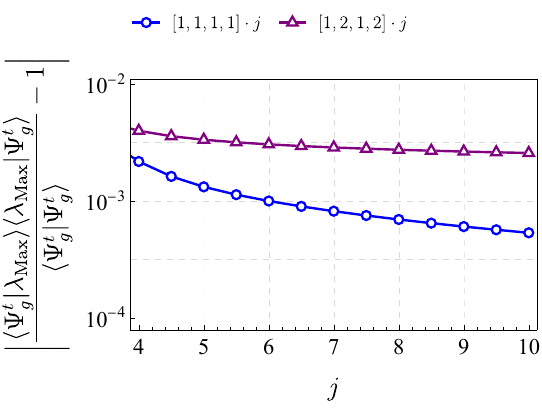} 
         \caption{Close-up image}
     \end{subfigure}
 \caption{Overlap between the eigenvector with the maximum eigenvalue and the coherent state as shown by the relative difference between $\langle\Psi^t_{\Gamma,\{g\}}|\lambda_{\mathrm{Max}}\rangle\langle\lambda_{\mathrm{Max}}|\Psi^t_{\Gamma,\{g\}}\rangle$ and $\langle\Psi^t_{\Gamma,\{g\}}|\Psi^t_{\Gamma,\{g\}}\rangle=\sum\limits_{\lambda}\langle\Psi^t_{\Gamma,\{g\}}|\lambda\rangle\langle\lambda|\Psi^t_{\Gamma,\{g\}}\rangle$.}\label{V4Overlap}
 \end{figure}

Finally, the probability distribution mapping the overlap transition amplitudes between the pure eigenvectors of $\widehat{V}_v$ and the macroscopic coherent state is analyzed. Fig. \ref{V4Overlap} monitors the fractional deficit defined by $\left| \frac{\langle\Psi^t_{\Gamma,\{g\}}|\lambda_{\mathrm{Max}}\rangle\langle\lambda_{\mathrm{Max}}|\Psi^t_{\Gamma,\{g\}}\rangle}{\langle\Psi^t_{\Gamma,\{g\}}|\Psi^t_{\Gamma,\{g\}}\rangle} - 1 \right|$. This observable essentially measures how the single isolated eigenstate $|\lambda_{\mathrm{Max}}\rangle$ (corresponding to the maximal eigenvalue $\lambda_{\mathrm{Max}}$) saturates the total identity resolution $\sum_{\lambda}\langle\Psi|\lambda\rangle\langle\lambda|\Psi\rangle$.

It is clearly illustrated in Fig. \ref{V4Overlap}(a) that as the macroscopic spin boundary $j$ increases, the isolated inner product $\langle\Psi|\lambda_{\mathrm{Max}}\rangle\langle\lambda_{\mathrm{Max}}|\Psi\rangle$ asymptotically converges towards the full state normalization $||\Psi||^2$ across all four geometric configurations. The highest degree of probability saturation is recorded for the fully symmetric case $\vec{j}=[1,1,1,1]\cdot j$, whereas the concentration becomes increasingly smeared for the more irregular cases ($[1,2,1,2]\cdot j$, $[1,3,1,3]\cdot j$, and $[1,4,1,4]\cdot j$). Furthermore, Fig. \ref{V4Overlap}(b) provides a magnified observation window constrained to $4\leq j\leq 10$. Within this region, a logarithmic-linear trajectory begins to emerge. However, definitively resolving the asymptotic functional form necessitates simulating significantly larger $j$ limits. Because the dimension of the intertwiner basis scales unfavorably, pushing $j$ to extreme limits numerically imposes severe computational bottlenecks. Thus, the precise functional derivation of this decay rate is left open for future study.

Physically, these computational results imply the following conclusion: the probability distribution governing the quantum geometric transition from discrete spin-networks to classical continua is overwhelmingly peaked around the single specific eigenstate harboring the absolute maximum volume.
This localization is most pronounced in highly symmetric graph.
By exploiting this spectral localization, it is feasible to construct faster truncation algorithms in the future—calculating only the tight neighborhood surrounding $|\lambda_{\mathrm{Max}}\rangle$ while safely discarding the exponentially suppressed tail—thereby allowing macroscopic LQG dynamics to be simulated at a fraction of the current computational cost without sacrificing physical fidelity.

\section{Summary and Outlook}

In this work, a comprehensive computational architecture \cite{LQG-Volume} is developed to evaluate the quantum action of the volume operator in canonical LQG, enabling the calculation of its expectation values. Several physically distinct scenarios are explicitly studied. These include the expectation values and non-diagonal coherent state matrix elements formulated on gauge-variant 3-bridges, the expectation values of regular and heavily irregular geometric tetrahedra constructed from gauge-invariant 4-bridges, and the macroscopic volume of a cube evaluated on a gauge-variant 3-flower graph containing a complex 6-valent vertex. By comparing these discrete numerical results with the analytical expansions obtained in coherent state representations, several key physical results are established:

Firstly, we verified our proposed numerical state-sum algorithm. It is shown that the numerically computed state normalization factors and the pure matrix elements of the higher-order $\widehat{Q}^q$ operators coincide with the semiclassical expansions with high accuracy (relative errors bounded strictly below $10^{-10}$). The fundamental advantage of this non-perturbative methodology is that the behavior of the volume operator can be evaluated directly in the deep quantum regime. Consequently, the calculation no longer completely relies on the semiclassical operator expansion techniques.

Secondly, it is verified that our new semi-classical operator expansion series (\ref{VexpandNew}) remains valid for the matrix elements of the volume operator. For both gauge-variant and gauge-invariant spin-network graphs, the low-order analytical Taylor expansion reliably reproduces the quantum expectation values and matrix elements across a considerably wide semi-classical transition region ($0.5 \leq t \leq 2$). Combining this result with the analytical proof in our companion paper \cite{Liliu:202603A}, the validity of the new expansion formula is fully examined.

Finally, the macroscopic geometric correspondence bounding the maximum absolute eigenvalue of the discrete volume matrix is mapped to its classical continuous polyhedron counterpart. It is demonstrated that an asymptotic correspondence is achieved even for some small $j$ representations. Furthermore, by calculating the phase-space overlap between the macroscopic coherent state and the eigenvectors of the volume operator, it is observed that the overlap of coherent states becomes increasingly concentrated on the maximal-volume eigenstate, particularly for highly symmetric networks. Physically, this strong probability localization has the potential for truncating and optimizing future numerical algorithms.
We also observed a non-trivial change in the relative magnitudes of volume observables in the deep quantum regime, where strongly deformed configurations can yield larger volumes than more symmetric ones. This suggests that the quantitative hierarchy of quantum geometric volumes is not a simple continuation of its semiclassical counterpart.

Based on these results, a viable new pathway to extract the semiclassical effective dynamics of full LQG is proposed. Specifically, the fundamental actions of foundational quantum geometric operators—including both the volume and holonomy operators—are implemented natively onto the intertwiner basis. This ultimately enables the unrestricted and accurate computation of the scalar Hamiltonian constraint acting on given spin-network boundaries. By evaluating the transformation matrices bridging the spin-network states and the corresponding coherent state configurations, the quantum gravitational dynamics can be extracted by deploying reduced phase-space quantization schemes and coherent state path integral.

Looking ahead, this numerical framework opens a viable route toward extracting effective dynamics directly from full canonical LQG. A natural next step is to extend the present algorithm to the Lorentzian Hamiltonian operator and to more general spin-network configurations, thereby moving beyond symmetry-reduced models. On the computational side, the principal bottleneck lies in the large tensor contractions and repeated diagonalizations; these are natural targets for GPU acceleration and large-scale parallelization. With such improvements, the present framework should become capable of probing substantially more complicated quantum geometries and of providing first-principles numerical input for the dynamics of non-symmetric black holes, deparametrized cosmological models, and coherent-state path-integral formulations of LQG.

\bibliographystyle{jhep}
\bibliography{sample}

@book{Thiemann:2007pyv,
    author = "Thiemann, Thomas",
    title = "{Modern Canonical Quantum General Relativity}",
    doi = "10.1017/CBO9780511755682",
    isbn = "978-0-511-75568-2, 978-0-521-84263-1",
    publisher = "Cambridge University Press",
    series = "Cambridge Monographs on Mathematical Physics",
    year = "2007"
}

@article{Ashtekar:2005qt,
    author = "Ashtekar, Abhay and Bojowald, Martin",
    title = "{Quantum geometry and the Schwarzschild singularity}",
    eprint = "gr-qc/0509075",
    archivePrefix = "arXiv",
    reportNumber = "IGPG-05-09-01, AEI-2005-132",
    doi = "10.1088/0264-9381/23/2/008",
    journal = "Class. Quant. Grav.",
    volume = "23",
    pages = "391--411",
    year = "2006"
}

@article{Gambini:2013ooa,
    author = "Gambini, Rodolfo and Pullin, Jorge",
    title = "{Loop quantization of the Schwarzschild black hole}",
    eprint = "1302.5265",
    archivePrefix = "arXiv",
    primaryClass = "gr-qc",
    reportNumber = "LSU-REL-022113",
    doi = "10.1103/PhysRevLett.110.211301",
    journal = "Phys. Rev. Lett.",
    volume = "110",
    number = "21",
    pages = "211301",
    year = "2013"
}

@article{Perez:2012wv,
    author = "Perez, Alejandro",
    title = "{The Spin Foam Approach to Quantum Gravity}",
    eprint = "1205.2019",
    archivePrefix = "arXiv",
    primaryClass = "gr-qc",
    doi = "10.12942/lrr-2013-3",
    journal = "Living Rev. Rel.",
    volume = "16",
    pages = "3",
    year = "2013"
}

@article{Ashtekar:2006rx,
    author = "Ashtekar, Abhay and Pawlowski, Tomasz and Singh, Parampreet",
    title = "{Quantum nature of the big bang}",
    eprint = "gr-qc/0602086",
    archivePrefix = "arXiv",
    reportNumber = "IGPG-06-2-1",
    doi = "10.1103/PhysRevLett.96.141301",
    journal = "Phys. Rev. Lett.",
    volume = "96",
    pages = "141301",
    year = "2006"
}

@article{Li:2022dei,
    author = "Li, Haida and Li, Shengzhi and Ma, Yongge",
    title = "{Connection Dynamics of Reduced 5-dimensional Kaluza-Klein Theory and Its Deparametrization}",
    eprint = "2207.14726",
    archivePrefix = "arXiv",
    primaryClass = "gr-qc",
    month = "7",
    year = "2022"
}

@article{Han:2020iwk,
    author = "Han, Muxin and Li, Haida and Liu, Hongguang",
    title = "{Manifestly gauge-invariant cosmological perturbation theory from full loop quantum gravity}",
    eprint = "2005.00883",
    archivePrefix = "arXiv",
    primaryClass = "gr-qc",
    doi = "10.1103/PhysRevD.102.124002",
    journal = "Phys. Rev. D",
    volume = "102",
    number = "12",
    pages = "124002",
    year = "2020"
}

@article{Brunnemann:2004xi,
    author = "Brunnemann, Johannes and Thiemann, Thomas",
    title = "{Simplification of the spectral analysis of the volume operator in loop quantum gravity}",
    eprint = "gr-qc/0405060",
    archivePrefix = "arXiv",
    reportNumber = "PI-2004-002",
    doi = "10.1088/0264-9381/23/4/014",
    journal = "Class. Quant. Grav.",
    volume = "23",
    pages = "1289--1346",
    year = "2006"
}

@article{Brown:1994py,
    author = "Brown, J. David and Kuchar, Karel V.",
    title = "{Dust as a standard of space and time in canonical quantum gravity}",
    eprint = "gr-qc/9409001",
    archivePrefix = "arXiv",
    doi = "10.1103/PhysRevD.51.5600",
    journal = "Phys. Rev. D",
    volume = "51",
    pages = "5600--5629",
    year = "1995"
}

@article{Giesel:2006um,
    author = "Giesel, K. and Thiemann, T.",
    title = "{Algebraic quantum gravity (AQG). III. Semiclassical perturbation theory}",
    eprint = "gr-qc/0607101",
    archivePrefix = "arXiv",
    reportNumber = "AEI-2006-60",
    doi = "10.1088/0264-9381/24/10/005",
    journal = "Class. Quant. Grav.",
    volume = "24",
    pages = "2565--2588",
    year = "2007"
}

@book{Rovelli:2014ssa,
    author = "Rovelli, Carlo and Vidotto, Francesca",
    title = "{Covariant Loop Quantum Gravity}: {An Elementary Introduction to Quantum Gravity and Spinfoam Theory}",
    isbn = "978-1-107-06962-6, 978-1-316-14729-0",
    publisher = "Cambridge University Press",
    series = "Cambridge Monographs on Mathematical Physics",
    month = "11",
    year = "2014"
}

@article{Ashtekar:2011ni,
    author = "Ashtekar, Abhay and Singh, Parampreet",
    title = "{Loop Quantum Cosmology: A Status Report}",
    eprint = "1108.0893",
    archivePrefix = "arXiv",
    primaryClass = "gr-qc",
    doi = "10.1088/0264-9381/28/21/213001",
    journal = "Class. Quant. Grav.",
    volume = "28",
    pages = "213001",
    year = "2011"
}

@book{Ashtekar:2017yom,
    editor = "Ashtekar, Abhay and Pullin, Jorge",
    title = "{Loop Quantum Gravity}: {The First 30 Years}",
    doi = "10.1142/10445",
    isbn = "978-981-320-992-3, 978-981-322-001-0, 978-981-320-993-0",
    publisher = "World Scientific",
    series = "100 Years of General Relativity",
    volume = "4",
    year = "2017"
}

@article{Ashtekar:2008zu,
    author = "Ashtekar, Abhay",
    title = "{Loop Quantum Cosmology: An Overview}",
    eprint = "0812.0177",
    archivePrefix = "arXiv",
    primaryClass = "gr-qc",
    doi = "10.1007/s10714-009-0763-4",
    journal = "Gen. Rel. Grav.",
    volume = "41",
    pages = "707--741",
    year = "2009"
}

@article{Thiemann:1996aw,
    author = "Thiemann, T.",
    title = "{Quantum spin dynamics (QSD)}",
    eprint = "gr-qc/9606089",
    archivePrefix = "arXiv",
    reportNumber = "HUTMP-96-B-351A",
    doi = "10.1088/0264-9381/15/4/011",
    journal = "Class. Quant. Grav.",
    volume = "15",
    pages = "839--873",
    year = "1998"
}

@article{Thiemann:1996av,
    author = "Thiemann, T.",
    title = "{Quantum spin dynamics (qsd). 2.}",
    eprint = "gr-qc/9606090",
    archivePrefix = "arXiv",
    reportNumber = "HUTMP-96-B-352",
    doi = "10.1088/0264-9381/15/4/012",
    journal = "Class. Quant. Grav.",
    volume = "15",
    pages = "875--905",
    year = "1998"
}

@article{Zhang:2019dgi,
    author = "Zhang, Cong and Lewandowski, Jerzy and Li, Haida and Ma, Yongge",
    title = "{Bouncing evolution in a model of loop quantum gravity}",
    eprint = "1904.07046",
    archivePrefix = "arXiv",
    primaryClass = "gr-qc",
    doi = "10.1103/PhysRevD.99.124012",
    journal = "Phys. Rev. D",
    volume = "99",
    number = "12",
    pages = "124012",
    year = "2019"
}

@article{Han:2019vpw,
    author = "Han, Muxin and Liu, Hongguang",
    title = "{Effective Dynamics from Coherent State Path Integral of Full Loop Quantum Gravity}",
    eprint = "1910.03763",
    archivePrefix = "arXiv",
    primaryClass = "gr-qc",
    doi = "10.1103/PhysRevD.101.046003",
    journal = "Phys. Rev. D",
    volume = "101",
    number = "4",
    pages = "046003",
    year = "2020"
}

@article{Dapor:2017rwv,
    author = "Dapor, Andrea and Liegener, Klaus",
    title = "{Cosmological Effective Hamiltonian from full Loop Quantum Gravity Dynamics}",
    eprint = "1706.09833",
    archivePrefix = "arXiv",
    primaryClass = "gr-qc",
    doi = "10.1016/j.physletb.2018.09.005",
    journal = "Phys. Lett. B",
    volume = "785",
    pages = "506--510",
    year = "2018"
}

@article{Han:2020uhb,
    author = "Han, Muxin and Liu, Hongguang",
    title = "{Improved effective dynamics of loop-quantum-gravity black hole and Nariai limit}",
    eprint = "2012.05729",
    archivePrefix = "arXiv",
    primaryClass = "gr-qc",
    doi = "10.1088/1361-6382/ac44a0",
    journal = "Class. Quant. Grav.",
    volume = "39",
    number = "3",
    pages = "035011",
    year = "2022"
}

@article{giesel2010algebraic,
  author = "Giesel, K. and Thiemann, T.",
    title = "{Algebraic quantum gravity (AQG). IV. Reduced phase space quantisation of loop quantum gravity}",
    eprint = "0711.0119",
    archivePrefix = "arXiv",
    primaryClass = "gr-qc",
    reportNumber = "AEI-2007-152",
    doi = "10.1088/0264-9381/27/17/175009",
    journal = "Class. Quant. Grav.",
    volume = "27",
    pages = "175009",
    year = "2010"
}

@article{thiemann2006reduced,
  author = "Thiemann, Thomas",
    title = "{Reduced phase space quantization and Dirac observables}",
    eprint = "gr-qc/0411031",
    archivePrefix = "arXiv",
    reportNumber = "AEI-2004-103",
    doi = "10.1088/0264-9381/23/4/006",
    journal = "Class. Quant. Grav.",
    volume = "23",
    pages = "1163--1180",
    year = "2006"
}

@article{Thiemann:1996au,
    author = "Thiemann, T.",
    title = "{Closed formula for the matrix elements of the volume operator in canonical quantum gravity}",
    eprint = "gr-qc/9606091",
    archivePrefix = "arXiv",
    reportNumber = "HUTMP-96-B-353",
    doi = "10.1063/1.532259",
    journal = "J. Math. Phys.",
    volume = "39",
    pages = "3347--3371",
    year = "1998"
}

@article{PhysRevLett.69.237,
  title = {Weaving a classical metric with quantum threads},
  author = {Ashtekar, Abhay and Rovelli, Carlo and Smolin, Lee},
  journal = {Phys. Rev. Lett.},
  volume = {69},
  issue = {2},
  pages = {237--240},
  numpages = {0},
  year = {1992},
  month = {Jul},
  publisher = {American Physical Society},
  doi = {10.1103/PhysRevLett.69.237},
  url = {https://link.aps.org/doi/10.1103/PhysRevLett.69.237}
}

@article{HALL1994103,
title = {The Segal-Bargmann "Coherent State" Transform for Compact Lie Groups},
journal = {Journal of Functional Analysis},
volume = {122},
number = {1},
pages = {103-151},
year = {1994},
issn = {0022-1236},
doi = {https://doi.org/10.1006/jfan.1994.1064},
url = {https://www.sciencedirect.com/science/article/pii/S0022123684710640},
author = {B.C. Hall},
}

@article{Thiemann:2000bw,
    author = "Thiemann, Thomas",
    title = "{Gauge field theory coherent states (GCS): 1. General properties}",
    eprint = "hep-th/0005233",
    archivePrefix = "arXiv",
    reportNumber = "AEI-2000-027",
    doi = "10.1088/0264-9381/18/11/304",
    journal = "Class. Quant. Grav.",
    volume = "18",
    pages = "2025--2064",
    year = "2001"
}

@article{Thiemann:2000ca,
    author = "Thiemann, T. and Winkler, O.",
    title = "{Gauge field theory coherent states (GCS). 2. Peakedness properties}",
    eprint = "hep-th/0005237",
    archivePrefix = "arXiv",
    reportNumber = "AEI-2000-028",
    doi = "10.1088/0264-9381/18/14/301",
    journal = "Class. Quant. Grav.",
    volume = "18",
    pages = "2561--2636",
    year = "2001"
}

@article{Thiemann:2000bx,
    author = "Thiemann, T. and Winkler, O.",
    title = "{Gauge field theory coherent states (GCS): 3. Ehrenfest theorems}",
    eprint = "hep-th/0005234",
    archivePrefix = "arXiv",
    reportNumber = "AEI-2000-029",
    doi = "10.1088/0264-9381/18/21/315",
    journal = "Class. Quant. Grav.",
    volume = "18",
    pages = "4629--4682",
    year = "2001"
}

@article{Thiemann:2000by,
    author = "Thiemann, T. and Winkler, O.",
    title = "{Gauge field theory coherent states (GCS) 4: Infinite tensor product and thermodynamical limit}",
    eprint = "hep-th/0005235",
    archivePrefix = "arXiv",
    reportNumber = "AEI-2000-030",
    doi = "10.1088/0264-9381/18/23/302",
    journal = "Class. Quant. Grav.",
    volume = "18",
    pages = "4997--5054",
    year = "2001"
}

@article{Sahlmann:2001nv,
    author = "Sahlmann, H. and Thiemann, T. and Winkler, O.",
    title = "{Coherent states for canonical quantum general relativity and the infinite tensor product extension}",
    eprint = "gr-qc/0102038",
    archivePrefix = "arXiv",
    reportNumber = "AEI-2001-011",
    doi = "10.1016/S0550-3213(01)00226-7",
    journal = "Nucl. Phys. B",
    volume = "606",
    pages = "401--440",
    year = "2001"
}

@article{Thiemann:2002vj,
    author = "Thiemann, Thomas",
    title = "{Complexifier coherent states for quantum general relativity}",
    eprint = "gr-qc/0206037",
    archivePrefix = "arXiv",
    reportNumber = "AEI-2002-045",
    doi = "10.1088/0264-9381/23/6/013",
    journal = "Class. Quant. Grav.",
    volume = "23",
    pages = "2063--2118",
    year = "2006"
}

@article{Bahr:2007xa,
    author = "Bahr, Benjamin and Thiemann, Thomas",
    title = "{Gauge-invariant coherent states for Loop Quantum Gravity. I. Abelian gauge groups}",
    eprint = "0709.4619",
    archivePrefix = "arXiv",
    primaryClass = "gr-qc",
    doi = "10.1088/0264-9381/26/4/045011",
    journal = "Class. Quant. Grav.",
    volume = "26",
    pages = "045011",
    year = "2009"
}

@article{Bahr:2007xn,
    author = "Bahr, Benjamin and Thiemann, Thomas",
    title = "{Gauge-invariant coherent states for loop quantum gravity. II. Non-Abelian gauge groups}",
    eprint = "0709.4636",
    archivePrefix = "arXiv",
    primaryClass = "gr-qc",
    doi = "10.1088/0264-9381/26/4/045012",
    journal = "Class. Quant. Grav.",
    volume = "26",
    pages = "045012",
    year = "2009"
}

@article{PhysRevD.92.104023,
  title = {Coherent state operators in loop quantum gravity},
  author = {Alesci, Emanuele and Dapor, Andrea and Lewandowski, Jerzy and M\"akinen, Ilkka and Sikorski, Jan},
  journal = {Phys. Rev. D},
  volume = {92},
  issue = {10},
  pages = {104023},
  numpages = {21},
  year = {2015},
  month = {Nov},
  publisher = {American Physical Society},
  doi = {10.1103/PhysRevD.92.104023},
  url = {https://link.aps.org/doi/10.1103/PhysRevD.92.104023}
}

@article{Rovelli:1994ge,
    author = "Rovelli, Carlo and Smolin, Lee",
    title = "{Discreteness of area and volume in quantum gravity}",
    eprint = "gr-qc/9411005",
    archivePrefix = "arXiv",
    reportNumber = "CGPG-94-11-1",
    doi = "10.1016/0550-3213(95)00150-Q",
    journal = "Nucl. Phys. B",
    volume = "442",
    pages = "593--622",
    year = "1995",
    note = "[Erratum: Nucl.Phys.B 456, 753--754 (1995)]"
}

@article{Ashtekar:1997fb,
    author = "Ashtekar, Abhay and Lewandowski, Jerzy",
    title = "{Quantum theory of geometry. 2. Volume operators}",
    eprint = "gr-qc/9711031",
    archivePrefix = "arXiv",
    reportNumber = "CGPG-97-11-1",
    doi = "10.4310/ATMP.1997.v1.n2.a8",
    journal = "Adv. Theor. Math. Phys.",
    volume = "1",
    pages = "388--429",
    year = "1998"
}

@book{hormander2015analysis,
  title={The analysis of linear partial differential operators I: Distribution theory and Fourier analysis},
  author={H{\"o}rmander, Lars},
  year={2015},
  publisher={Springer}
}

@article{Flori:2008nw,
    author = "Flori, C. and Thiemann, T.",
    title = "{Semiclassical analysis of the Loop Quantum Gravity volume operator. I. Flux Coherent States}",
    eprint = "0812.1537",
    archivePrefix = "arXiv",
    primaryClass = "gr-qc",
    month = "12",
    year = "2008"
}

@article{Ashtekar:1996eg,
    author = "Ashtekar, Abhay and Lewandowski, Jerzy",
    title = "{Quantum theory of geometry. 1: Area operators}",
    eprint = "gr-qc/9602046",
    archivePrefix = "arXiv",
    reportNumber = "CGPG-96-2-4",
    doi = "10.1088/0264-9381/14/1A/006",
    journal = "Class. Quant. Grav.",
    volume = "14",
    pages = "A55--A82",
    year = "1997"
}

@article{Ashtekar:2004eh,
    author = "Ashtekar, Abhay and Lewandowski, Jerzy",
    title = "{Background independent quantum gravity: A Status report}",
    eprint = "gr-qc/0404018",
    archivePrefix = "arXiv",
    doi = "10.1088/0264-9381/21/15/R01",
    journal = "Class. Quant. Grav.",
    volume = "21",
    pages = "R53",
    year = "2004"
}

@article{Bojowald:2001xe,
    author = "Bojowald, Martin",
    title = "{Absence of singularity in loop quantum cosmology}",
    eprint = "gr-qc/0102069",
    archivePrefix = "arXiv",
    reportNumber = "CGPG-01-2-1",
    doi = "10.1103/PhysRevLett.86.5227",
    journal = "Phys. Rev. Lett.",
    volume = "86",
    pages = "5227--5230",
    year = "2001"
}

@article{Banerjee:2011qu,
    author = "Banerjee, Kinjal and Calcagni, Gianluca and Martin-Benito, Mercedes",
    title = "{Introduction to loop quantum cosmology}",
    eprint = "1109.6801",
    archivePrefix = "arXiv",
    primaryClass = "gr-qc",
    doi = "10.3842/SIGMA.2012.016",
    journal = "SIGMA",
    volume = "8",
    pages = "016",
    year = "2012"
}

@article{Ashtekar:2018lag,
    author = "Ashtekar, Abhay and Olmedo, Javier and Singh, Parampreet",
    title = "{Quantum Transfiguration of Kruskal Black Holes}",
    eprint = "1806.00648",
    archivePrefix = "arXiv",
    primaryClass = "gr-qc",
    doi = "10.1103/PhysRevLett.121.241301",
    journal = "Phys. Rev. Lett.",
    volume = "121",
    number = "24",
    pages = "241301",
    year = "2018"
}

@article{Taveras:2008ke,
    author = "Taveras, Victor",
    title = "{Corrections to the Friedmann Equations from LQG for a Universe with a Free Scalar Field}",
    eprint = "0807.3325",
    archivePrefix = "arXiv",
    primaryClass = "gr-qc",
    reportNumber = "IGC-08-7-2",
    doi = "10.1103/PhysRevD.78.064072",
    journal = "Phys. Rev. D",
    volume = "78",
    pages = "064072",
    year = "2008"
}

@inbook{Agullo:2016tjh,
    author = "Agullo, Ivan and Singh, Parampreet",
    editor = "Ashtekar, Abhay and Pullin, Jorge",
    title = "{Loop quantum cosmology.}",
    booktitle = "{Loop Quantum Gravity}: {The First 30 Years}",
    eprint = "1612.01236",
    archivePrefix = "arXiv",
    primaryClass = "gr-qc",
    reportNumber = "LSU-REL-12160416",
    doi = "10.1142/9789813220003_0007",
    publisher = "WSP",
    pages = "183--240",
    year = "2017"
}

@article{Husain:2011tk,
    author = "Husain, Viqar and Pawlowski, Tomasz",
    title = "{Time and a physical Hamiltonian for quantum gravity}",
    eprint = "1108.1145",
    archivePrefix = "arXiv",
    primaryClass = "gr-qc",
    doi = "10.1103/PhysRevLett.108.141301",
    journal = "Phys. Rev. Lett.",
    volume = "108",
    pages = "141301",
    year = "2012"
}

@article{Brunnemann:2007ca,
    author = "Brunnemann, Johannes and Rideout, David",
    title = "{Properties of the volume operator in loop quantum gravity. I. Results}",
    eprint = "0706.0469",
    archivePrefix = "arXiv",
    primaryClass = "gr-qc",
    reportNumber = "IMPERIAL-TP-2007-DR-01",
    doi = "10.1088/0264-9381/25/6/065001",
    journal = "Class. Quant. Grav.",
    volume = "25",
    pages = "065001",
    year = "2008"
}

@article{DePietri:1996tvo,
    author = "De Pietri, Roberto and Rovelli, Carlo",
    title = "{Geometry eigenvalues and scalar product from recoupling theory in loop quantum gravity}",
    eprint = "gr-qc/9602023",
    archivePrefix = "arXiv",
    reportNumber = "UPRF-96-444",
    doi = "10.1103/PhysRevD.54.2664",
    journal = "Phys. Rev. D",
    volume = "54",
    pages = "2664--2690",
    year = "1996"
}

@article{Brunneman:2007as,
    author = "Brunneman, Johannes and Rideout, David",
    title = "{Properties of the volume operator in loop quantum gravity. II. Detailed presentation}",
    eprint = "0706.0382",
    archivePrefix = "arXiv",
    primaryClass = "gr-qc",
    reportNumber = "IMPERIAL-TP-2007-DR-02",
    doi = "10.1088/0264-9381/25/6/065002",
    journal = "Class. Quant. Grav.",
    volume = "25",
    pages = "065002",
    year = "2008"
}

@article{Bianchi:2010gc,
    author = "Bianchi, Eugenio and Dona, Pietro and Speziale, Simone",
    title = "{Polyhedra in loop quantum gravity}",
    eprint = "1009.3402",
    archivePrefix = "arXiv",
    primaryClass = "gr-qc",
    doi = "10.1103/PhysRevD.83.044035",
    journal = "Phys. Rev. D",
    volume = "83",
    pages = "044035",
    year = "2011"
}

@article{Han:2019feb,
    author = "Han, Muxin and Liu, Hongguang",
    title = "{Improved $\overline{\mu}$-scheme effective dynamics of full loop quantum gravity}",
    eprint = "1912.08668",
    archivePrefix = "arXiv",
    primaryClass = "gr-qc",
    doi = "10.1103/PhysRevD.102.064061",
    journal = "Phys. Rev. D",
    volume = "102",
    number = "6",
    pages = "064061",
    year = "2020"
}

@article{Han:2020chr,
    author = "Han, Muxin and Liu, Hongguang",
    title = "{Semiclassical limit of new path integral formulation from reduced phase space loop quantum gravity}",
    eprint = "2005.00988",
    archivePrefix = "arXiv",
    primaryClass = "gr-qc",
    doi = "10.1103/PhysRevD.102.024083",
    journal = "Phys. Rev. D",
    volume = "102",
    number = "2",
    pages = "024083",
    year = "2020"
}

@article{Bodendorfer:2020ovt,
    author = "Bodendorfer, Norbert and Han, Muxin and Haneder, Fabian and Liu, Hongguang",
    title = "{Path integral renormalization in loop quantum cosmology}",
    eprint = "2012.02068",
    archivePrefix = "arXiv",
    primaryClass = "gr-qc",
    doi = "10.1103/PhysRevD.103.126021",
    journal = "Phys. Rev. D",
    volume = "103",
    number = "12",
    pages = "126021",
    year = "2021"
}

@article{Han:2021cwb,
    author = "Han, Muxin and Liu, Hongguang",
    title = "{Loop quantum gravity on dynamical lattice and improved cosmological effective dynamics with inflaton}",
    eprint = "2101.07659",
    archivePrefix = "arXiv",
    primaryClass = "gr-qc",
    doi = "10.1103/PhysRevD.104.024011",
    journal = "Phys. Rev. D",
    volume = "104",
    number = "2",
    pages = "024011",
    year = "2021"
}

@misc{WignerSymbols.jl,
  author = {Jutho Haegeman},
  title  = {WignerSymbols.jl: A {Julia} package for computing {Wigner} symbols and related quantities, https://github.com/Jutho/WignerSymbols.jl},
  year   = {2021},
  url    = {https://github.com/Jutho/WignerSymbols.jl},
}

@article{Liliu:202603A,
    author = "Li, Haida and Liu, Hongguang",
    title = "{Beyond Expectation Values: Generalized Semiclassical Expansions for Matrix Elements of Gauge Coherent States}",
    year = "2026"
}

@misc{LQG-Volume,
  author = {Li, Haida and Liu, Hongguang},
  title  = {LQG-Volume, https://github.com/QG-Westlake/LQG-Volume},
  year   = {2026},
  url    = {https://github.com/QG-Westlake/LQG-Volume},
}

@article{Berezin:1974du,
    author = "Berezin, F. A.",
    title = "{General Concept of Quantization}",
    reportNumber = "ITF-74-20E-MC",
    doi = "10.1007/BF01609397",
    journal = "Commun. Math. Phys.",
    volume = "40",
    pages = "153--174",
    year = "1975"
}

\end{document}